\documentclass[11pt]{gsasthesis} 

\usepackage{lmodern}
\usepackage{amsthm}
\usepackage{etex} 

\usepackage[margin={1.2in}]{geometry}

\usepackage[titletoc]{appendix}

\usepackage{rotating}

\usepackage{longtable}

\usepackage{natbib}

\usepackage{courier}

\usepackage[utf8]{inputenc}
\usepackage[T1]{fontenc}

\usepackage[english]{babel}
\usepackage{blindtext}

\usepackage{amsmath}

\usepackage{microtype}

\usepackage[stable,multiple]{footmisc}

\RequirePackage[font=small,format=plain,labelfont=bf,textfont=it]{caption}
\usepackage{multirow}
\usepackage{multicol}
\usepackage{hhline}
\usepackage{color, colortbl}
\usepackage{amssymb}
\usepackage{enumerate}

\usepackage{algorithmic}
\usepackage[ruled,linesnumbered, noend]{algorithm2e}
\usepackage[bookmarks=false, hidelinks]{hyperref}

\usepackage{mathtools}
\usepackage{color}

\usepackage{mathrsfs} 
\usepackage{bm}
\usepackage{graphicx}
\usepackage{caption}
\usepackage{subcaption}
\usepackage{mathabx}
\usepackage{dsfont}

\DeclareMathOperator{\Normal}{Normal}
\DeclareMathOperator{\Multinom}{Multinom}
\DeclareMathOperator{\Bernoulli}{Bern}
\DeclareMathOperator{\Poisson}{Poisson}
\DeclareMathOperator{\Dirichlet}{Dirichlet}
\DeclareMathOperator{\logit}{logit}

\title{Causal Inference Under Network Interference: \\ A Framework for Experiments on Social Networks} 
\author{\normalsize Edward K. Kao} 
\degreename{Doctor of Philosophy}
\degreefield{Statistics} 
\department{\normalsize The Department of Statistics} 
\degreemonth{January} 
\degreeyear{2017} 
\principaladvisor{Edoardo M. Airoldi}
\secondadvisor{Donald B. Rubin}

\begin{document}


\pagenumbering{roman} 



\thesistitlepage
\copyrightpage

\begin{abstract}
No man is an island, as individuals interact and influence one another daily in our society. When social influence takes place in experiments on a population of interconnected individuals, the treatment on a unit may affect the outcomes of other units, a phenomenon known as interference. This thesis develops a causal framework and inference methodology for experiments where interference takes place on a network of influence (i.e.\ network interference). In this framework, the network potential outcomes serve as the key quantity and flexible building blocks for causal estimands that represent a variety of primary, peer, and total treatment effects. These causal estimands are estimated via principled Bayesian imputation of missing outcomes. The theory on the unconfoundedness assumptions leading to simplified imputation highlights the importance of including relevant network covariates in the potential outcome model. Additionally, experimental designs that result in balanced covariates and sizes across treatment exposure groups further improve the causal estimate, especially by mitigating potential outcome model mis-specification. The true potential outcome model is not typically known in real-world experiments, so the best practice is to account for interference and confounding network covariates through both balanced designs and model-based imputation. A full factorial simulated experiment is formulated to demonstrate this principle by comparing performance across different randomization schemes during the design phase and estimators during the analysis phase, under varying network topology and true potential outcome models. Overall, this thesis asserts that interference is not just a nuisance for analysis but rather an opportunity for quantifying and leveraging peer effects in real-world experiments.
\end{abstract}

\renewcommand{\contentsname}{\protect\centering\protect\Large Contents}
\renewcommand{\listtablename}{\protect\centering\protect\Large List of Tables}
\renewcommand{\listfigurename}{\protect\centering\protect\Large List of Figures}

\setcounter{secnumdepth}{3}
\setcounter{tocdepth}{3}
\tableofcontents 

\listoftables
\listoffigures
\begin{acknowledgments}

I would like to first thank my advisors, Professor Edoardo Airoldi and Professor Donald Rubin. Professor Airoldi was the reason I came to Harvard Statistics and he led me into the exciting research area of causal inference and experimental design on networks. Through him, I met wonderful collaborators such as David Kim and Professor Nicholas Christakis who conduct cutting edge real-world experiments on networks. Professor Rubin educated me on the core insights of causal inference and spurred me on to define the fundamental quantities and theories that drive the framework and the experiments in this thesis. His questions and guidance gave me clarity and pointed me to the essence of the problem.

I also want to thank Professor Tirthankar Dasgupta for sharing his research on Bayesian experimental design, Professor Joe Blitzstein for discussion on network models and inference, and Professor Alan Agresti for educating me on GLMs. Panos Toulis is a dear colleague and friend to me. We shared not only our work but also our lives and countless memorable (often silly) moments. :-) Members of the Airoldi Lab over the years also gave me valuable feedback and ideas, especially Alexander Blocker, Alex D'Amour, Hossein Azari Soufiani, Simon Lunagomez, Daniel Sussman, Alex Franks, Alexander Volfovsky, and Guillaume Basse. I also cherish the friendships and many discussions with other PhD students, especially Samuel Wang, Yang Chen, Yang Li, Hyungsuk Tak, Avi Feller, and Viktoriya Krakovna.

MIT Lincoln Laboratory is my gracious employer who supported my study at Harvard financially through Lincoln Scholarship. Even more importantly was the research interest and training I received at Lincoln that led me to this study at Harvard. I would especially like to thank my mentors Michael Hurley, Gary Condon, Kenneth Senne, and Steven Smith for their support and guidance. Each of them played a critical role, without which I would not be able to have this opportunity to enter and complete this study. My work with Steve on networks continued throughout my time at Harvard and it is my honor to have him as a chair on my thesis committee, alongside my advisors.

I would also like to thank the National Defense Science and Engineering Graduate (NDSEG) Fellowship for funding the first three semesters of my study.

Daniel Morris, Professor John Sheppard, and Jack Riddle are my past mentors who each played an important role that led me into research in machine learning and statistical inference. 

I would be remiss not to thank the important people in my personal life. The folks at Antioch Baptist Church are my family here in Boston. Their love and prayers carried me through all the ups and downs. From them, I am nourished with the spiritual truth that sets me free and points me foremost to the eternal things. I am especially indebted to Pastor Paul, Becky Jdsn, Pastor Heechin, Pastor David, Angela Smn, Pastor Thomas, Pastor Sang, and Songae Jdsn.

My father Hwa-Perng and mother Hu-Shan instilled in me the value of learning and scientific curiosity. For my education and opportunity to grow, they gave up their lives and security in Taiwan and brought our family to the US. From observing their lives, I learned what it means to be passionate and to persevere for what one believes in.  

I am thankful for my wife Amy's support, prayers, and sacrifices that allowed me to finish this study, including carrying the family burdens and bearing with my statistics rants.

Last but not least, my deepest gratitude goes to God, whose love for me through Jesus gives me identity and worth through it all.

\end{acknowledgments}

\begin{dedication}
To my Lord and Savior, Jesus Christ, who is with me through it all.
\end{dedication}

\pagenumbering{arabic} 


\newcommand{\Like}{\mathcal{L}}
\newcommand{\FisherInfo}{\bm{\mathcal{I}}}
\newcommand{\N}{\mathcal{N}}
\newcommand{\G}{\mathcal{G}}
\newcommand{\VSet}{\mathcal{V}} 
\newcommand{\ESet}{\mathcal{E}}
\newcommand{\ASet}{\mathcal{A}}
\newcommand{\ASetClosed}{\mathcal{A}_{\N_{i}}}
\newcommand{\ASetOpen}{\mathcal{A}_{\N_{-i}}}
\newcommand{\ZkSet}{\mathcal{Z}_{\N_{-i}}^k}
\newcommand{\ZlSet}{\mathcal{Z}_{\N_{-i}}^l}
\newcommand{\ZSetClosed}{\mathcal{Z}_{\N_{i}}}
\newcommand{\ZSetOpen}{\mathcal{Z}_{\N_{-i}}}
\newcommand{\A}{\bm{A}}
\newcommand{\B}{\bm{B}}
\newcommand{\Z}{\bm{Z}}
\newcommand{\Zclosed}{\Z_{\N_{i}}}
\newcommand{\Zopen}{\Z_{\N_{-i}}}
\newcommand{\YSet}{\mathds{Y}}
\newcommand{\YmisSet}{\mathds{Y}_\text{\normalfont mis}}
\newcommand{\YobsSet}{\mathds{Y}_\text{\normalfont obs}}
\newcommand{\Y}{\bm{Y}}
\newcommand{\z}{\bm{z}}
\newcommand{\TSet}{\mathcal{T}}
\newcommand{\KSet}{\mathcal{K}}
\newcommand{\M}{\mathcal{M}}
\newcommand{\DD}{\mathcal{D}}

\newcommand{\w}{\textbf{w}}
\newcommand{\ave}{\text{\normalfont ave}}
\newcommand{\fix}{\text{\normalfont fix}}
\newcommand{\mis}{\text{\normalfont mis}}
\newcommand{\obs}{\text{\normalfont obs}}
\newcommand{\expected}{\text{\normalfont exp}}
\newcommand{\Assumption}{\textbf{Assumption}}
\newcommand{\Theorem}{\textbf{Theorem}}
\newcommand{\Notation}{\textbf{Notation}}
\newcommand{\Definition}{\textbf{Definition}}
\newcommand{\Zall}{\Z(\N_i; k)}
\newcommand{\Zone}{\Z_1(\N_i; k)}
\newcommand{\Zo}{\Z_0(\N_i; k)}
\newcommand{\allo}{\textbf{0}}
\renewcommand\refname{}

\newcommand{\y}{\bm{y}}
\newcommand{\gi}{\textbf{g}_i}
\newcommand{\ai}{\bm{a}_i}
\newcommand{\bb}{\bm{\beta}}
\newcommand{\X}{\bm{X}}
\renewcommand{\SS}{\bm{S}}
\newcommand{\kb}{\bm{\kappa}}

\newcommand{\inrc}{INR^{0}}
\newcommand{\inrm}{INR^{0.6}}
\newcommand{\inrp}{INR^{1.0}}
\newcommand{\lm}{LMR}
\newcommand{\lmopt}{LMO}
\newcommand{\inrx}{\texttt{INR}^x}
\newcommand{\EDIT}{\textbf{xx EDIT xx } }

\newcommand{\bv}{\begin{array}}
\newcommand{\myfig}[1]{Figure~\ref{fig:#1}}
\newcommand{\mysec}[1]{Section~\ref{sec:#1}}
\newcommand{\myeq}[1]{Equation~\ref{eq:#1}}
\newcommand{\g}{\, | \,}
\newcommand{\mybox}[4]{\centering
\colorbox{lightgrey}{\parbox{#1\textwidth}{\centering \vskip #3
\begin{minipage}[t]{#2\textwidth} #4 \end{minipage} \vskip #3}}}

\newcommand{\diag}[1]{\mathrm{diag}\left(#1\right)}
\newcommand{\tr}[1]{\mathrm{tr}\left(#1\right)}
\newcommand{\diagvec}[1]{\mathrm{\overrightarrow{diag}}\left(#1\right)}
\newcommand{\unitmat}[0]{\vec{1}\vec{1}^T}
\newcommand{\dd}{\; \mathrm{d}}
\newcommand{\range}[2]{\left\{#1 \ldots #2\right\}}
\newcommand{\fnDef}[3]{#1 \,:\, #2 \rightarrow #3}

\newtheorem{example}{Example}[chapter]
\newtheorem{conjecture}{Conjecture}[chapter]
\newtheorem{remark}{Remark}[chapter]
\newtheorem{theorem}{Theorem}[chapter]
\newtheorem{lemma}{Lemma}[chapter]
\newtheorem{fact}{Fact}[chapter]
\newtheorem{note}{Note}[chapter]
\newtheorem{definition}{Definition}[chapter]
\newtheorem{assumption}{Assumption}[chapter]
\newtheorem{problem}{Problem}[chapter]

\renewcommand\floatpagefraction{.9}
\renewcommand\topfraction{.9}
\renewcommand\bottomfraction{.9}
\renewcommand\textfraction{.1} 
\renewcommand{\vec}[1]{\boldsymbol{\bm{#1}}}
\newcommand{\mat}[1]{\boldsymbol{\bm{#1}}}

\newcommand{\specialcell}[2]{%
  \begin{minipage}{#1 cm} \vspace{0.3cm} \centering #2 \vspace{0.2cm} \end{minipage}}


\def\Deltab{{\bm{\Delta}}}
\def\Gammab{{\bm{\Gamma}}}
\def\Psib{{\bm{\Psi}}}
\def\Pib{{\bm{\Pi}}}
\def\Lambdab{{\bm{\Lambda}}}
\def\Sigmab{{\bm{\Sigma}}}
\def\Xib{{\bm{\Xi}}}
\def\Thetab{{\bm{\Theta}}}
\def\Upsilonb{{\bm{\Upsilon}}}
\def\Omegab{{\bm{\Omega}}}

\def\alphab{{\bm{\alpha}}}
\def\gammab{{\bm{\gamma}}}
\def\thetab{{\bm{\theta}}}
\def\etab{{\bm{\eta}}}
\def\deltab{{\bm{\delta}}}
\def\lambdab{{\bm{\lambda}}}
\def\rhob{{\bm{\rho}}}
\def\phib{{\bm{\phi}}}
\def\pib{{\bm{\pi}}}
\def\psib{{\bm{\psi}}}
\def\varthetab{{\bm{\vartheta}}}
\def\xib{{\bm{\xi}}}
\def\omegab{{\bm{\omega}}}

\def\thetahatb{{\hat\thetab}}
\def\phihatb{{\hat\phib}}

\def\Ab{{\A}}
\def\Bb{{\bm{B}}}
\def\Cb{{\bm{C}}}
\def\Db{{\bm{D}}}
\def\Eb{{\bm{E}}}
\def\Hb{{\bm{H}}}
\def\Gb{{\bm{G}}}
\def\Ib{{\bm{I}}}
\def\Jb{{\bm{J}}}
\def\Kb{{\bm{K}}}
\def\Lb{{\bm{L}}}
\def\Mb{{\bm{M}}}
\def\Pb{{\bm{P}}}
\def\Qb{{\bm{Q}}}
\def\Rb{{\bm{R}}}
\def\Sb{{\bm{S}}}
\def\Tb{{\bm{T}}}
\def\Ub{{\bm{U}}}
\def\Vb{{\bm{V}}}
\def\Wb{{\bm{W}}}
\def\Xb{{\bm{X}}}
\def\Yb{{\bm{Y}}}
\def\Zb{{\Z}}

\def\ab{{\bm{a}}}
\def\bb{{\bm{b}}}
\def\cb{{\bm{c}}}
\def\db{{\bm{d}}}
\def\eb{{\bm{e}}}
\def\gb{{\bm{g}}}
\def\hb{{\bm{h}}}
\def\lb{{\bm{l}}}
\def\nb{{\bm{n}}}
\def\pb{{\bm{p}}}
\def\qb{{\bm{q}}}
\def\rb{{\bm{r}}}
\def\sb{{\bm{s}}}
\def\tb{{\bm{t}}}
\def\ub{{\bm{u}}}
\def\vb{{\bm{v}}}
\def\wb{{\bm{w}}}
\def\xb{{\bm{x}}}
\def\yb{{\bm{y}}}
\def\zb{{\bm{z}}}

\def\zerob{{\bm{0}}}
\def\oneb{{\bm{1}}}

\def\by{\ifmmode $\hbox{-by-}$\else \leavevmode\hbox{-by-}\fi}

\definecolor{Pink}{rgb}{1,0.85,0.85}
\definecolor{Yellow}{rgb}{1,1,0}

\chapter*{Introduction}\label{ch:intro}
\addcontentsline{toc}{chapter}{Introduction}
Principled causal experiments have been the gold standard for testing the efficacy of treatment in many real-world applications, such as medicine, public and health policy, epidemiology, education, etc. Typically, causal analysis assumes that the outcome of a unit is only affected by its own treatment, and not by the treatments on other units. However, this assumption does not hold in the presence of social influence between units. This phenomenon, known as interference, happens when the outcome of a unit $i$ is affected not only by its own treatment but also by the treatments on other units. Commonly, interference is assumed away or dealt with as a nuisance. However, with the rise of social networks and media, there is an ever increasing number of experiments conducted on social networks where some information about the network is available. This trend presents a unique challenge as well as opportunity to quantify and leverage interference. In addition to the commercial and civil applications such as marketing, propaganda spreading on social media by terrorist organizations and nation states \citep{TerrorOnTwitter, RussianTempering} presents an emerging threat to national security. Understanding and measuring such social influence with rigorous experiments and analysis is a crucial step towards harnessing the social effects through conducting or countering campaigns on social networks.  

This thesis presents a novel causal inference framework that accounts for interference on a social network, as illustrated in Figure~\ref{fig:networkInterferenceDiagram}. 
\begin{figure}[ht]
     \centering
     \includegraphics[width=1\textwidth]{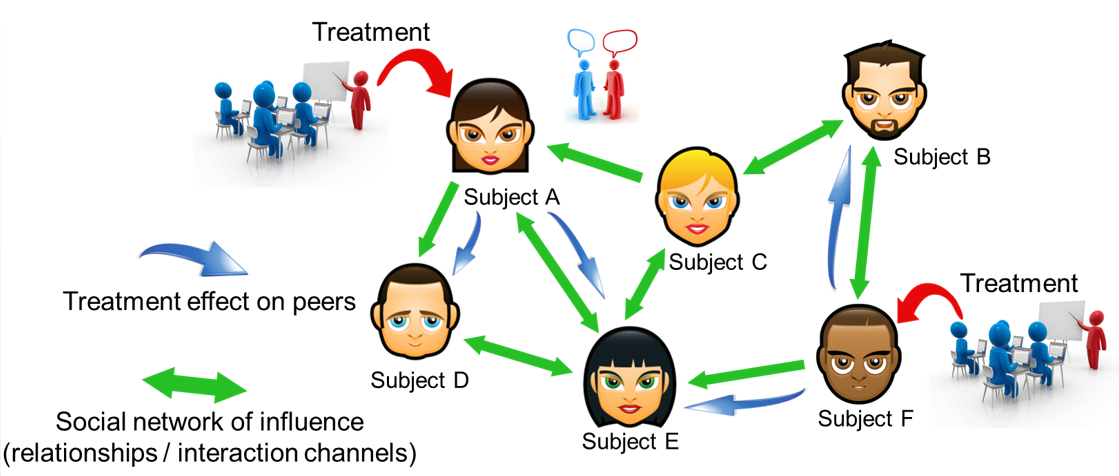}
     \caption[Network interference illustration]{Experiment where interference takes place on a social network. The treatment, an educational program in this example, affects not only those who receive it but also their peers through word of mouth.}
     \label{fig:networkInterferenceDiagram}
\end{figure}
Under this framework,  interference is explicitly represented and its effect quantified, providing causal estimates of peer effects and ways to leverage social influence in achieving treatment objectives, such as greater overall effect on the population. Building upon the widely used Rubin causal framework \citep{ImbensRubin2015, neyman1923, rubin1974estimating}, Chapter \ref{ch:1} proposes the network potential outcome framework and its rich variety of causal estimands that capture the primary, peer, and overall effects. It also develops a theory for principled Bayesian imputation of missing outcomes for estimating the causal estimands, accounting for the relevant network covariates as confounders. In practice, this can be best accomplished by both experimental designs that result in covariate balance across treatment exposure groups, and by including the relevant confounders in the potential outcome model for imputation. To demonstrate this point, Chapter \ref{ch:2} lays out a full factorial design for a simulation study to highlight the advantage of addressing interference and network confounders at both the design phase and the analysis phase, especially in the presence of potential outcome model mis-specifications. Robustness to mis-specification is important because the true potential outcome model is not typically known in real-world experiments. Chapter \ref{ch:3} takes a detour from the core of the causal framework and develops a rich and realistic candidate model for the influence network. Its focus is more computational and specifically on the estimation of the network covariates that play an important role in the principled causal estimation procedures laid out in Chapter \ref{ch:1} and \ref{ch:2}. Each chapter either performs or proposes simulated experiments to evaluate and demonstrate the proposed methods. Almost all of the methods are Bayesian and the evaluation is on their frequentist properties in repeated experiments, such as truth coverage, average posterior interval width, and expected mean square error.    

\chapter[Network Potential Outcome Framework with Bayesian Imputation]{\huge Network Potential Outcome Framework \\ with Bayesian Imputation}\label{ch:1}
\section{Overview}
In causal experiments, interference takes place when the treatment on a unit affects not only its own outcome (i.e.\ primary treatment effect) but also the outcomes of other units due to their interactions and relationships (i.e.\ peer treatment effect). A common example is when the units are individuals in a social network. Existing causal inference procedures typically assume away interference. This chapter introduces a framework that accounts for interference between units in a network of influence, known as network interference. First, network potential outcomes are defined as the flexible building blocks for causal estimands that represent a variety of primary, peer, and total treatment effects. Under network interference, most of the potential outcomes will be unobserved and a complete randomization over all possible peer treatment assignment vectors is infeasible, making the estimation of the causal estimands challenging. A natural solution is to perform Bayesian imputation of missing potential outcomes. This chapter develops a theoretical framework, based on the critical assumptions of unconfounded treatment assignment mechanism and unconfounded influence network, to greatly simplify the imputation by allowing it to ignore the neighborhood treatment assignment mechanism. From these theoretical properties, a rigorous analytical procedure is proposed for Bayesian imputation under network interference. Finally, simulated experiments demonstrate how this framework leads to causal estimates with good frequentist coverage properties. Overall, the framework demonstrates that interference is not just a nuisance but rather an opportunity for quantifying and leveraging peer effects in many real-world experiments, including observational studies.

\section {Introduction}
\addtocontents{toc}{\protect\enlargethispage{2\baselineskip}}
\label{sec:introduction1}

Interference, in the context of causal inference, refers to the situation when the outcome of a unit $i$ is affected not only by its own treatment, but also by the treatments of other units. This may take place due to interactions and relationships between the units. Common examples are experiments conducted on a social network where the treatment or intervention affects the outcomes of not only the individuals who receive it, but also the individuals who are nearby in the social network through peer influence, spread of knowledge, or social benefits, etc. This phenomenon is also known as spillover effects, peer influence effects, and social contagions. This section briefly surveys the existing work in application as well as in methodology in this area of research. Although there is a rising number of applications, methodological work is still emerging and remains an open area of research. This section concludes by summarizing the contributions in this chapter and how they address the open methodological challenges.

\subsection{Background}
Causal inference under interference has been drawing an increasing amount of interest, especially over the last ten years, due to an ever increasing number of applications and real-world experiments. Correspondingly, methodological work has been emerging to address the technical challenges of interference, of which many remain open.

\subsubsection{Rising Applications}
Traditionally, interference has been viewed as a nuisance in causal experiments. However, its presence can not be ignored in an increasing number of real-world experiments. Moreover, in many such cases, interference effects are the main quantity of interest, especially in experiments conducted on social networks. The application areas are numerous, for example, in public health, education , and policy \citep{Basse2016, Kim2015, Christakis, Sobel2006}, in social media and marketing \citep{Gui2015, bond2012, Bakshy2012, Parker}, in network security \citep{Coronges, Shah}, in economics \citep{Banerjee, Acemoglu, Manski}, and even in social propaganda threatening national security \citep{TerrorOnTwitter, RussianTempering}. To highlight a few, in \cite{Kim2015}, health intervention and education is distributed in small Honduras villages through a few seed participants. Social network information is collected beforehand through relationship nomination surveys. The experiment aims to determine the presence of peer effects and the optimal seeding strategy to maximize the overall effect.  In \cite{bond2012}, a massive experiment on Facebook aims to quantify peer influence on voting turnout through informational messages that simply encourage voting and social messages that also inform who in your friendship circles have voted. Estimating social effects have long been a topic of interest in social science. \cite{Manski} highlights the identifiability issues that arise when various forms of social effects are present together. The presence of interference in these experiments provide opportunities for answering interesting and practical questions, but also gives rise to additional challenges during the experimental design and analysis phase. Furthermore, interference when not properly accounted for, leads to biased causal estimates as warned and demonstrated in \cite{Sobel2006}.

\subsubsection{Methodological Work} 


Methodological work for causal inference under interference is still emerging, with some limited earlier works. Currently, there are still many open challenges and a lack of rigorous methodology for many real-world experiments, especially when the proper randomization may not be available or achieved (e.g. in observational studies).

Earlier works propose experimental designs that render the interference effects ignorable, but restrict the interference to simple block structures \citep{david1996designs, Azais1993}. However, most social networks exhibit more complicated structures, demanding an approach that can be applied to more general interference structure. Moreover, these designs focus on estimating the primary effect by controlling the interference, although many applications today are also interested in estimating the interference  effects (i.e.\ peer effects). Lastly, in many experiments such as observational studies, interference cannot be effectively controlled, and needs to be addressed in the analysis phase. 

More recent works attempt to tackle the presence of interference in more realistic settings, with some of them focusing specifically on the cases where interference takes place on a social network (i.e.\ network interference). Many of them build on the potential outcome framework \citep{ImbensRubin2015, neyman1923, rubin1974estimating} and relax the Stable Unit Treatment Value Assumption (SUTVA) to allow for each unit's outcome to depend on the treatments of other units. Extending the idea of Fisher's exact test to the more complicated setting under interference, some recent works propose null hypothesis testing for the presence of primary and peer treatment effects. Other recent works attempt to estimate causal estimands corresponding to the primary treatment effect as well as specific types of peer effects, via special randomization strategies. Some of these works are briefly summarized below.

Hypothesis testing of treatment effects is complicated by the presence of interference because most of the null hypotheses are no longer sharp. As a result, the usual Fisher exact tests with randomization are not directly applicable, and the randomization tests need to be tailored. \citet{rosenbaum2007} takes a non-parametric approach for testing the presence of primary effect and any effect at all, basing the null hypotheses on a uniformity trial. Using statistics that count the number of times treated units have higher responses than control units, this work accepts or rejects the null hypotheses without assuming any structure on the interference. \cite{Bowers2013} propose a general hypothesis testing framework under interference, on potential outcome models that hypothesizes the presence of certain primary effect and peer effects, especially under network interference. Also focusing on the cases under network interference, \cite{Athey2015} propose exact p-value tests for the presence of primary effect as well as peer effects of different orders (i.e.\ how far they spread). For each test, an artificial experiment is defined on the original experimental units such that the null hypothesis is sharp.

In many experiments, researchers want to quantify the causal effects instead of just testing for the presence of effects. There has been some methodological work on estimating various causal estimands under interference. \cite{hudgens2008} propose a two stage randomization method (i.e.\ group-level then individual-level) for estimating the average primary, peer, and total effects under a specific treatment strategy. Estimates of peer and total effects exist only at the population level. \cite{Aronow2012} propose a general estimation method using the inverse probability weighting on the various types of exposure to primary and peer treatments. The challenge with this method is that often it is difficult to compute the probability of a certain exposure to treatments for each unit, and the estimates can be unstable when the probability is very small. \cite{tchetgen2012} propose a model-based approach for estimating the probability of exposures, and \cite{Ugander2013} develop a cluster randomization approach that leads to a closed-form solution on the probabilities of certain neighborhood exposures under network interference.

Another methodological challenge is caused by the presence of confounders. Unidentifiability of the peer effects in the presence of social confounders (i.e.\ covariates shared by a group of units that impact the outcomes) has been shown earlier by \cite{Manski} under linear outcome models. In the context of causal inference under interference, more recent work has shown how confounding social covariates may cause unidentifiability and biased estimates of causal effects \citep{Vanderweele2014}, especially when the social effects take place on a social network \citep{ogburn2014, shalizi2011}. Some mitigation strategies have been proposed such as network clustering \citep{shalizi2011}, additional information from longitudinal outcome data \citep{omalley2014}, and sensitivity analysis \citep{Vanderweele2014}.

\subsection{Contributions}

The framework in this chapter targets causal inference under network interference, where interference arises from a network of influence between the experimental units. The network potential outcomes are introduced as the flexible building blocks for a wide range of causal estimands capturing various types of primary effects and peer effects, including the causal estimands proposed by existing work. An estimation framework is developed through model-based Bayesian imputation of missing potential outcomes. Unlike existing estimation methods for interference, it does not require specific randomizations nor estimating the probability of exposures to peer treatments. The key theoretical properties for Bayesian imputation are developed to resolve the issue of possible confounders. These properties, specifically the assumptions of unconfounded treatment assignment mechanism and unconfounded influence network, greatly simplify the imputation by allowing it to ignore the neighborhood treatment mechanism. These insights provide the theoretical framework for a principled approach for Bayesian imputation under network interference. Our framework is flexible and enables researchers, even in observational studies, to quantify a rich variety of primary, peer, and total effects down to the individual level. 

The chapter proceeds by introducing in Section \ref{sec:POUnderNetworkInteference} the network potential outcomes and the exclusion restriction assumption under network interference. Section \ref{sec:CEUnderNetworkInteference} constructs a few categories of causal estimands using the network potential outcomes. To estimate these estimands, the key theoretical properties for  Bayesian imputation of missing outcomes are developed in Section \ref{sec:BayesianImputationUnfoundedAssumptions}. Simulated experiments are conducted in Section \ref{sec:resultsOnSimulatedExperiments} to demonstrate the proposed causal framework and how estimation procedure motivated by the key theoretical properties leads to posterior causal estimates with good frequentist coverage properties.

\section {Potential Outcomes Under Network Interference}
\label{sec:POUnderNetworkInteference}

Causal inference under network interference is complicated by the fact that the outcomes of individuals are affected not only by their own treatment, but also by treatments to those ``around them'' (i.e.\ in their neighborhood). This is intuitive in many of the real-world applications where the treatment is applied to a population where individuals relate and interact with one another (e.g. a social network), such that the effects of the treatment may spill over as a result of peer influence. This thesis approaches this problem by building upon Rubin's framework \citep{ImbensRubin2015, rubin2005, neyman1923, rubin1974estimating}, where the potential outcomes serve as the basic units that compose the various causal estimands corresponding to the quantities of interest for each experiment. This section begins with a brief review of the regular potential outcomes and the Stable Unit Treatment Value Assumption (SUTVA) in the absence of interference. The potential outcome definition must be extended under general interference but can be simplified through the introduction of the Stable Unit Treatments on Neighborhood Value Assumption (SUTNVA). Finally, the network potential outcomes are introduced for interference that takes place on a network of influence between units. The network potential outcomes are the basic building blocks for a rich variety of causal estimands. They are also the key quantities to be estimated and imputed during the analysis phase.

\subsection{Potential Outcomes Under SUTVA}
Most existing works in causal inference assume an absence of interference and a simple binary treatment assignment, which is the SUTVA \citep{Rubin1980} as stated below:

\begin{assumption} [Stable Unit Treatment Value Assumption, i.e. SUTVA]
\label{as:SUTVA}
The potential outcomes for any units do not vary with the treatment assigned to other units, and, for each unit, there are no different forms or versions of each treatment level, which lead to different potential outcomes.
\end{assumption}

\noindent Under SUTVA, the potential outcome $Y_i$ of each unit $i$ in an experiment only depends on whether it receives the treatment ($Z_i=1$) or not ($Z_i=0$), with the widely used notation below:

\begin{definition}[Potential Outcomes Under SUTVA] 
\label{def:POUnderSUTVA}
Under SUTVA, the potential outcomes of a unit $i$ is denoted as $Y_i(Z_i)$, where $Z_i$ is a binary indicator for whether the treatment is assigned to unit $i$. The potential outcomes of all $N$ units in an experiment, $\Y$, can be partitioned into two vectors of $N$ components: $\bm{Y}(0)$ for all outcomes under control and $\bm{Y}(1)$ for all outcomes under treatment. The potential outcomes can also be partitioned according to whether it is observed. In an experiment, a unit is either under control or under treatment. Therefore, half of all potential outcomes are observed, denoted as $\bm{Y}_\obs$. The other half of the potential outcomes are unobserved, denoted as $\bm{Y}_\mis$.
\end{definition}

\noindent Under SUTVA, a natural causal estimand of interest is the population average treatment effect:
\begin{equation}
\label{eq:causalEstimandUnderSUTVA}
\tau^\ave = \frac{1}{N} \sum_{i=1}^N Y_i(1) - Y_i(0)
\end{equation}
In an ideal world, both $Y_i(1)$ and $ Y_i(0)$ are observed for each unit $i$. However in reality, as mentioned in Definition \ref{def:POUnderSUTVA}, one can only observe either $Y_i(1)$ or $ Y_i(0)$. This makes causal inference inherently a missing data problem, as half of the potential outcomes in Equation \eqref{eq:causalEstimandUnderSUTVA} are missing. For example, Table \ref{twoByTwoPOTable} lists the potential outcomes for an experiment with two units where unit one is in control and unit two under treatment.
\\
\begin{table}[h]
\begin{center}
\renewcommand{\arraystretch}{1.3}
\begin{tabular}{c | c c |  c c } \hline
Unit  & \multicolumn{2}{c}{``Ideal'' world} & \multicolumn{2}{c}{Real world} \\ \hline
 & $Z_i=0$ & $Z_i=1$ &  $Z_i=0$ & $Z_i=1$ \\
\hline
1 & $Y_1(0)$ & $Y_1(1)$ & $Y_1(0)$ & $ ?$ \\
2 & $Y_2(0)$ & $Y_2(1)$ & $?$ & $ Y_2(1)$ \\
\hline
\end{tabular}
\end{center}
\caption[Causal inference as a missing data problem]{Observed potential outcomes in an experiment with two units. The question marks denote missing outcomes.}
\label{twoByTwoPOTable}
\end{table}

\noindent This missing data problem is classically addressed through completely randomized treatment assignment which makes the Neyman estimator unbiased \citep{neyman1923}:
\begin{equation}
\label{eq:neymanEstimator}
\widehat{\tau}^\ave = \widebar{Y}^t_\obs - \widebar{Y}_\obs^c
\end{equation}
where $\widebar{Y}^t_\obs$ is the mean of the observed potential outcomes under treatment and $\widebar{Y}^c_\obs$ is the mean of the observed potential outcomes under control. Because each unit is equally likely to receive the treatment, the missing potential outcomes under treatment and control are directly imputed from the observed ones. The missing data problem can also be addressed through model-based Bayesian imputation, often in the context of observational studies. This alternative approach will be discussed later in Section \ref{sec:unconfoundedAssignmentRegular}.

\subsection{Potential Outcomes Under Interference and SUTNVA}
Under interference, a unit's outcome may be affected by the treatments on other units, due to peer influence. This breaks the commonly assumed SUTVA. In the most general case, the potential outcomes on a unit may be a function of the treatments on the entire population:
\begin{definition} [Potential Outcomes Under General Interference] 
\label{def:POGeneral}
Under general interference, the potential outcomes of a unit $i$ is denoted as $Y_i(\Z)$, a function of the treatment vector $\Z$ on the entire population.
\end{definition}
\noindent Including the treatments on the entire population in the potential outcomes notation for each unit $i$ is the most general, but does not take into account any of the structure of the peer influence. Often in reality, only the treatments on those in social proximity to unit $i$ may have an influence on it. The units inside this ``sphere of influence'' are members of the set $\N_i$, which is the closed neighborhood of unit $i$.  By definition, the closed neighborhood of $i$ includes $i$. Adding this constraint may simplify the potential outcomes to a function of much fewer treatment assignments, and therefore may greatly reduce the number of potential outcomes for each unit $i$. Formally, this can be expressed through the following exclusion restriction assumption:

\begin{assumption}[Stable Unit Treatments on Neighborhood Value Assumption, i.e. SUTNVA]
\label{as:SUTNVA}
The potential outcomes of a unit $i$ can only be affected by the treatments on units in its closed neighborhood $\N_i$:
\begin{equation}
	Y_i(\Z = \bm{z}) = Y_i(\Z = \bm{z'})  \;\;\;\; \text{for~all } (\bm{z}, \bm{z'}) \text{, if } z_j = z'_j \text{ for~all } j \in \N_i
\end{equation}
\end{assumption}

\noindent Note that SUTNVA does not restrict treatment assignments to binary (i.e.\ treatment and control). However, this thesis assumes the binary treatment case going forward in order to focus on the issue of interference. Our framework can be generalized for multi-valued treatments as future work, quite possibly by applying existing methods such as \cite{Imai2004}. Because under SUTNVA the potential outcomes for each unit $i$ do not depend on the treatments on units outside of its closed neighborhood $\N_i$, the following notation is used for potential outcomes with much fewer treatment arguments:

\begin{definition}[Potential Outcomes Under SUTNVA] 
\label{def:POUnderSUTNVA}
Under SUTNVA, the potential outcomes of a unit $i$ is denoted as $Y_i(\Z_{\N_i})$, a function of the treatment vector $\Z_{\N_i}$ on units inside $\N_i$, the closed neighborhood of influence for unit $i$.
\end{definition}

\noindent Sometimes, it is clearer to denote the treatment on certain units separately. For example, one may want to denote the treatment on unit $i$ itself. In such cases, the following notational convention is adopted:  $Y_i(\Z_{\N_i}) \equiv Y_i(Z_i, \Z_{\N_{-i}}) \equiv Y_i(Z_i, Z_j, \Z_{\N_{-i}-j})$, where $\N_{-i}$ is the open neighborhood of $i$  (note: open neighborhood of $i$ excludes $i$) and $\N_{-i}-j$ is the open neighborhood of $i$ excluding unit $j$, where unit $j \in \N_i$ .

In general, $\N_i$ may be the set of units with a certain relationship or context with $i$. For example: all the students in the same class as $i$. In the context of network interference where the structure of peer influence between units is captured by a network, $\N_i$ can be defined based on the topology of the network graph, for example: immediate neighbors of $i$, $n$-hop neighbors of $i$, etc. Formal definition and examples will be provided in Section \ref{subsec:POUnderSUTNVA}.

\subsection{Network Potential Outcomes Under SUTNVA}
\label{subsec:POUnderSUTNVA}
In many real-world experiments, interference arises from the influence between units due to their relationships and interactions. Such interference mechanism can be represented as a network graph where the nodes represent the units in the experiment and the edges represent the direction and strength of the influence. The causal framework in this thesis aims at exactly this type of experiments. This section introduces the network potential outcomes that serve as the basic building blocks going forward. First, the population of units under network interference is defined as below:

\begin{definition}[Finite Population in a Network of Influence]
\label{def:networkPopulation}
The experiment takes place on a finite population of $N$ units where the units' influence on each other is represented as a $N \times N$ influence matrix $\A$. Each element of the influence matrix, $A_{ij} \in \mathbb{R}$, represents the strength of influence unit $i$ has on unit $j$. One may visualize this population network as a graph, $\G=\{\VSet, \ESet\}$, of which the node set $\VSet$ consists of the units of the experiment ($|\VSet| = N$) and the edge set $\ESet$ represents the none-zero entries of the influence matrix $\A$.
\end{definition}

\noindent In this definition, the network of influence has weighted and directed edges because, in reality, influence is often not symmetrical and the strength of influence varies from one relationship to another. Of course, this definition also captures undirected as well as binary networks as special cases. The strength of influence can be both positive and negative because peer influence may take the positive form of persuasion and promotion as well as the negative form of suppression and prevention. The influence matrix, $\A$, is basically the weighted and directed adjacency matrix familiar to the network inference community. The network graph defined here for causal inference under network interference represents the influence between units, and does not imply causal relationships. In other words, it is not a causal graph.

Recall that under the Stable Unit Treatments on Neighborhood Value Assumption (SUTNVA), the potential outcomes are simplified as: $Y_i(\Z_{\N_i})$, a function of the treatments $\Z_{\N_i}$ on the closed neighborhood of $i$. Having defined the population on the network of influence, we can now define the neighborhood set $\N_i$. Under network interference, the network of influence is the mechanism and structure on which interference takes place, and therefore, naturally determines the closed neighborhood set in SUTNVA. An intuitive definition of $\N_i$, the ``sphere of influence'' on unit $i$, is to include all the units within a certain graph distance to $i$ (i.e.\ the $n$-hop neighborhood):

\begin{definition}[$n$-Hop Neighborhood for SUTNVA]
\label{def:nHopNeighborhood}
Given a finite population with the influence network defined previously by the influence matrix $\A$. The neighborhood $\N_i$ of unit $i$ under SUTNVA is the n-hop neighborhood $\N_i^{\leq n} = \{\N_i^0, \N_i^1, ... ,\N_i^n\}$, defined recursively as: $\N_i^0 = \{i\}, \; \N_i^{n+1} = \{j : \exists l \in \N_i^{n}\; \text{s.t.}\; A_{jl}\neq 0 \; \text{and} \; j \notin \N_i^{\leq n} \}$. Here $\N_i^n$, the $n$-hop neighbors, is the set of nodes with graph distance of exactly $n$ to node $i$. The $n$-hop neighborhood of $i$, $\N_i^{\leq n}$, includes all the nodes with a graph distance of $n$ or smaller to node $i$ (i.e.\ within $n$ hops to $i$).
\end{definition}

\noindent Choosing $n=0$ recovers the conventional SUTVA case where unit $i$'s potential outcome only depends on its own treatment. Choosing $n=1$ results in the potential outcomes on unit $i$ as a function of its own treatment and the treatments on its immediate neighbors, $\N_i^1$ (those that influence $i$). In similar fashion, increasing $n$ to two further adds the treatments on those units, $\N_i^2$, which influence $i$'s immediate neighbors, $\N_i^1$. With each increase of $n$ adds to $i$'s potential outcome arguments another layer of treatments on units with distance $n$ to $i$. One can visualize the effects of the treatment propagating on the network of influence, for example when $n\geq 2$, a treatment on unit $j$ may have an effect on unit $i$ because unit $j$ influences unit $l$ ($A_{jl}\neq 0$) and unit $l$ influences unit $i$ ($A_{li}\neq 0$). The choice on $n$ depends on how far the treatment effects may propagate in the experiment. The proper choice of $n$ is application specific. In some of the real-world applications, peer effects are shown to be significant only through immediate neighbors \citep{Goel2012}, while in others, peer effects are shown to propagate over multiple hops on the social network \citep{Fowler2010}. In practice, the choice of $n$ can be viewed as a model selection issue and should be verified in the analysis phase. Lastly, the $n$-hop neighborhood is not the only possible definition for the closed neighborhood under SUTNVA. This thesis adopts it going forward as a reasonable definition under network interference.

Following the above definitions, a unit $i$ may receive an effect not only from its own treatment, $Z_i$, but also from a treatment on any unit $j$ in its closed neighborhood of influence, $\N_i$, through a path of influence from $j$ to $i$ indicated by the influence subnetwork represented by $\A_{\N_i}$. The matrix $\A_{\N_i}$ is a submatrix of the population influence matrix $\A$, keeping only the rows and columns corresponding to the units in the neighborhood $\N_i$. In other words, $\A_{\N_i}$ represents a subgraph of the population network graph in Definition \ref{def:networkPopulation} by keeping only the nodes in $\N_i$ and the edges between them. Now we are finally ready to introduce the network potential outcomes:

\begin{definition}[Network Potential Outcomes Under SUTNVA] 
\label{def:networkPOUnderSUTNVA}
Under network interference and SUTNVA, the outcomes of a unit $i$ change according to its exposure to the treatments, $\Z_{\N_i}$, on its closed neighborhood of influence, $\N_i$, through the network of influence, $\A_{\N_i}$, among them. The network potential outcomes of $i$ are denoted as $Y_i(\Z_{\N_i}, \A_{\N_i})$.
\end{definition}

\noindent Compared to potential outcomes under SUTNVA in Definition \ref{def:POGeneral}, network potential outcomes have the influence network $\A_{\N_i}$ as the additional argument because a unit's exposure to the treatments on peers is determined also by the influence network. Under network interference, the mechanism of interference is through the influence network. Like the treatment vector $\Z$, the influence network $\A$ is typically considered a fixed quantity, representing the social influence that takes place during the experiment. However, unlike the treatment vector, the influence network is often not fully known and imputed during the analysis phase from prior information on the social network as well as the observed outcomes. Depending on how one models the propagation of exposures to peer treatments, multiple networks may be imputed representing the influence at each propagation wave. For parsimony, this detail is not included in the notation because it is not relevant to the developments in this chapter.

Like potential outcomes under SUTNVA, sometimes it is clearer to denote the treatment on certain units separately. For example, one may want to denote the treatment on unit $i$ itself. In such cases, the following notational convention is adopted:  $Y_i(\Z_{\N_i}, \A_{\N_i}) \equiv Y_i(Z_i, \Z_{\N_{-i}}, \A_{\N_i}) \equiv Y_i(Z_i, Z_j, \Z_{\N_{-i}-j}, \A_{\N_i})$, where $\N_{-i}$ is the open neighborhood of $i$  (note: open neighborhood of $i$ excludes $i$) and $\N_{-i}-j$ is the open neighborhood of $i$ excluding unit $j$, where unit $j \in \N_i$.

With the additional arguments indicating the exposure to treatments on peers, there are numerous network potential outcomes for each unit $i$ and for the entire population. Sometimes, it is convenient to denote the entire set of network potential outcomes as the following:
\begin{definition}[Network Potential Outcome Sets]
\label{def:networkPOSet}
The entire set of network potential outcomes for unit $i$ is $\YSet_i = \{Y_i(\Z_{\N_i}=\z, \A_{\N_i}=\bm{a})\}$ for~all $\z \in \ZSetClosed , \text{ for~all } \bm{a} \in \ASetClosed$ where $\ZSetClosed$ is the set of all possible assignment vectors on the closed neighborhood of $i$ and $\ASetClosed$ is the set of all possible influence networks on the closed neighborhood of $i$. The set of all network potential outcomes on the finite population with $N$ units is $\YSet = \{\YSet_1, \YSet_2, ... , \YSet_N\}$. The network potential outcomes, $\YSet$, can also be partitioned based on whether it is observed. In an experiment, for each unit, only one of the numerous possible neighborhood treatments in $\ZSetClosed$ is realized. The set of observed network potential outcomes, typically of size $N$, is denoted as $\YobsSet$. Most of the network potential outcomes will be unobserved. The set of unobserved outcomes, its size depending on the size of each unit's closed neighborhood, is denoted as $\YmisSet$.
\end{definition}

\noindent The network potential outcome notation has the flexibility to express all kinds of exposures to treatments on the neighborhood of influence. Below are some example network potential outcomes corresponding to various treatment patterns of interest:
\begin{enumerate}
\item When only unit $i$ is treated, the potential outcome of $i$ is $Y_i(Z_i = 1, \Z_{\N_{-i}} = \bm{0}, \A_{\N_i})$

\item When only the most influential immediate neighbor is treated, the potential outcome of $i$ is $Y_i(Z_j = 1, \Z_{\N_i-j} = \bm{0}, \A_{\N_i})$ where $j = \underset{j}{\text{argmax}} (A_{ji})$ 

\item When \emph{exactly} $k$ of $i$'s neighbors are treated, but $i$ is in control, the set of potential outcomes of $i$ are $\{Y_i(Z_i = 0, \Z_{\N_{-i}} = \z, \A_{\N_i})\} :  \z \in \ZkSet$ the set of all possible treatment assignment vectors on $\N_{-i}$ that sums to \emph{exactly} $k$ (i.e.\ $\sum_{\Z_{\N_{-i}}} = k$). If the $n$-hop neighborhood of influence, $\N_i = \N_i^{\leq n}$, is defined with $n \geq 2$, it may make sense to differentiate the neighbors at different distance to unit $i$ by having multiple $k$'s (e.g.\ $\sum_{\Z_{\N_i^1}} = k_1, \sum_{\Z_{\N_i^2}} = k_2, ... \sum_{\Z_{\N_i^n}} = k_n$).

\item When assignment strategy $\gimel$ (pronounced as ``gimel'') is applied, the expected potential outcome is $\; \sum_{\bm{z} \in \mathcal{Z}_{\N_i}}{Y_i(\Z_{\N_i} = \bm{z}, \A_{\N_i}) p(\bm{z}|\gimel, \YSet, \bm{X}, \A)}$, over all possible assignment vectors under $\gimel$ where $\mathcal{Z}_{\N_i}$ is the set of all possible assignment vectors on the closed neighborhood $\N_i$. Note that an assignment strategy specifies the probability of any assignment vector given the potential outcomes ($\YSet$), unit covariates ($\bm{X}$), and the influence network ($\A$): $P(\Z | \bm{\YSet, X, A})$.
\end{enumerate}

Section \ref{sec:CEUnderNetworkInteference} will demonstrate how these network potential outcomes can be used as building blocks for a wide range of causal estimands that quantify different treatment effects of interest under network interference. In an experiment, most of the network potential outcomes will be unobserved. Furthermore, a completely randomized assignment of the different exposures to neighborhood treatments is typically infeasible, as the structure of the influence network impacts each unit's chance to receive a certain exposure \citep{Toulis2013}. These make the estimation of causal estimands challenging under network interference. A natural solution is to perform model-based Bayesian imputation of missing potential outcomes, which will be discussed later in Section \ref{sec:BayesianImputationUnfoundedAssumptions}.

\section {Causal Estimands Under Network Interference}
\label{sec:CEUnderNetworkInteference}

Having defined the network potential outcomes, we now have the building blocks to define the appropriate causal estimands to answer various causal questions under network interference. This section gives some example causal estimands including those in existing literature, each focusing on quantifying the effect of a particular kind of exposure to treatment. Many more relevant causal estimands may be defined using the network potential outcomes, but these hopefully demonstrate the flexibility of the network potential outcomes in expressing causal quantities under network interference.

To make the examples in this section more concrete, let us consider a causal experiment on an educational program where some of the individuals receive education on a particular subject matter. These individuals receive the ``treatment''. Following the previously defined network potential outcome notations, $Z_i=1$ if individual $i$ receives the education and $Z_i=0$ otherwise. The outcome of interest ($Y_i$ for individual $i$) is each individual's knowledge on the subject at the end of the experiment, in terms of a score on the subject test. The individuals may pass on information from the education program through a network of influence ($\A$). 

\subsection {Primary Causal Effect Estimands}
\label{sec:primaryCausalEstimand}
If the primary treatment causal effect is the quantity of interest (i.e.\ want to separate it from the peer influence effects), an appropriate set of conditional causal estimands for each unit $i$ are:
\begin{equation}
\label{eq:primaryEffectConditionalCausalEstimand}
	\xi_i(\z) \equiv Y_i(Z_i = 1, \Zopen  = \z , \A_{\N_i}) -  Y_i(Z_i = 0,  \Zopen = \z ,\A_{\N_i})
\end{equation}
where $\z \in \ZSetOpen$, is a member of the set of all possible assignment vectors on the open neighborhood $\N_{-i}$. This set of conditional estimands capture the causal effect of the treatment on unit $i$ conditioning on a particular treatment assignment vector $\z$ on $i$'s neighbors. Going back to the example of the educational program experiment, this estimand focuses on the causal effect of receiving the education itself  by fixing the exposure to the education through peers (i.e.\ the social effect). In the absence of any exposure to treatments on peers, $\xi_i(\bm{0})$, the classical (i.e.\ under SUTVA) unit level causal effect, $Y_i(1)-Y_i(0)$, is recovered. If one wants to estimate the average primary treatment causal effect on unit $i$ under all possible neighborhood treatments, the causal estimand becomes: 
\begin{equation}
	\xi_i^\ave \equiv \frac{1}{2^{|\N_{-i}|}} \sum_{\z \in \ZSetOpen} {\xi_i(\z)}
\end{equation}
For a population of size $N$, the average primary treatment causal effect on the population is simply:
\begin{equation}
\label{eq:populationAveragePrimaryEffect}
	\xi^\ave \equiv \frac{1}{N} \sum_{i=1}^N{\xi_i^\ave}
\end{equation}

\subsection {$k$ Treated Neighbors Causal Effect Estimands}
\label{sec:kTreatedNeighborCausalEstimand}
If the peer influence effect is of interest, a natural quantity to consider is the causal effect of having $k$ of the neighbors of unit $i$ treated (assuming unit $i$ has at least $k$ neighbors), an appropriate set of conditional causal estimands are:
\begin{equation}
\label{eq:kNeighborIndividualCausalEstimand}
	\delta_{i, k} (z) \equiv {|\N_{-i}| \choose  k}^{-1} 
	\sum_{\z \in \ZkSet} {Y_i(Z_i=z,\Zopen = \z, \A_{\N_i}) - Y_i(Z_i = z, \Zopen = \bm{0}, \A_{\N_i})}
\end{equation}
where $z \in \{0, 1\}$ are the possible assignments on unit $i$, and $\ZkSet$ is the set of all treatment assignment vectors on the open neighborhood, $\N_{-i}$, that sums to \emph{exactly} $k$ (i.e.\ $\sum_{\Z_{\N_{-i}}} = k$). This set of conditional estimands capture the average causal effect on unit $i$ for having $k$ of its neighbors treated, while conditioning on the treatment assignment $z$ on $i$ itself. Going back to the educational program example, this causal estimand quantifies how much higher an individual scores if $k$ of his or her neighbors receive the education, by focusing on the effect of exposure to treatments on peers. For simplicity, the notation here does not distinguish neighbors by their distance to node $i$. Depending on the application, one may want to specify $k_n$ for each $n$-hop neighborhood when the neighborhood of influence is beyond the immediate neighbors.   

Similar to the previous example, the $k$ treated neighbor causal effect averaged over unit $i$'s own treatment can be expressed in the following estimand:
\begin{equation}
	\delta_{i, k}^\ave \equiv \frac{1}{2} \sum_{z=\{0,1\}} {\delta_{i, k} (z)}
\end{equation}
The population here is a bit more nuanced, because not all units have at least $k$ neighbors and therefore can not possibly receive such peer treatments. Therefore, the population average effect should only be averaged over the units that have at least $k$ neighbors. The set $\VSet_k$ contains units that have \emph{exactly} $k$ neighbors, and $\VSet_{\geq k}$ contains the units with \emph{at least} $k$ neighbors. The average $k$ treated neighbor causal effect on the population is:
\begin{equation}
	\delta_ k^\ave \equiv \frac{1}{|\VSet_{\geq k}|} \sum_{i \in \VSet_{\geq k} }{\delta_{i, k}^\ave}
\end{equation}

The idea of capturing causal peer effects based on the number of neighbors being treated has been proposed by other work. \cite{Ugander2013} define a similar but different condition called the ``absolute $k$-neighborhood exposure'' where unit $i$ meets this neighborhood treatment condition if $i$ is treated and \emph{at least} $k$ of $i$'s neighbors are treated. Adopting Ugander et al.'s peer treatment condition gives the following estimand for unit $i$:
\begin{equation}
	\tilde{\delta}_{i, k} \equiv \left[ \sum_{l=k}^{|\N_{-i}|} {|\N_{-i}| \choose  l}\right]^{-1} \;
	\sum_{l=k}^{|\N_{-i}|} \; \sum_{\z \in \ZlSet} {Y_i(Z_i=1,\Zopen = \z, \A_{\N_i}) - Y_i(Z_i = 1, \Zopen = \bm{0}, \A_{\N_i})}
\end{equation}
 where $|\N_{-i}|$ is the number of $i$'s neighbors and $\ZlSet$ is the set of all treatment assignment vectors on $\N_{-i}$ that sums to \emph{exactly} $l$ (i.e.\ $\sum_{\Z_{\N_{-i}}} = l$).

One may want to know the causal effect of having a certain fraction of the neighbors treated (e.g. 30\% of the neighbors treated), instead of the absolute number of neighbors. Ugander et al. define a version of such treatment condition called the ``fractional q-neighborhood exposure''. A causal estimand for fractional neighborhood treatments can be defined by simply mapping the fractional criterion to an absolute number for each unit $i$. For example, a $q\%$ neighborhood treatments for unit $i$ could map to a $k$ neighbor treatments with $k=\lceil\frac{q}{100} |\N_{-i}|)\rceil$. In any case, the individual $k$ treated neighbor causal estimand in equation  (\ref{eq:kNeighborIndividualCausalEstimand}) serves as the basic building block for these types of neighborhood treatment causal estimands.

\subsection {Fixed Assignment Causal Effect Estimands}

In many real-world experiments, researchers are interested in the causal effects with a specific treatment assignment vector. This helps one answer questions such as: ``What are the causal effects of the treatment assignment vector implemented in this experiment?'' and ``Which units should be treated in order to maximize certain effects on the population?''. Going back to the educational program example, the estimands below quantify various primary and peer causal effects when a particular group of individuals receive the education. Denoting the fixed treatment assignment vector as $\bm{z}$ and the set of treated units as $\TSet$, an appropriate primary effect conditional causal estimand is:
\begin{equation}
\label{eq:primaryEffectAverageCausalEstimand}
	\xi^\fix(\bm{z}) \equiv \frac{1}{\sum_i{z_i}} \sum_{i \in \TSet} \xi_i(\z_{\N_{-i}})
\end{equation}
$\xi_i(\z_{\N_{-i}})$ here is the primary effect causal estimand defined earlier in Equation \ref{eq:primaryEffectConditionalCausalEstimand} , conditional on the fixed treatment assignment vector for the open neighborhood of $i$.
This estimand quantifies the average primary treatment effect in the presence of peer effects. However, if one is only interested in the primary effect in the absence of peer effects (e.g. the primary effect may take place before any peer effects), the appropriate conditional causal estimand is:  
\begin{equation}
	\xi^\fix_{\delta_0}(\bm{z}) \equiv \frac{1}{\sum_i{z_i}} \sum_{i \in \TSet} \xi_i(\bm{0})
\end{equation}
Because the peer effects arise out of exposure to the primary treatments, the peer effect causal estimand is naturally in the presence of primary effects:
\begin{equation}
\label{eq:fixedAssignmentPeerEffect}
	\delta^\fix(\bm{z}) \equiv \frac{1}{N} \sum_{i = 1}^N {Y_i(Z_i = z_i, \Zopen  =  \bm{z}_{\N_{-i}}, \A_{\N_i} ) -  Y_i(Z_i = z_i,  \Zopen = \bm{0}, \A_{\N_i} )}
\end{equation}
This estimand quantifies the average amount of peer effects on the population as a result of the fixed treatment assignment vector.

Naturally, experimenters are often interested in the total effect of the treatment, which can be captured by the following causal estimand on the population average total effect:  
\begin{equation}
\label{eq:totalEffectCausalEstimand}
	\zeta^\fix(\bm{z}) \equiv \frac{1}{N} \sum_{i =1}^N {Y_i(Z_i = z_i, \Zopen  = \bm{z}_{\N_{-i}}, \bm{A}_{\N_i} ) -  Y_i(Z_i = 0,  \Zopen = \bm{0}, \bm{A}_{\N_i} )}
\end{equation}
Sometimes in a given experiment, researchers are interested in how far the treatment effects propagate on the social network, before diminishing to a negligible quantity. This question can be answered through the following average total effect causal estimands on the $n$-hop neighbors of the treated units:
\begin{equation}
\label{eq:totalEffectByHopCausalEstimand}
	\zeta^\fix_n(\bm{z}) \equiv \frac{1}{|\N^n_\TSet|} \sum_{i \in \N^n_\TSet} {Y_i(Z_i = z_i, \Zopen  = \bm{z}_{\N_{-i}}, \bm{A}_{\N_i} ) -  Y_i(Z_i = 0,  \Zopen = \bm{0}, \bm{A}_{\N_i} )}
\end{equation}
where $\N^n_\TSet$ denotes the set of units exactly $n$ hops away from the closest treated units. Typically, this quantity diminishes with increasing $n$, as the social impact of the treatment gets discounted with each hop away from the treated units.

\subsection {Assignment Strategy Causal Effect Estimands}
If the causal effects under certain treatment assignment strategies is of interest (e.g. randomly treat 30\% of the population), one can express the causal estimands in terms of expected potential outcomes. Going back to the educational program example, the estimands below can be used to evaluate different strategies in selecting individuals for enrollment, such as random selection, need-based selection, and targeting the most influential individuals, etc. Note that an assignment strategy specifies the probability of any assignment vector given the potential outcomes ($\YSet$), unit covariates ($\bm{X}$), and the influence network ($\A$): $P(\Z | \bm{\YSet, X, A})$.

An individual's primary treatment effect under assignment strategy $\gimel$ (pronounced as ``gimel'') can be expressed with the following conditional estimand: 
\begin{align}
	\xi_i^\expected(\gimel) \equiv \sum_{\z \in \ZSetOpen}&{\left[ Y_i(Z_i = 1, \Zopen = \z, \A_{\N_i}) \;  p(Z_i = 1, \Zopen=\z |\gimel, \bm{\YSet, X, A}) \right. } \nonumber \\
                                                                                              &\left. - Y_i(Z_i = 0, \Zopen = \z, \A_{\N_i})  \; p(Z_i = 0, \Zopen=\z |\gimel, \bm{\YSet, X, A})\right]
\end{align}
where $\ZSetOpen$ is the set of all possible assignment vectors on the open neighborhood of $i$. Hudgens and Halloran refer to this estimand as the ``individual average direct causal effect'' under assignment strategy $\gimel$ \citep{hudgens2008}. Similarly to the previous estimands, one can simply take an average of this estimand over the population for a population average primary causal effect estimand.

For comparing the causal \emph{peer} effect of assignment strategy $\daleth$ (pronounced as ``daleth'') against strategy $\gimel$, one may consider the following estimand for unit $i$: 
\begin{align}
	\delta_i^\expected(\daleth,\gimel) \equiv \frac{1}{2}\sum_{z=\{0,1\}} \;\; \sum_{\z \in \ZSetOpen} \; &{\left[ Y_i(Z_i = z, \Zopen = \z_, \A_{\N_i}) \;\; p(Z_i = z, \Zopen=\z |\daleth,\bm{\YSet, X, A}) \right. } \nonumber \\
                                                                                              &\left. - Y_i(Z_i = z, \Zopen = \z, \A_{\N_i}) \; p(Z_i = z, \Zopen=\z |\gimel,\bm{\YSet, X, A})\right]
\end{align}
Hudgens and Halloran refer to this estimand as the ``individual average indirect causal effect'' of assignment strategy $\daleth$ compared to assignment strategy $\gimel$. Causal estimands for total effects can be constructed in similar fashion.

\subsection {Influence Network Manipulation Causal Estimands}

All the causal estimands listed above capture the effect of the treatment by comparing the potential outcomes of one treatment assignment vector versus another on a fixed influence network. However, if the influence network can be manipulated, one can capture the causal effect of such manipulation using the network potential outcomes as building blocks. This may seem counterintuitive at first because the influence network itself is not a ``treatment''. However, under network interference, the influence network leads to exposure to treatments on peers, so manipulating the influence network alters the ``social treatment''. So in a way, the influence network and the treatment assignment vector together can be viewed as an assignment of ``social treatment''. Consider the causal estimand below on the average total effect of manipulating the influence network from $\A$ to $\A'$, given a particular treatment assignment vector $\z$,:
\begin{equation}
	\zeta_{\A}(\bm{z}) \equiv \frac{1}{N} \sum_{i=1}^N {Y_i(Z_i = z_i, \Zopen  = \bm{z}_{\N_{-i}}, \bm{A}'_{\N_i} ) -  Y_i(Z_i = z_i,  \Zopen = \bm{z}_{\N_{-i}}, \bm{A}_{\N_i} )}
\end{equation}
Going back to the educational program example, this estimand may quantify the benefit of strengthening the influence network by setting up study groups prior to the experiment. In other examples where the outcome is not desirable, such as in the cases of infectious disease and propaganda spreading over social networks, one may be interested in quantifying the effect of weakening the influence network through quarantine. This type of estimands highlight the flexibility and expressiveness of the network potential outcomes as the basic building block for causal inference under network interference.


\section {Bayesian Imputation and Unconfounded Assumptions}
\label{sec:BayesianImputationUnfoundedAssumptions}

Under network interference, most of the potential outcomes in the causal estimands will be unobserved. Furthermore, a completely randomized assignment over all possible exposures to neighborhood treatments is typically infeasible, as the structure of the influence network impacts each unit's chance to receive a certain exposure \citep{Toulis2013}. These make the estimation of causal estimands challenging under network interference. A natural solution is to perform Bayesian imputation of missing potential outcomes. Also known as the Bayesian predictive inference for causal effects, this method has been well established for causal inference in the absence of interference (i.e.\ SUTVA holds) \citep{Rubin1991, Rubin1978, Rubin1975}. 

As the potential outcomes serve as the building blocks for each causal estimand, computing the posterior distribution of the unobserved potential outcomes also gives the posterior distribution of any causal estimands. In the regular case under SUTVA, the posterior distribution of interest is $P(\bm{Y}_\mis | \bm{X,Z},\bm{Y}_\obs)$. Inference of the unobserved potential outcomes from the observed potential outcomes, the unit covariates $\X$, and the treatment assignment vector $\Z$, is typically done with a potential outcome model. Modeling and inference of this posterior distribution is greatly simplified when the treatment assignment mechanism can be ignored (i.e.\ $\Z$ can be dropped from the posterior distribution). Rubin shows how the unconfounded treatment assignment assumption leads to this simplification \citep{Rubin1991, Rubin1978, Rubin1975}. 

In the case under network interference, the posterior distribution is on the expanded sets of network potential outcomes and includes the influence network: $P(\YmisSet | \bm{X,Z,A},\YobsSet)$. Similar to the regular case under SUTVA, this posterior distribution can also be greatly simplified when the neighborhood treatment mechanism can be ignored (i.e.\ both $\Z$ and $\A$ can be dropped from the posterior distribution). This section shows how the assumptions of unconfounded treatment assignment and unconfounded influence network lead to this simplification. 

Beginning with Section \ref{sec:unconfoundedAssignmentRegular} is a review of Rubin's work under SUTVA on the regular unconfounded treatment assignment assumption and how it enables simplified Bayesian imputation of missing potential outcomes by ignoring the treatment assignment mechanism. Moving onto the network interference case in Section \ref{sec:unconfoundedAssignmentNetworkInterference}, the unconfounded treatment assignment assumption and the unconfounded influence network assumption are presented. Together they enable simplified Bayesian imputation of missing network potential outcomes by ignoring the neighborhood treatment mechanism. For each unconfounded assumption, a toy example is constructed in Section \ref{sec:violatingRegularAssumption} and \ref{sec:violatingNetworkInterferenceAssumption} to show how violating the assumption leads to biased causal estimates unless the treatment mechanism is explicitly modeled. This section also shows how the unconfounded assumptions can be met in real-world experiments under network interference, and presents a theorem that guides the design of network potential outcome models with respect to the unconfounded influence network assumption. From these theoretical properties, a rigorous analytical procedure is proposed in Section \ref{sec:BayesianImputationProcedure} for Bayesian imputation under network interference. Overall, this section aims to motivate that under network interference, Bayesian imputation of missing network potential outcomes needs to respect the key unconfounded assumptions in order to avoid incorrect causal estimates or having to model the complicated neighborhood treatment mechanism.

\subsection {Unconfounded Assumption and Simplified Imputation Under SUTVA}
\label{sec:unconfoundedAssignmentRegular}
Beginning with the regular case under SUTVA in the absence of interference, below is a brief review on Rubin's work \citep{ImbensRubin2015, Rubin1991, Rubin1978, Rubin1975} on the unconfounded treatment assignment assumption leading to simplified Bayesian imputation of missing outcomes where the treatment assignment mechanism can be ignored:

\begin{assumption}[Unconfounded Treatment Assignment Assumption Under SUTVA]
\label{as:unconfoundedTreatmentAssumptionSUTVA}
Conditional on the relevant unit covariates $\X$, the treatment assignment vector $\Z$ does not depend on the potential outcomes:
\begin{equation}
\label{eq:regularUnconfounded}
P(\Z | \X, \Y) = P(\Z | \X, \Y') \;\;\;\; \text{for~all } \Z, \X, \Y, \text{and } \Y'
\end{equation}
\end{assumption}

\noindent This assumption obviously holds in the completely randomized experiment. However, in randomized experiments with non-uniform assignment probability (i.e.\ some units are more likely to be treated than the others) and in observational studies where randomizing the treatment assignment is not an option, it is very important to condition on all confounding covariates $\X$ that are correlated to both the assignment vector $\Z$ and the potential outcomes $\Y$. This can be done by including the confounding covariates $\X$ in the potential outcome model. Recalling Definition \ref{def:POUnderSUTVA}, the potential outcomes under SUTVA can be partitioned in halves according to treatment assignment vector, $\Y = (\Y(0), \Y(1))$, or according to whether it is observed in an experiment, $\Y = (\Y_\obs, \Y_\mis)$. Rubin shows the critical role of Assumption~\ref{as:unconfoundedTreatmentAssumptionSUTVA} in Bayesian imputation of causal effects through the resulting simplification:

\begin{theorem}[Simplified Imputation Under SUTVA]
\label{thm:simplifiedImputationUnderSUTVA}
If the unconfounded treatment assignment assumption is met, the treatment assignment mechanism does not enter the posterior distribution of the missing potential outcomes:
\begin{equation}
	 P(\Y_\mis | \bm{X,Z},\Y_\obs) = P(\Y_\mis | \bm{X},\Y_\obs)
\end{equation}
\end{theorem}

\noindent Theorem \ref{thm:simplifiedImputationUnderSUTVA} is simply proved through a factorization, an application of Assumption \ref{as:unconfoundedTreatmentAssumptionSUTVA}, and a cancellation:
\begin{align}
	 P(\Y_\mis | \X,\Z,\Y_\obs) &= \frac{ P(\Z | \X, \Y) P(\Y | \X)}{\int  P(\Z | \X,\Y) P(\Y | \X) d \Y_\mis} \nonumber \\
					    &= \frac{ P(\Z | \X) P(\Y | \X)}{P(\Z | \X) \int P(\Y | \X) d \Y_\mis} \nonumber \\
                                                         &= \frac{P(\Y | \X)}{\int P(\Y | \X) d \Y_\mis}  \nonumber \\
                                                         &=  P(\Y_\mis | \X,\Y_\obs) 
\end{align}
Ignoring the treatment assignment mechanism simplifies the posterior distribution for the missing potential outcomes, making it easier to model and infer $\Y_\mis$, simply from the unit covariates $\X$ and the observed potential outcomes $\Y_\obs$. 

\subsubsection {Consequence of Violating the Unconfounded Assumption}
\label{sec:violatingRegularAssumption}
Violation of the unconfounded treatment assignment assumption leads to incorrect estimates of the causal estimands, unless the treatment assignment mechanism is explicitly modeled. This consequence is true for both the classical estimators such as the Neyman estimate in Equation \eqref{eq:neymanEstimator} and the model-based Bayesian imputation estimators. Let us consider the following toy example that demonstrates this intuition.

In a treatment study with five individuals, the outcome of interest is a measure on their quality of life, potentially through a survey at the end of the experiment. Individuals enter the experiment with a measure of their level of health, with zero being the least healthy and one being the healthiest. The level of health is a pre-treatment covariate with the following values: $\bm{x} = [0.1, \; 0.9, \; 0.6, \; 0.2, \; 0.5, \; ]^T$. Individual one and four receive the treatment in this experiment, because they are the least healthy entering into the study and therefore need the treatment the most. For a simple demonstration, assume the potential outcomes follow the linear model below for each unit $i$: 
\begin{equation}
Y_i(Z_i) = \tau Z_i + \beta x_i +  \epsilon_i
\end{equation}
where $\tau=5$ is the treatment effect, $\beta=10$ is the effect of the initial level of health, and $\epsilon_i \sim \Normal(0,\sigma^2=0.1)$ adds a small variation between the individuals. The observed outcomes are: $Y_1(1) = 5.44$, $Y_2(0) = 9.18$, $Y_3(0) = 5.94$, $Y_4(1) = 7.1$, and $Y_5(0) = 5.11$. The causal estimand of interest is the population average treatment effect in Equation \eqref{eq:causalEstimandUnderSUTVA}. Under this simple linear additive effect model, the true value of this estimand is simply: $\tau^\ave = \frac{1}{5} \sum_{i=1}^5 Y_i(1) - Y_i(0) = \tau = 5$

Note that in this experiment, the unconfounded treatment assignment assumption is violated, unless the level of health covariate is conditioned on. The outcome and the treatment assignments are correlated through the initial level of health. Because the treatment assignments are not completely randomized, the Neyman estimate in Equation \eqref{eq:neymanEstimator} for the population average treatment effect is badly biased: $\widehat{\tau}^\ave = \widebar{Y}^t_\obs - \widebar{Y}_\obs^c = \frac{Y_1(1)+Y_4(1)}{2} - \frac{Y_2(0)+Y_3(0)+Y_5(0)}{3} = -0.473$. This gross underestimate happens because the treated individuals have worse initial level of health, and therefore lower potential outcomes. Bayesian imputation of the causal estimand using Bayesian regression without conditioning on $\bm{x}$ (i.e.\ not including $\bm{x}$ in the regression) leads to similarly bad estimate with the $90\%$ posterior interval: $\widehat{\tau}^\ave = [-0.94, \; 0.01]$. However, if $\bm{x}$ is conditioned on and included in the regression, the unconfounded treatment assignment assumption is met and an adequate estimate of the causal estimand is recovered with the $90\%$ posterior interval: $\widehat{\tau}^\ave = [3.97, \; 5.99]$. 

This toy example demonstrates the consequence of violating the unconfounded treatment assignment assumption and how an adequate estimate of the causal estimand can be recovered by conditioning on the confounding covariate and therefore meeting the unconfounded treatment assumption.

\addtocontents{toc}{\protect\newpage} 

\subsection {Unconfounded Assumptions and Simplified Imputation Under Network Interference}
\label{sec:unconfoundedAssignmentNetworkInterference}
The previous section highlights the importance of unconfounded treatment assumption in causal inference for the regular case under SUTVA. Expanding to the case under network interference, this section lays out the key unconfounded assumptions that enables the imputation to ignore the neighborhood treatment mechanism. As one of the main contributions of this chapter, the unconfounded assumptions and the follow-on theorems provide the theoretical framework that guides the principled approach to causal inference under network interference in the presence of possible confounders. Let us begin with the unconfounded treatment assignment assumption under network interference.

\begin{assumption}[Unconfounded Treatment Assignment Assumption Under Network Interference]
\label{as:unconfoundedTreatmentAssignmentUnderNetworkInterference}
Conditional on the relevant unit covariates $\X$ and the influence network $\A$, the treatment assignment vector $\Z$ does not depend on the potential outcomes:
\begin{equation}
\label{eq:networkInterferenceUnconfoundedTreatment}
	 P(\Z | \bm{X,A, \YSet}) =  P(\Z | \bm{X,A, \YSet'}) \;\;\;\; \text{for~all } \Z, \X, \A, \YSet, \text{and } \YSet'
\end{equation} 
\end{assumption}

\noindent This assumption can be met in real-world experiments in similar ways as the regular unconfounded treatment assignment mechanism under SUTVA (i.e.\ Assumption \ref{as:unconfoundedTreatmentAssumptionSUTVA}), such as complete randomization of $\Z$ or including possible confounders in the conditional unit covariates $\X$. Sometimes, experiments on a social network target units with certain network characteristics for treatment, in order to achieve a desirable overall outcome. For example, a researcher may target the most influential units (e.g. high degree nodes) in order to maximize peer influence effects. Assumption \ref{as:unconfoundedTreatmentAssignmentUnderNetworkInterference} holds under such treatment assignment strategies because the influence network, $\bm{A}$, is conditioned on.

Under network interference, each unit may be exposed to treatments through peer influence. The influence network serves as a vehicle through which exposures to peer treatments are delivered. In this sense, $\Z$ and $\A$ together form the mechanism for neighborhood treatments. So naturally, the influence network is another source for confoundedness which is addressed by the next unconfounded assumption. The influence network is a result of some underlying mechanism governing the social influence, and will be treated as a random variable with a distribution going forward. Later on in Theorem \ref{thm:unconfoundedInfluenceNetworkByConditioning}, the distribution of the influence network will be characterized by a network model. 

\newpage

\begin{assumption}[Unconfounded Influence Network Assumption Under Network Interference]
\label{as:unconfoundedInfluenceNetworkUnderNetworkInterference}
Conditional on the relevant unit covariates $\X$, the influence network $\A$ does not depend on the potential outcomes:
\begin{equation}
\label{eq:networkInterferenceUnconfoundedInfluenceNetwork}
	 P(\A | \bm{X,\YSet}) =  P(\A | \bm{X, \YSet'}) \;\;\;\; \text{for~all } \A, \X, \YSet, \text{and } \YSet'
\end{equation}
\end{assumption}

\noindent This assumption is met if the mechanism of the influence network has no correlation with the potential outcomes. However, this is often not true in real-world experiments \citep{shalizi2011}. Intuitively, correlation between the potential outcomes and the influence network may arise from certain characteristics of a unit that are correlated with both its outcomes as well as its relationships with other units on the network. Activity level and group memberships are examples of such characteristics. For example, in an educational program experiment where the potential outcomes are the level of knowledge on a certain subject, a unit's activity level may be positively correlated with its potential outcomes because the more active nodes in the influence network turn out to be community leaders with higher education. Similarly, a unit's membership to a highly educated group may be positively correlated with its potential outcomes and at the same time result in more interactions and influence with other members in that group. Therefore, meeting Assumption \ref{as:unconfoundedInfluenceNetworkUnderNetworkInterference} will likely require including such possible confounding characteristics in the conditional unit covariates $\X$. A method to achieve this will be formally proposed later in Theorem \ref{thm:unconfoundedInfluenceNetworkByConditioning}. But first, let us introduce the simplified imputation as a result of Assumption \ref{as:unconfoundedTreatmentAssignmentUnderNetworkInterference} and \ref{as:unconfoundedInfluenceNetworkUnderNetworkInterference}, because it is the main motivation for the theoretical developments here.

\begin{theorem}[Simplified Imputation Under Network Interference]
\label{thm:simplifiedImputationUnderNetworkInterference}
If the unconfounded treatment assignment assumption and the unconfounded influence network assumption are both met, the neighborhood treatment mechanism does not enter the posterior distribution of the missing potential outcomes:
\begin{equation}
	 P(\YmisSet | \bm{X,Z,A},\YobsSet) = P(\YmisSet | \bm{X},\YobsSet) 
\end{equation}
\end{theorem}

\newpage

\noindent Theorem \ref{thm:simplifiedImputationUnderNetworkInterference} is simply proved through a factorization, an application of Assumption \ref{as:unconfoundedTreatmentAssignmentUnderNetworkInterference}, a cancellation, another factorization, an application of Assumption \ref{as:unconfoundedInfluenceNetworkUnderNetworkInterference}, and another cancellation:

\begin{align}
	 P(\YmisSet | \bm{X,Z,A},\YobsSet) &= \frac{ P(\Z | \bm{X,A, \YSet}) P(\YSet | \bm{X,A})}{\int  P(\Z | \bm{X,A, \YSet}) P(\YSet | \bm{X,A}) d \YmisSet} \nonumber \\
					    &= \frac{ P(\Z | \bm{X,A}) P(\YSet | \bm{X,A})}{P(\Z | \bm{X,A}) \int  P(\YSet | \bm{X,A}) d \YmisSet} \nonumber \\
                                                         &= \frac{P(\YSet | \bm{X,A})}{\int P(\YSet | \bm{X,A}) d \YmisSet}  \nonumber \\
                                                         &=  P(\YmisSet | \bm{X, A},\YobsSet)  \nonumber \\
					    &= \frac{ P(\A | \bm{X,\YSet}) P(\YSet | \bm{X})}{\int  P(\A | \bm{X,\YSet}) P(\YSet | \bm{X}) d \YmisSet} \nonumber \\
					    &= \frac{ P(\A | \bm{X}) P(\YSet | \bm{X})}{P(\A | \bm{X}) \int  P(\YSet | \bm{X}) d \YmisSet} \nonumber \\
                                                         &= \frac{P(\YSet | \bm{X})}{\int P(\YSet | \bm{X}) d \YmisSet}  \nonumber \\
                                                         &=  P(\YmisSet | \bm{X},\YobsSet) 
\end{align}

The posterior distribution for the missing network potential outcomes is greatly simplified when the neighborhood treatment mechanism can be ignored (i.e.\ both $\Z$ and $\A$ can be dropped from the posterior distribution). Similar to the regular case under SUTVA, this makes it easier to model and infer $\Y_\mis$, simply from the unit covariates $\X$ and the observed potential outcomes $\Y_\obs$. More importantly, this theoretical result provides a rigorous guideline for which unit covariates should be included in the potential outcome model. Specifically, all the relevant covariates should be included in $\X$ to meet the unconfounded assumptions. As mentioned earlier, the unconfounded influence network assumption is likely violated in real-world experiments unless the estimation procedure conditions on the relevant unit covariates that are correlated with both the potential outcomes and the influence network. This can be formalized with the following theorem by modeling the distribution of the influence network:

\newpage

\begin{theorem}[Unconfounded Influence Network by Conditioning on Network Parameters]
\label{thm:unconfoundedInfluenceNetworkByConditioning}
The unconfounded influence network assumption in Assumption \ref{as:unconfoundedInfluenceNetworkUnderNetworkInterference}, $P(\A | \bm{X,\YSet}) =  P(\A | \bm{X, \YSet'}) \; \text{for~all } \A, \X, \YSet, \text{and } \YSet'$, is met if:
\begin{enumerate}
\item The distribution of the influence network $\A$ can be characterized by a model $H_G$ with nodal parameters $\X_G$ and population parameters $\bm{\Theta}_G$: $\A \sim H_G(\X_G, \bm{\Theta}_G)$
\item The influence network $\A$ correlates with the potential outcomes only through a subset of the nodal parameters $\widetilde{\X}_G \in \X_G$ and population parameters $\widetilde{\bm{\Theta}}_G \in \bm{\Theta}_G$
\item The unit covariates $\X$ contain these network parameters $\widetilde{\X}_G$ and $\widetilde{\bm{\Theta}}_G$
\end{enumerate}
\end{theorem}

\noindent Theorem \ref{thm:unconfoundedInfluenceNetworkByConditioning} leverages the assumption that the influence network $\A$ can be characterized by a model with nodal (i.e.\ unit-specific) parameters $\X_G$ (e.g. expected degree, community membership, position in the latent space, etc.) and population parameters $\bm{\Theta}_G$ (e.g. inter-community interaction, sparsity, etc.). Most if not all of the currently well known network models in the network inference community can be described in this way, including the latent space models ~\citep{Hoff2002}, the latent class models such as the membership blockmodels ~\citep{Wang1987, Airoldi2008}, degree distribution models ~\citep{Aiello2001, Chung2002}, and the graphon ~\citep{Lovasz2006}. Because the distribution of the influence network is completely characterized by the parameters $\X_G$ and $\bm{\Theta}_G$, including all of them in the unit covariates $\X$ during imputation simply fulfills the unconfounded influence network assumption. However, in reality only a subset of the nodal parameters $\widetilde{\X}_G$ may be correlated with the potential outcomes, like the unit-specific characteristics such as activity level and community membership in the educational program example described earlier. Similarly, most likely only some of the population parameters, $\widetilde{\bm{\Theta}}_G$ (e.g.\ sparsity), may be correlated with the potential outcomes. For parsimony and simplicity of the potential outcome model, including only these subsets of parameters break the correlation between the potential outcomes and the influence network, therefore meeting the unconfounded influence network assumption. Often, these network model parameters are not readily observed, but can be estimated from the social network data collected in the experiment, as shown in Chapter \ref{ch:3}.

\subsubsection {Bayesian Imputation Procedure Guided by Theory}
\label{sec:BayesianImputationProcedure}
So how exactly does one perform the analysis using the proposed framework in a real-world experiment? Driven by the theoretical framework for causal inference under network inference, the analytical procedure for Bayesian imputation of missing potential outcomes and causal estimands is summarized in the following steps:
\begin{enumerate}
\item Define the population, the treatment, and the network  potential outcomes. Collect prior information on the underlying influence network. This can be from data on interactions between the units (e.g. emails, tweets, phone calls, etc.) or a survey on the social network (e.g. list of relationships). One may only have partial or prior information on the influence network, in which case the network will need to be imputed during analysis.
\item Propose an appropriate network model and estimate the model parameters using the observed influence network or the prior distribution on the influence network.
\item Propose an appropriate potential outcome model including the unit's own treatment, its exposure to peer treatments, and all the possible confounding covariates, guided by Assumption \ref{as:unconfoundedTreatmentAssignmentUnderNetworkInterference}, \ref{as:unconfoundedInfluenceNetworkUnderNetworkInterference}, and Theorem \ref{thm:unconfoundedInfluenceNetworkByConditioning}, so that the neighborhood treatment mechanism can be ignored during imputation as specified in Theorem \ref{thm:simplifiedImputationUnderNetworkInterference}. Some of these may be the estimated network parameters from the previous step and the possible treatment assignment confounders. Perform Bayesian inference to compute the posterior distributions of the potential outcome model parameters as well as the influence network if it is not fully known, as typically is the case.
\item Using the posterior distributions from the step above, impute the missing network potential outcomes in the causal estimands of interest, and finally arrive at the posterior distribution of the causal estimands, the quantity of interest. 
\end{enumerate}
This procedure is a principled approach that conditions on all the necessary covariates to meet the unconfounded assumptions. The rest of the chapter demonstrates the consequence of not doing so through toy examples and simulated experiments.

\subsubsection {Consequence of Violating the Unconfounded Assumptions}
\label{sec:violatingNetworkInterferenceAssumption}
Violation of the unconfounded treatment assignment assumption and the unconfounded influence network assumption under network interference leads to incorrect estimates of the causal estimands, unless the neighborhood treatment assignment mechanism is explicitly modeled. Let us consider the following toy example that demonstrates this intuition.

In an education program study with five individuals, the outcome of interest is a measure on their knowledge in the subject, potentially through a test at the end of the experiment. Individuals enter the experiment with a measure of their level of education, with zero being the least educated and two being the most educated. The level of education is a pre-treatment covariate with the following values: $\bm{x} = [0.1, \; 0.9, \; 0.6, \; 0.2, \; 0.5]^T$. Individual one and four receive the ``treatment'' (i.e.\ enrollment in the education program) in this experiment, because they are the least educated entering into the study and therefore need the program the most. So the treatment vector is $\bm{Z} = [1, \; 0, \;0, \; 1, \; 0]^T$. These two individuals who receive the education may pass on their newly acquired knowledge through the influence network shown in Figure \ref{fig:toyExampleInfluenceNetwork}. The influence network is $\bm{A} = \left[\begin{smallmatrix} 0 & 1 & 2 & 0 & 0\\ 1 & 0 & 2 & 0 & 0\\ 2 & 3 & 0 & 1 & 1 \\ 0 & 0 & 2 & 0 & 0 \\ 0 & 0 & 1 & 0 & 0\end{smallmatrix}\right]$. For simplicity, the neighborhood of influence under SUTNVA is assumed to include only the immediate neighbors (i.e.\ $\N_i = \N_i^{\leq 1}$). For demonstration, assume the potential outcomes follow the linear model below for each unit $i$: 
\begin{equation}
Y_i(\Z_{\N_i},\A_{\N_i}) = \tau Z_i + \gamma \sum_{j \in \N_{-i}} A_{ji} Z_j + \beta x_i +  \epsilon_i
\end{equation}
where $\tau=30$ is the primary treatment effect, $\gamma=5$ is the peer effect coefficient, $\sum_{j \in \N_{-i}} A_{ji} Z_j$ is the amount of exposure to treatments on neighbors, $\beta=10$ is the effect of the initial level of education, and $\epsilon_i \sim \Normal(0,\sigma^2=0.1)$ adds a small variation between the individuals. A more detailed description of the linear potential outcome model is given later in Section \ref{linearPOModel}. The observed outcomes here are: $Y_1(\Z_{\N_1},\A_{\N_1}) = 31.06$, $Y_2(\Z_{\N_2},\A_{\N_2}) = 13.55$, $Y_3(\Z_{\N_3},\A_{\N_3}) = 25.48$, $Y_4(\Z_{\N_4},\A_{\N_4}) = 32.14$, and $Y_5(\Z_{\N_5},\A_{\N_5}) = 4.51$. The causal estimands of interest are the population average primary effect, $\xi^\ave$ in Equation \eqref{eq:populationAveragePrimaryEffect}, and the fixed-assignment peer effects, $\delta^\fix(\bm{z})$ in Equation \eqref{eq:fixedAssignmentPeerEffect}. Under this simple linear additive effect model, the true values of these estimands are $\xi^\ave = \tau = 30$ and $\delta^\fix(\bm{Z}) = \gamma \times \frac{1}{N} \sum_{i=1}^N \sum_{j \in \N_{-i}} A_{ji} Z_j  = 5$.

\begin{figure}[ht]
     \centering
     \includegraphics[width=0.3\textwidth]{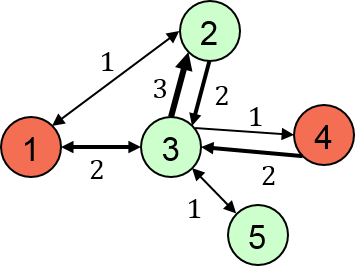}
     \caption[Influence network in toy example]{The influence network in this example, with treated nodes colored in red (darker nodes on a black and white print). Directed and weighted edges represent the magnitude and direction of influence.}
     \label{fig:toyExampleInfluenceNetwork}
\end{figure}

Note that in this experiment, the unconfounded treatment assignment assumption is violated, unless the level of education covariate is conditioned on. The outcome and the treatment assignments are correlated through the initial level of education. Bayesian imputation of the causal estimands using Bayesian regression without conditioning on $\bm{x}$ (i.e.\ not including $\bm{x}$ in the regression) leads to bad estimate of the population average primary effect causal estimand with the $90\%$ posterior interval: $\widehat{\xi}^\ave = [24.74, \; 25.83]$. This underestimate happens because the treated individuals have worse initial level of education, and therefore lower network potential outcomes. However, if $\bm{x}$ is conditioned on and included in the regression, the unconfounded treatment assignment assumption is met and an adequate estimate of the causal estimand is recovered with the $90\%$ posterior interval: $\widehat{\xi}^\ave = [29.37, \; 31.55]$.

Now suppose a different experiment is conducted on the same five individuals but individual one and four were chosen randomly to receive the treatment. The level of education now has the following values: $\bm{x} = [0.4, \; 0.7, \; 1.4, \; 0.3, \; 0.2 ]^T$ which is correlated with how active each individual is in the influence network. Specifically, the more educated individuals are more active in the network. All the other parameters stay the same and the true values of the causal estimands stay the same. The observed outcomes are: $Y_1(\Z_{\N_1},\A_{\N_1}) = 34.02$, $Y_2(\Z_{\N_2},\A_{\N_2}) = 11.91$, $Y_3(\Z_{\N_3},\A_{\N_3}) = 33.79$, $Y_4(\Z_{\N_4},\A_{\N_4}) = 32.93$, and $Y_5(\Z_{\N_5},\A_{\N_5}) = 2.07$. Note that in this experiment, the unconfounded influence network assumption is violated, unless the level of education covariate is conditioned on. The outcome and the influence network are correlated through the initial level of education. Bayesian imputation of the causal estimands using Bayesian regression without conditioning on $\bm{x}$ (i.e.\ not including $\bm{x}$ in the regression) leads to bad estimate of the fixed-assignment peer effect causal estimand with the $90\%$ posterior interval: $\widehat{\delta}^\fix(\bm{Z}) = [7.59, \; 7.95]$. This overestimate happens because the individuals who receive higher exposure to treatments on peers also have higher level of education, and therefore higher network potential outcomes. However, if $\bm{x}$ is conditioned on and included in the regression, the unconfounded influence network assumption is met and an adequate estimate of the causal estimand is recovered with the $90\%$ posterior interval: $\widehat{\delta}^\fix(\bm{Z}) = [4.22, \; 5.97]$.

This toy example demonstrates the consequence of violating the unconfounded assumptions under network interference and how an adequate estimate of the causal estimands can be recovered by conditioning on the confounding covariate and therefore meeting the unconfouned assumptions. A more extensive study on the effects of different types of confounders with a larger population and a realistic influence network model is conducted in the following section.


\section{Result on Simulated Experiments}
\label{sec:resultsOnSimulatedExperiments}
To demonstrate our proposed framework and imputation procedure for causal inference under network interference in the presence of confounders, simulated experiments are conducted using a realistic network model and a simple linear network potential outcome model. Different types of confounders are introduced to highlight the importance of satisfying the theoretical properties developed in Section \ref{sec:unconfoundedAssignmentNetworkInterference}. The results show how, by violating the unconfounded assumptions, different confounders lead to incorrect causal estimates on the causal estimands for the population average primary effect, $\xi^\ave$ in Equation \eqref{eq:populationAveragePrimaryEffect}, and the fixed-assignment peer effects, $\delta^\fix(\bm{z})$ in Equation \eqref{eq:fixedAssignmentPeerEffect}. The proposed procedure in Section \ref{sec:BayesianImputationProcedure} include these confounders as covaraites in the potential outcome model, satisfying the unconfounded assumptions (Assumption \ref{as:unconfoundedTreatmentAssignmentUnderNetworkInterference} and \ref{as:unconfoundedInfluenceNetworkUnderNetworkInterference}) which allow the neighborhood treatment mechanism to be ignored during imputation (Theorem \ref{thm:simplifiedImputationUnderNetworkInterference}). The results show how the proposed imputation procedure is able to account for the confounders and to arrive at Bayesian posterior estimates with good frequentist coverage properties over repeated experiments. To focus on the importance of including different types of confounders in the imputation, the results here assume knowledge of the true potential outcome model during the analysis phase, which often is not the case in real-world experiments. The more realistic scenario with model mis-specification will be investigated in the simulation study described in Chapter \ref{ch:2}, with mitigation by balancing randomizations in the design phase.

\subsection {Linear Model for the Network Potential Outcomes}
\label{linearPOModel}
The Generalized Linear Models (GLMs) provide a broad class of linear models for different types of potential outcomes (e.g. binary, count data, continuous response, etc.) and different relationships (e.g. identity, logit, log, etc) between the effects (i.e.\ linear predictors) and the potential outcomes. Under the GLM, the effects' contribution to the outcome may be additive (with identity link), multiplicative (with log link), and discounted at small and large values (with logit link), etc. For simplicity and clarity, the simulated experiments here are conducted using an ordinary linear model with additive effects, which is a special case of the GLM. The additivity assumption of various effects is also made by previous works, such as Manski's ``linear in means'' model~\citep{Manski} and Parker's linear network effect model ~\citep{Parker}. Also for simplicity, the neighborhood of influence under SUTNVA is assumed to include only the immediate neighbors (i.e.\ $\N_i = \N_i^{\leq 1}$). In real-world experiments, the model for the network potential outcomes and the diameter of the neighborhood of influence should be designed according to the phenomenology of the given application. The possible models, by no means, are limited to the ordinary linear model or even the general class of GLMs. No matter the model choice, the possible confounding covariates should be included in order to avoid badly biased causal estimates. The experiment here aims to demonstrate this point with a simple network potential outcome model, for each unit $i$:
\begin{equation}
\label{eq:potentialOutcomeModel}
	Y_i(\Z_{\N_i},\A_{\N_i}) =  \tau Z_i + \gamma \sum_{j \in \N_{-i}} A_{ji} Z_j + \bm{\beta}^T \bm{x}_i + \mu + \epsilon_i
\end{equation}
The first term, $\tau Z_i$, represents the primary treatment effect. The second term, $\gamma \sum_{j \in \N_{-i}} A_{ji} Z_j$, represents the peer effects of exposure to treatments on neighbors. Note that each neighbor $j$ who receives the treatment contributes $A_{ji}$ units of exposure to peer treatments, so the exposure is proportional to the amount of influence $j$ has on $i$. The total accumulative exposure from all the neighbors is multiplied by $\gamma$ which is a coefficient for the effect of the exposure. Although this is a simple and intuitive model for the peer effects, by no means is this the only one. The network potential outcomes, described in Section \ref{subsec:POUnderSUTNVA}, are defined to allow much flexibility in modeling the exposures to treatments on peers, which can be any function of the treatment vector and the influence network. The third term $\bm{\beta}^T \bm{x}_i$  is the effect of the unit covariates $\bm{x}_i$. This is where confounders are introduced in this experiment. The fourth term, $\mu$, is the baseline effect on the entire population. The last term, $\epsilon_i \sim \Normal(0,\sigma_\epsilon^2)$, gives independent and identically distributed variation for heterogeneity between the units.

\subsection {Hybrid Mixed-Membership Blockmodels for Influence Network}
\label{hybridMembershipBlockmodel}
In real-world experiments, the model for the influence network (i.e.\ influence matrix $\A$) should be designed according to the process of social influence in the given application. The model choice informs the relevant network model parameters which should be estimated and included in the network potential outcome model, as laid out in the imputation procedure in Section \ref{sec:BayesianImputationProcedure}. Although our proposed causal inference framework is not limited to any particular network model, the simulated experiments in this section adopt an influence network model that captures characteristics of real-world networks. 

Real-world networks display topological characteristics such as a Power-Law degree distribution (i.e.\ the ``small world'' property)~\citep{Chakrabarti2004}, membership-based community structure (i.e.\ individuals interact based on their community membership)~\citep{Wasserman1994}, and sparsity~\citep{Newman2006}. Each well-known canonical network model captures one or a subset of these traits. For example, the Chung-Lu model~\citep{Aiello2001, Chung2002} captures the Power-Law degree distribution. The mixed-membership stochastic blockmodel~\citep{Airoldi2008} provides sparsity and community structure, and the  Erd\H{o}s-R\'enyi model~\citep{Erdos1960} captures sparsity only. To capture these characteristics of the real-world networks, the experiment in this section adopts the ``hybrid mixed-membership blockmodels'' (HMMB), which is developed in detail in Chapter \ref{ch:3}. The HMMB is an aggregate of the following models and their features: Chung-Lu for Power-Law degree distribution, the mixed-membership stochastic blockmodel for community structure, and Erd\H{o}s-R\'enyi for sparsity. Other aggregate network models exist in the open literature such as \cite{Karrer2011} that combines the single-membership stochastic blockmodel with the Chung-Lu model for degree correction. Recall, from Definition \ref{def:networkPopulation}, that each edge in the influence network, $A_{ij}$, represents the amount of influence individual $i$ has on $j$. As influence typically arises from social interactions, the amount of influence represented as each edge, $A_{ij}$, is modeled as a Poisson random variable with expected value $\lambda_{ij}T$. The symbol $T$ is the time span during which the social influence takes place, typically from the time of treatment to the time of the outcomes of interest. The rate $\lambda_{ij}$ can be viewed as the frequency of influential interaction from individual $i$ to $j$, given by the product below:
\begin{equation}
\lambda_{ij} = (\lambda_i \lambda_j)  \times (\bm{\pi}_i^T \bm{B} \bm{\pi}_j) \times I_{ij}
\end{equation}
The first (Chung-Lu) term $\lambda_i\lambda_j$ gives each node $i$ an activity level proportional to $\lambda_i$ which is drawn from a Power-Law distribution. The second (mixed-membership blockmodel) term $\bm{\pi}_i^T\Bb\bm{\pi}_j$ determines the strength of interaction between $i$ and $j$ based on their memberships to the $K$ communities indicated by vectors $\bm{\pi}_i$ and $\bm{\pi}_j$ of $K$ components, and the inter-community interaction strength indicated by the $K \times K$ block matrix $\Bb$. The third (Erd\H{o}s-R\'enyi) term $I_{ij}$ is the binary indicator function switched on with probability $s$. For a detailed description of the ``hybrid mixed-membership blockmodels'' (HMMB) and the inference procedure for estimating the model parameters, please refer to Chapter \ref{ch:3}. Using the notations from Theorem \ref{thm:unconfoundedInfluenceNetworkByConditioning}, one may denote the HMMB network model as $H_G(\X_G, \bm{\Theta}_G)$ with the nodal parameters $\X_G = \{\bm{\lambda}, \bm{\Pi}\}$ and the population parameters $\bm{\Theta_G} = \{\bm{B}, s\}$. When some of these network parameters are correlated with the network potential outcomes, as discussed in the examples right under Assumption \ref{as:unconfoundedInfluenceNetworkUnderNetworkInterference}, they become confounders that should be estimated and included as covariates in the potential outcome model. The reason for this is already explained by the theoretical framework in Section \ref{sec:unconfoundedAssignmentNetworkInterference} and the consequence of not accounting for these confounders during the imputation of missing network potential outcomes is demonstrated in the simulation results below.

\subsection{Experimental Setup: Different Confounders and Inclusions}
\label{sec:confounderExperimentSetup}

To demonstrate the importance of satisfying the theoretical properties developed in Section \ref{sec:unconfoundedAssignmentNetworkInterference}, four different types of confounders are considered in the simulated experiments. For each type, estimation results are generated both with the exclusion and the inclusion of the confounder in the network potential outcome model. Excluding the confounder in the potential outcome model breaks one of the unconfounded assumptions (i.e.\ Assumption \ref{as:unconfoundedTreatmentAssignmentUnderNetworkInterference} or \ref{as:unconfoundedInfluenceNetworkUnderNetworkInterference}), while including it satisfy the unconfounded assumptions which allow the neighborhood treatment mechanism to be ignored during imputation (Theorem \ref{thm:simplifiedImputationUnderNetworkInterference}). Each confounder type demonstrates how violating the unconfounded assumptions leads to incorrect causal estimates on the causal estimands for the population average primary effect, $\xi^\ave$ in Equation \eqref{eq:populationAveragePrimaryEffect}, and the fixed-assignment peer effect, $\delta^\fix(\bm{z})$ in Equation \eqref{eq:fixedAssignmentPeerEffect}. In each type of confounder, the covariate, $\bm{x}_i$, affects the potential outcomes, $Y_i(\Z_{\N_i}, \A_{\N_i})$, through the $\bm{\beta}^T \bm{x}_i$  term in Equation \eqref{eq:potentialOutcomeModel}. Here are the four types of confounders in the experiment:

\begin{enumerate}
  \item {\bf Independent covariate}: The covariate $x_i$ is scalar and independent of the treatment assignment vector as well as the influence network. In this experiment, each $x_i$ is drawn independently from the $\Normal(1,1)$ distribution. Because $x_i$ is only correlated with the potential outcomes, it is not a confounder. Whether this covariate is included or excluded from the potential outcome model does not break the unconfounded assumptions, therefore, the neighborhood treatment mechanism can be ignored during imputation either way. This case serves as a benign baseline for comparison.
  \item {\bf Treatment likelihood confounder}: The covariate $x_i$ is scalar and affects the likelihood for unit $i$ to receive the primary treatment. This is achieved by modeling the treatment likelihood for individual $i$ as proportional to $e^{-x_i}$ where each $x_i$ is drawn independently from the $\Normal(0,2)$ distribution. In real-world experiments, this can be the case where the treatment is given to units based on their needs which are inversely correlated to their potential outcome. For example, the patients who need the treatment the most tend to have worse potential outcomes than the average patients. This type of confounder causes the treatment assignment vector $\bm{Z}$ to be correlated with the network potential outcomes $\YSet$. This violates Assumption \ref{as:unconfoundedTreatmentAssignmentUnderNetworkInterference} and therefore breaks the condition in Theorem \ref{thm:simplifiedImputationUnderNetworkInterference}, unless $x_i$ is conditioned on and included as a unit covariate in the potential outcome model.
  \item {\bf Activity level confounder}: The scalar covariate $x_i$ is the activity level $\lambda_i$, for node $i$. In other words, the activity level of the nodes has an effect on the potential outcome. This can be the case in real-world experiments because more active individuals may have more resources (e.g. higher education background, income, etc.) and, therefore, better potential outcomes. This type of confounder causes the influence network $\A$ to be correlated with the network potential outcomes $\YSet$. This violates Assumption \ref{as:unconfoundedInfluenceNetworkUnderNetworkInterference}  and therefore breaks the condition in Theorem \ref{thm:simplifiedImputationUnderNetworkInterference}, unless $x_i$ is conditioned on and included as a unit covariate in the potential outcome model as laid out in Theorem \ref{thm:unconfoundedInfluenceNetworkByConditioning} and in the imputation procedure in Section \ref{sec:BayesianImputationProcedure}.
  \item {\bf Community membership confounder}: The covariate $\bm{x}_i$ is the community membership vector of individual $i$. In real-world experiments, members of a certain community may share a common attribute that affects their potential outcome (e.g. they all live in a particular area, have a particular diet, go to the same school, etc.). This type of confounder causes the influence network $\A$ to be correlated with the network potential outcomes $\YSet$. This violates Assumption \ref{as:unconfoundedInfluenceNetworkUnderNetworkInterference}  and therefore breaks the condition in Theorem \ref{thm:simplifiedImputationUnderNetworkInterference}, unless $\bm{x}_i$ is conditioned on and included as unit covariates in the potential outcome model as laid out in Theorem \ref{thm:unconfoundedInfluenceNetworkByConditioning} and in the imputation procedure in Section \ref{sec:BayesianImputationProcedure}.
\end{enumerate}

Except for the simulations on the treatment likelihood confounder (type 2), the treatment is randomly assigned in the experiment. A total of $25$ individuals or about $10\%$ of the population receive the treatment. The influence networks are generated according to the HMMB model outlined in Section \ref{hybridMembershipBlockmodel}. For each experimental setup, $2500$ simulated experiments are conducted over $25$ draws of the influence network each with $100$ draws of treatment assignment vector and observed potential outcomes. The network parameters are set for a population size of $256$ over four communities, with moderate mixed-membership, inter-community interactions, and realistic variation in activity level. For simplicity, the influence networks are observed in these experiments. An example simulated experiment on an influence network with completely randomized treatment assignment is shown in Figure \ref{fig:treatmentOnInfluenceNetwork}.
\begin{figure}[ht]
     \centering
     \includegraphics[width=0.6\textwidth]{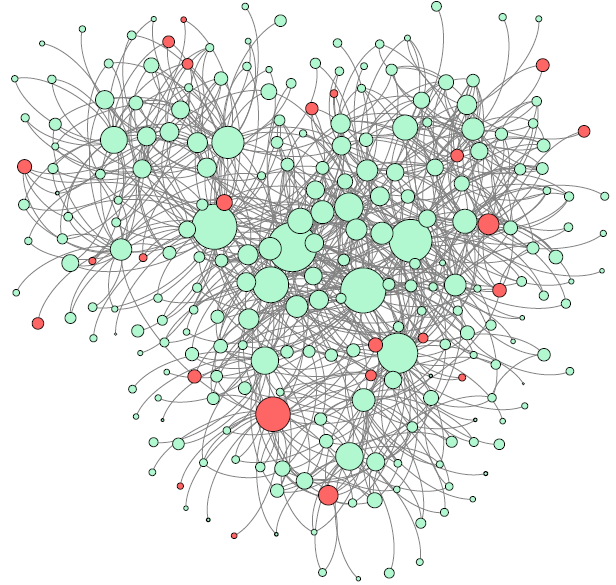}
     \caption[Influence network with completely randomized treatment assignment in a simulated experiment]{An influence network with treated nodes colored in red (darker nodes on a black and white print). Node sizes correspond to how active the nodes are. Edges represent influence between nodes. For visual clarity, only the more influential edges are drawn.}
     \label{fig:treatmentOnInfluenceNetwork}
\end{figure}

The potential outcomes are simulated using the linear model described in Section \ref{linearPOModel}, specifically in Equation \eqref{eq:potentialOutcomeModel}, with primary treatment effect coefficient $\tau=5$, peer treatment effect coefficient $\gamma=0.1$, covariate effect coefficient $\beta = 5$ for the scalar covariates described above and $\bm{\beta} = [0, \; 2, \; 4, \;  6]^T$ for the mixed-membership covariate vector, baseline effect $\mu=3$, and the variation variance $\sigma^2_\epsilon=1$ for heterogeneity between units. 

\subsection{Results and Discussion}

This section reports and discusses the results of the simulated experiments on how well the causal estimands of interest can be estimated in each of the experimental set up. Estimation is performed based on the Bayesian imputation procedure proposed in
Section \ref{sec:BayesianImputationProcedure}.The network model parameters, specifically the nodal parameters on activity level and community mixed-membership $\X_G = \{\bm{\lambda}, \bm{\Pi}\}$, are estimated in order to potentially be included as covariates in the potential outcome model. For a detailed description of the ``hybrid mixed-membership blockmodels'' (HMMB) and the inference procedure for estimating the model parameters, please refer to Chapter \ref{ch:3}. The potential outcome model parameters are estimated using Bayesian regression with relatively flat Normal-Inverse-Gamma priors. The resulting posterior distributions on the primary effect coefficient $\tau$ and the peer treatment effect coefficient $\gamma$ are sufficient for imputing the posterior distributions of the causal estimands of interest, $\xi^\ave$ and $\delta^\fix(\bm{z})$. This is because, under the ordinary linear outcome model in Equation \eqref{eq:potentialOutcomeModel}, the population average primary effect causal estimand simplifies to $\xi^\ave = \tau$, according to Equation \eqref{eq:populationAveragePrimaryEffect}, and the fixed-assignment peer effect causal estimand simplifies to a scalar product of $\gamma$: $\delta^\fix(\bm{z}) = \omega \gamma$, according to Equation \eqref{eq:fixedAssignmentPeerEffect}. Here, $\bm{z}$ is the treatment assignment vector and $\omega = \frac{1}{N} \sum_{i=1}^N \sum_{j \in \N_{-i}}A_{ji}z_j $ is the population average exposure to peer treatments. The average exposure $\omega$ varies from experiment to experiment, but hovers between $19$ and $30$ in $90\%$ of the simulations.

Table \ref{table:estimationPerformanceLargeBeta} summarizes the adequacy of the posterior estimates of the primary and peer effect causal estimands and coefficients under different confounder type and covariate inclusion case, in terms of their frequentist coverage properties averaged over $2500$ simulations. The properties are specifically the posterior mean(standard deviation) and the truth coverage probability by the $90\%$ posterior interval. The posterior mean and standard deviation characterize the location and width of the posterior intervals. Inadequate estimation results are highlighted in the corresponding cells. These results will now be discussed one confounder type at a time, to demonstrate the importance of satisfying the unconfounded assumptions and ultimately the condition laid out in Section \ref{sec:unconfoundedAssignmentNetworkInterference}:

\renewcommand{\tabcolsep}{3pt}
\renewcommand{\arraystretch}{1.6}
\begin{table*}[!t]
\begin{centering}
\setlength{\belowcaptionskip}{6pt}
\caption[Estimation performance under each confounder type and covariate inclusion case]{Posterior estimate of the primary and peer effect causal estimands and coefficients under different confounder type and covariate inclusion, reported in terms of the posterior mean(standard deviation) and the truth coverage by the $90\%$ posterior interval, averaged over $2500$ simulations. The true values of the causal estimands and coefficients are: $\xi^\ave = \tau = 5, \frac{\delta^\fix(\bm{z})}{\omega} = \gamma = 0.1$ where $\omega$ is the average exposure to peer treatments for a given simulation. Inadequate estimation results are highlighted in the corresponding cells.}
\label{table:estimationPerformanceLargeBeta}
\scalebox{0.84}{
\hspace*{-0.25 cm}
\begin{tabular}{|c|c|c|c|c|c|l}
\hhline{|-|-|-|-|-|-|} 
\multicolumn{2}{|c|} {Experimental setup} & \multicolumn{2}{c|}{ \begin{minipage}{5 cm} \centering \vspace*{0.2 cm} Posterior estimate mean(standard deviation) \end{minipage}} & \multicolumn{2}{c|}{ \begin{minipage}{3.0 cm} \centering \vspace*{0.2 cm} Truth coverage \\ by $90\%$ interval  \end{minipage}}\\ 
\vspace*{-0.3 cm}
& & $\widehat{\xi}^\ave = \hat{\tau}$ & $\frac{\widehat{\delta}^\fix(\bm{z})}{\omega} = \hat{\gamma}$  &  \begin{minipage}{1.5 cm} \centering \vspace*{0.5 cm} $\widehat{\xi}^\ave, \hat{\tau}$ \end{minipage} &  \begin{minipage}{1.5 cm} \centering \vspace*{0.5 cm} $\widehat{\delta}^\fix(\bm{z}), \hat{\gamma}$ \end{minipage} \\
\begin{minipage}{4.2 cm} \centering \vspace*{-0.4 cm} Confounder type \end{minipage} & \begin{minipage}{4.0 cm} \centering \vspace*{-0.4 cm}Covariate included \end{minipage} & \begin{minipage}{2.7 cm}\centering $\xi^\ave = \tau=5$ \end{minipage} & \begin{minipage}{3.2 cm}\centering $\frac{\delta^\fix(\bm{z})}{\omega} = \gamma=0.1$  \end{minipage} & & \\ \cline{1-6}

\multirow{2}{*} {\begin{minipage}{3.7 cm} \centering Independent covariate \end{minipage}}
& None & 5.01(1.06) & 0.102(0.013)  & 0.901 & 0.902  & \\ \hhline{|~|-|-|-|-|-|}
& Ind. cov. & 5.00(0.21) & 0.100(0.003)  & 0.900 & 0.888  & \\ \hhline{|-|-|-|-|-|-|} 
\multirow{2}{*}{\begin{minipage}{3.7 cm} \centering Treatment likelihood confounder \end{minipage}}
& None & $-$4.17(1.39) \cellcolor{Pink} & 0.100(0.018)  & \cellcolor{Pink}  0.000 & 0.881  & \\ \hhline{|~|-|-|-|-|-|}
& Treatment like. cov. & 5.00(0.23) & 0.100(0.003)  & 0.891 & 0.902 &\\ \hhline{|-|-|-|-|-|-|}
\multirow{4}{*}{\begin{minipage}{3.7 cm} \centering Activity level confounder \end{minipage}}
& None & 5.04(0.25) & 0.123(0.003) \cellcolor{Pink} & 0.892 & 0.00 \cellcolor{Pink} &\\ \hhline{|~|-|-|-|-|-|}
& Activity level & 5.00(0.21) & 0.101(0.003)  & 0.887 & 0.876  &\\ \hhline{|~|-|-|-|-|-|}
& Comm. mem. & 5.06(0.25) & 0.124(0.003) \cellcolor{Pink}  & 0.890 & 0.00  \cellcolor{Pink} &\\ \hhline{|~|-|-|-|-|-|}
& All network cov.& 5.00(0.21) & 0.100(0.004)  & 0.897 & 0.890  &\\ \hhline{|-|-|-|-|-|-|}
\multirow{4}{*}{\begin{minipage}{3.7 cm} \centering Community mixed-membership confounder \end{minipage}}
& None & 5.00(0.40) & 0.105(0.005) \cellcolor{Pink}  & 0.889  & 0.497 \cellcolor{Pink}  & \\ \hhline{|~|-|-|-|-|-|}
& Activity level & 5.01(0.40) & 0.107(0.006) \cellcolor{Pink} & 0.882  & 0.433 \cellcolor{Pink} &\\ \hhline{|~|-|-|-|-|-|}
& Comm. mem. & 5.00(0.22) & 0.100(0.003)  & 0.900 & 0.895  &\\ \hhline{|~|-|-|-|-|-|}
& All network cov. & 5.00(0.22) & 0.100(0.004)  & 0.904 & 0.905  &\\ \hhline{|-|-|-|-|-|-|}
\end{tabular}
}
\bigskip
\bigskip
\end{centering}
\end{table*}

\begin{enumerate}

\item {\bf Independent covariate}: The posterior estimates of the causal estimands are adequate in this case, whether the covariate is included in the potential outcome model or not. Because the covariate is independent of the treatment assignment vector and the influence network, the unconfounded assumptions (i.e.\ Assumption \ref{as:unconfoundedTreatmentAssignmentUnderNetworkInterference} and \ref{as:unconfoundedInfluenceNetworkUnderNetworkInterference}) hold and the condition for ignoring the neighborhood treatment mechanism in Theorem \ref{thm:simplifiedImputationUnderNetworkInterference} is met. In other words, this covariate is not a confounder and this case serves as a benign baseline. When the covariate is not included, its effect is absorbed by the baseline effect estimate, $\hat{\mu}$, and the variance estimate $\hat{\sigma}^2_\epsilon$. The inflation in the variance estimate causes the posterior estimates of the primary treatment effect coefficient, $\hat{\tau}$, and the peer treatment effect coefficient, $\hat{\gamma}$, to have inflated variances and larger posterior intervals that adequately cover the truth. When the covariate is included, the potential outcome model is correctly specified and the posterior estimates adequately cover the truth with the narrowest posterior interval, which is the best case scenario.

\item {\bf Treatment likelihood confounder}: Not including the confounding covariate in the potential outcome model breaks the unconfounded treatment assignment assumption (i.e.\ Assumption \ref{as:unconfoundedTreatmentAssignmentUnderNetworkInterference}) and therefore the condition for ignoring the neighborhood treatment mechanism in Theorem \ref{thm:simplifiedImputationUnderNetworkInterference} is not met. This leads to inadequate posterior estimates on the primary treatment effect causal estimand $\widehat{\xi}^\ave$ and coefficient $\hat{\tau}$. Specifically, $\hat{\tau}$ has a large negative bias because the treated individuals have lower potential outcomes. The effect of this covariate is absorbed by $\hat{\tau}$ and $\hat{\sigma}_\epsilon^2$. The inflation in $\hat{\sigma}_\epsilon^2$ causes $\hat{\tau}$ and $\hat{\gamma}$ to have inflated variances but $\hat{\gamma}$ and $\widehat{\delta}^\fix(\bm{z})$  still adequately covers the truth. Including this confounding covariate in the potential outcome model recovers the posterior estimates in the best case scenario because the unconfounded assumptions and the condition in Theorem \ref{thm:simplifiedImputationUnderNetworkInterference} are met and the potential outcome model is correctly specified.

\item {\bf Activity level confounder}: Not including the confounding covariate in the potential outcome model breaks the unconfounded influence network assumption (i.e.\ Assumption \ref{as:unconfoundedInfluenceNetworkUnderNetworkInterference}) and therefore the condition for ignoring the neighborhood treatment mechanism in Theorem \ref{thm:simplifiedImputationUnderNetworkInterference} is not met. This leads to inadequate posterior estimates on the peer treatment effect causal estimand $\widehat{\delta}^\fix(\bm{z})$ and coefficient $\hat{\gamma}$. Specifically, $\hat{\gamma}$ has a positive bias because the more active nodes tend to be exposed to more peer influence, so the effect of this covariate is absorbed by $\hat{\gamma}$. The primary treatment effect causal estimand and coefficient are still adequate in this case. Including this confounding covariate in the potential outcome model recovers the posterior estimates in the best case scenario by meeting the unconfounded assumptions and the condition in Theorem \ref{thm:simplifiedImputationUnderNetworkInterference}. Note that under the four different inclusion cases of the network model parameters (i.e.\ none, activity level, community mixed-memberships, or all) as covariates in the potential outcome model, estimation performance is adequate as long as the activity level is included. This matches the conditions laid out in Theorem \ref{thm:unconfoundedInfluenceNetworkByConditioning} which states that the unconfounded influence network assumption is met as long as the subset of the confounding network model parameters are conditioned on.

\item {\bf Community membership confounder}: Not including the confounding covariate in the potential outcome model breaks the unconfounded influence network assumption (i.e.\ Assumption \ref{as:unconfoundedInfluenceNetworkUnderNetworkInterference}) and therefore the condition for ignoring the neighborhood treatment mechanism in Theorem \ref{thm:simplifiedImputationUnderNetworkInterference} is not met. This leads to inadequate posterior estimates on the peer treatment effect causal estimand $\widehat{\delta}^\fix(\bm{z})$ and coefficient $\hat{\gamma}$. Specifically, $\hat{\gamma}$ is biased because units in  communities with stronger interactions (e.g. the community cluster in the middle of Figure \ref{fig:treatmentOnInfluenceNetwork}) tend to receive greater amounts of exposure to peer treatments, so the effect of this covariate is absorbed by $\hat{\gamma}$. Also, the inflation in $\hat{\sigma}_\epsilon^2$ causes $\hat{\tau}$ and $\hat{\gamma}$ to have inflated variances. The primary treatment effect causal estimand and coefficient are still adequate in this case. Including this confounding covariate in the potential outcome model recovers the posterior estimates in the best case scenario by meeting the unconfounded assumptions and the condition in Theorem \ref{thm:simplifiedImputationUnderNetworkInterference}. Note that under the four different inclusion cases of the network model parameters (i.e.\ none, activity level, community mixed-membership, or all) as covariates in the potential outcome model, estimation performance is adequate as long as the community mixed-membership is included. This matches the conditions laid out in Theorem \ref{thm:unconfoundedInfluenceNetworkByConditioning} which states that the unconfounded influence network assumption is met as long as the subset of the confounding network model parameters are conditioned on.
\end{enumerate}

Overall, these results demonstrate the importance of meeting the unconfounded assumptions and the condition in Theorem \ref{thm:simplifiedImputationUnderNetworkInterference}, by conditioning on the confounding covariates. This is done in practice by including them in the potential outcome model. The Bayesian imputation procedure proposed in Section \ref{sec:BayesianImputationProcedure} is shown to accomplish this, resulting in adequate posterior estimates on the causal estimands. The consequences of not including the confounding covariates in the potential outcome model (i.e.\ not conditioning on them during the Bayesian imputation) is demonstrated here assuming a simple ordinary linear model for the potential outcomes. The negative consequences is likely to impact more causal estimands and model parameters with a more complex potential outcome model of choice.

\section{Conclusion}
Interference is not just a nuisance but rather an opportunity for quantifying and leveraging peer effects in many real-world experiments, including observational studies. When the mechanism of the interference is on a network of influence, we propose a framework to infer the primary effect as well as a wide range of peer effects, using the network potential outcomes as the basic building blocks. A theoretical framework is developed to guide the design of a Bayesian imputation procedure for estimating the causal estimands. Through simulation results, this framework is demonstrated to arrive at adequate causal estimates by including exposure to peer treatments and possible confounders in the potential outcome model. However, the true potential outcome model in real-world experiments is typically not known, so relying on the estimator to have the exactly correct potential outcome model may be unrealistic. Imputation from representative data through careful experimental design is important because it relieves the burden from the model and mitigate possible model mis-specification. In this way, causal inference under network interference should be best conducted by accounting for peer exposure and confounding covariates at both the design phase and the analysis phase. Chapter \ref{ch:2} proposes a full-factorial simulation study to investigate and demonstrate this intuition.

\chapter[{Design and Analysis for Causal Experiments on Networks: A Full Factorial Simulation Study}]{\huge Design and Analysis for Causal Experiments on Networks:\\ A Full Factorial Simulation Study}\label{ch:2}
\section{Overview}

Causal inference on populations in which individuals influence one another has garnered more interest lately, with the rise in experiments on social networks and social media. To demonstrate the best practice for causal experiments on networks, this chapter presents a plan for a full factorial simulation study to evaluate different combinations of experimental design strategies and estimators for the primary and peer effect causal estimands. Evaluation is done for a realistic set of truth settings, specifically on the network topologies and the generators for the potential outcome. The network potential outcome framework and its supporting theory, described in Chapter \ref{ch:1}, presents the need to include relevant network parameters as covariates in the potential outcome model for Bayesian imputation. Intuitively, the preferred course is to impute missing outcomes from representative observed outcomes and covariates, to avoid extrapolation with little supporting data. Achieving this through experimental design is especially important because the true potential outcome model in real-world experiments is typically not known, and imputation from representative data relieves the burden from the model and mitigates possible model mis-specification. Therefore, causal inference under network interference is best conducted by accounting for peer exposure and confounding covariates at both the design phase and the analysis phase. The theory and results from Chapter \ref{ch:1} suggest that a combination of novel balancing designs across treatment exposure groups and novel model-based estimators for the analysis should achieve the best overall performance.

\section {Introduction}
\label{sec:introduction2}

With the rise in social networks and media, more and more experiments are conducted on populations where the individuals are expected to interact and influence one another. In such experiments, the ``treatment'' not only affects the outcomes of the individuals receiving it, but also others through social influence. Experiments on networks take place in numerous application areas, for example, in public health and policy, social media, marketing, network security, and economics \citep{Kim2015, Christakis, Sobel2006, Gui2015, bond2012, Bakshy2012, Parker}. Depending on the application, the treatment may take the form of medicine, advertisement, incentives, and education, etc. In fact, treatment effects on individuals due to social influence have been considered for quite some time, under the phenomenon known as interference. However, it has been treated as a nuisance until recently as information on the social networks becomes available in more experiments. With information on the underlying network structure of the interference effect, there is now an opportunity to quantify and leverage peer effects in experiments in addition to quantifying the primary effect of the treatment. Chapter \ref{ch:1} proposes a framework to perform such inference through Bayesian imputation of missing network potential outcomes. Adopting this framework, this chapter proposes a simulation study to demonstrate how to account for interference and the network confounders at both the design phase and at the analysis phase.

\subsection{Background}

For a brief survey on the background literature for causal inference in the presence of general interference and network interference, please refer to Chapter \ref{ch:1}. Much of the existing work, including the framework adopted here, is built upon the potential outcome framework for causal inference \citep{rubin2005, ImbensRubin2015}. Under network interference, Chapter \ref{ch:1} defines the network potential outcomes and causal estimands to capture primary and peer treatment effects. Because most of the network potential outcomes are not observed, Bayesian imputation is used to compute the posterior distribution of the missing outcomes and the causal estimands of interest. Bayesian imputation of missing potential outcomes was developed in \cite{Rubin1978, Rubin1975} with the key unconfounded assumption that allows the imputation to ignore the treatment assignment mechanism. Chapter \ref{ch:1} extends the Bayesian imputation methodology and the key unconfounded assumptions for causal inference under network interference.

Imputation of missing potential outcomes from observed outcomes is at the core of causal inference, which can be fundamentally viewed as a missing data problem. If the observed outcomes provide good representation of the missing outcomes in the causal estimand, imputation can be done directly through taking the mean of the observed outcomes in each treatment group. The classical Neyman estimator \citep{neyman1923} is an example of this for causal inference under SUTVA (Assumption \ref{as:SUTVA}) and complete randomization. Under network interference, observing outcomes that are representative of the missing outcomes can be challenging, due to the complexity of the exposure to peer treatments and the structural constraint by the influence network. \cite{Toulis2013} illustrate this challenge and propose sequential randomization schemes to observe some of the potential outcomes in the causal estimand. However, the observed outcomes for each treatment exposure group can be few and unrepresentative of the missing outcomes. The advantage of direct imputation through the mean estimator is that it makes no model assumption, but the resulting estimate is subject to bias and large variance from few and unrepresentative observed outcomes.

A natural solution to the problem of having few observed outcomes for each of the treatment exposure group in the causal estimand is to employ model-based imputation of missing outcomes. Under this approach, all of the observed outcomes may be used to estimate the parameters of the potential outcome model, from which missing outcomes are imputed. This alleviates the need to observe outcomes that match precisely in the treatment exposure condition  to those outcomes in the causal estimands, which is especially critical for observational studies where the treatment assignment vector is not determined by the experimenter. The Generalized Linear Model (GLM) is a flexible class of models that can accommodate different types of outcomes (e.g.\ binary, continuous, and count data) and different relationships between the linear predictor and the outcomes. The GLM is well established and widely used in statistical modeling \citep{Agresti2015}, and has been used for modeling outcomes in causal inference by \cite{Bang2005}. Parameters of the GLM can be estimated with Bayesian methods for computing the posterior distributions of the missing potential outcomes \citep{Gelman2014, Gelman2015, Polson2013}. For model selection and evaluation, a natural measure on the goodness of fit is the posterior predictive check \citep{Gelman1996} which evaluates how representative the observed data is under its predictive posterior distribution under the model. Following the theoretical development in Chapter \ref{ch:1} that results in the imputation procedure in Section \ref{sec:BayesianImputationProcedure}, the unit's own treatment, its exposure to peer treatments, and the possible confounding covariates (e.g.\ treatment assignment and network confounders) should be included in the potential outcome model. 

Under network interference, the exposure to peer treatments is a critical part of the network potential outcome model. Social influence has been modeled as a propagation or cascades on the social network, originating from works in social sciences and economics \citep{Bikhchandani1992}. Often known as the Independent Cascade Model, recent works in computer science continue to build on this idea by proposing ways to estimate the model parameters \citep{Goyal2010} as well as to maximize social influence through optimal selection of the treatment receivers \citep{Kempe2003}. The network potential outcome models may incorporate exposures to treatments on peers using this idea of propagation on social networks, up to an arbitrary number of hops according to the size of the neighborhood of influence in Definition \ref{def:nHopNeighborhood}.

Intuitively, the preferred course is to impute missing outcomes from representative observed outcomes and covariates, to avoid extrapolation with little supporting data \citep{ImbensRubin2015}. Achieving this through experimental design is especially important because the true potential outcome model in real-world experiments is typically not known, and imputation from representative data relieves the burden from the model and mitigates potential model mis-specification. This principle is demonstrated in \cite{Ho2007} for observational studies using matching and parametric estimators. Similarly, causal inference under network interference is best conducted by accounting for peer exposure and confounding network covariates not only at the analysis phase but also at the design phase. Balancing the covariates on the observed outcomes between different treatment groups make them more representative of the missing outcomes. In the absence of interference (i.e.\ under SUTVA), balancing the covariates through rerandomization has been shown to lead to causal estimates that are closer to the truth \citep{Morgan2012}. When balancing all the covariates is difficult due to competition between the large number of covariates, \cite{Morgan2015} propose a multi-tier method to prioritize covariates according to their importance. The proposed simulation study in this chapter will apply multi-tier rerandomization to experiments under network interference, to balance the relevant covariates and the number of observed potential outcomes under each discretized exposure to peer treatment group. Although not in the scope of this simulation study, the same principle may be applied to observational studies under network interference, through matching \citep{Stuart2010} to produce the same kind of balance between each peer treatment group.

\subsection{Contributions}

Adopting the network potential outcome framework in Chapter \ref{ch:1}, this chapter proposes procedures to account for exposures to peer treatments and confounding network covariates at both the experimental design and the analysis phase. At the experimental design phase, various randomization schemes are considered including novel schemes of rerandomization for units in a network. The novel rerandomization schemes attempt to observe balanced outcomes and covariates in order to provide a good representation of the outcomes across different treatment exposure groups in the causal estimands. They are compared with the classical complete randomization as well as other randomization schemes from recent literatures on experiments on networks. At the analysis phase, various estimators are considered including the classical difference in means estimator and novel model-based Bayesian estimators that incorporate various possible confounding covariates and assume different relationships between the predictors and the outcome. 

To demonstrate the best practice for causal experiments on networks, this chapter presents a plan for a full factorial simulation study to evaluate all combinations of these randomization schemes and estimators for the primary and peer effect causal estimands. Evaluation is done for a realistic set of truth settings, specifically on the network topologies and the generators for the potential outcome. Full factorial experiments provide the statistics for rigorous analysis to identify the key factors and their interactions that influence the performance \citep{Box2005}. The factors in this simulation study capture the three main aspects of causal experiments: 1.) the true process that determines the outcomes given the treatment vector (i.e.\ the science),  2.) the experimental design that determines the treatment assignment vector, and 3.) the analysis that produces estimates of the causal estimand. Similar to the study by \cite{Rubin1979}, analysis of variance (i.e\ ANOVA) can be first performed to identify influential factors and interactions at a high level, before taking closer looks at specific factors using performance tables to compare the efficacy of combinations of design strategies and estimators under different truth settings. 

Section \ref{sec:simulationStudySetup} describes the set up of the full factorial simulation study, including the causal estimands of interest and the factors relating to the science, the experimental design, and the analysis. Section \ref{sec:simulationStudyAnalysis} describes the analysis plan for the simulation study to identify the key factors that drive the performance of the causal inference and the best practice for experiments on networks under realistic settings, specifically in the presence of confounders and with imperfect knowledge of the true potential outcome model.

\section{Full Factorial Simulation Study Setup}
\label{sec:simulationStudySetup}

The simulation study plan in this chapter is full factorial, meaning all combinations of levels across each factor will be realized,  in order to explore the interaction effects between factors. Some of these interactions may be of higher order, because certain combinations of randomization scheme and estimator may address the challenges from certain truth settings, for example, the presence of confounding network covariates, network topologies that make it difficult to realize outcomes in the causal estimand exactly, etc. This section describes in detail the setup of the simulation study, beginning with the causal estimands of interest, then all the factors in each of the three aspects of the causal experiments: the science, the experimental design, and the analysis.

\subsection{Causal Estimands of Interest}
\label{sec:causalEstimandsOfInterest}

One of the key advantage of the framework from Chapter \ref{ch:1} is its flexibility to represent causal effects from both the primary treatment as well as the exposure to treatments on peers. For primary effect, the population average causal estimand is a natural choice for this simulation study, as it represents the classical average treatment effect under network interference:

\begin{equation}
\label{eq:populationAveragePrimaryEffectCh2}
	\xi^\ave \equiv \frac{1}{N} \sum_{i=1}^N \left[ \frac{1}{2^{|\N_{-i}|}} \sum_{\z \in \ZSetOpen} {Y_i(Z_i = 1, \Zopen  = \z , \A_{\N_i}) -  Y_i(Z_i = 0,  \Zopen = \z ,\A_{\N_i})} \right]
\end{equation}

\noindent This estimand is averaged over the population as well as all possible neighborhood treatment assignment vectors for each unit. For a more detailed description of this estimand, please refer to Section \ref{sec:primaryCausalEstimand}.

For peer effect, the population average $k$ treated neighbors causal estimand is chosen for its straightforward peer effect representation of having exactly $k$ neighbors being treated.

\begin{equation}
	\delta_ k^\ave \equiv \frac{1}{|\VSet_{\geq k}|} \sum_{i \in \VSet_{\geq k} }\left[{\frac{1}{2} \sum_{z=\{0,1\}} {\delta_{i, k} (z)}}\right]
\end{equation}

\noindent where

\begin{equation}
\delta_{i, k} (z) \equiv {|\N_{-i}| \choose  k}^{-1} 
	\sum_{\z \in \ZkSet} {Y_i(Z_i=z,\Zopen = \z, \A_{\N_i}) - Y_i(Z_i = z, \Zopen = \bm{0}, \A_{\N_i})}
\end{equation}

\noindent This estimand is averaged over the qualified population, $\VSet_{\geq k}$ (i.e.\ those units with at least $k$ neighbors), and over each unit's own treatment and its neighborhood treatments, $\ZkSet$, that have exactly $k$ neighbors treated. In this simulation study, for simplicity, the neighborhood of influence for each node $i$ only consists of its immediate neighbors (i.e.\ one-hop neighborhood). For a more detailed description of this estimand, please refer to Section \ref{sec:kTreatedNeighborCausalEstimand}.

In this simulated study, the true values for these estimands can be computed on each generated network population from the true potential outcome model and its parameters. The knowledge of the science (i.e.\ true process) is only used for evaluation purpose. The inference procedures at the design and analysis phase of course do not assume any such knowledge. 

\subsection{Factors Under Study}
\label{sec:factorsUnderStudy}

The factors in this simulation study capture the three main aspects of causal experiments: 1.) the true process that determines the outcomes given the treatment vector (i.e.\ the science),  2.) the experimental design that determines the treatment assignment vector, and 3.) the analysis that produces estimates of the causal estimand. The factors and their levels are designed for exploring the best practice for causal experiments on networks, specifically, which combinations of randomizations and estimators perform well under different truth settings. Overall, the full factorial study consists of $3 \times 3 \times 3 \times 3 \times 3 \times 3 \times 2 \times 2 \times 2 \times 8 \times 12 = 559872$ cases. For each case, ten replications are performed for each of the two causal estimands of interest, making a total of about $11.2$ million simulated experiments on networks.  

\subsubsection{Truth Factors}
\label{sec:truthFactors}

To explore causal inference performance over a range of realistic settings, the truth factors below capture the two aspects of the generating process (i.e.\ the science) for outcomes of network experiments: 1.) the network of peer influence, and 2.) the potential outcome model. Overall, the truth factors consist of $3 \times 3 \times 3 \times 3 \times 3 \times 3 \times 2 \times 2 = 2916$ truth settings.

\paragraph{Influence Network Topology Factors:}
The structure of the influence network impacts the performance of the inference procedure because it determines how challenging it is to find a treatment assignment vector that renders observed outcomes that represent the missing outcomes in the casual estimands. The influence networks will be generated using the Hybrid Mixed-Membership Blockmodels (HMMB) presented in Chapter \ref{ch:3}, for its capacity to capture key characteristics of real-world networks. The influence network may not be fully known by the experimenter but some a-prior information on the rate of influence is typically collected (e.g. nature of relationship, frequency of interactions, etc.). The HMMB provides parameters that enable easy adjustment on different factors of the network topology. Examples of such adjustment can be found in Section \ref{sec:estimationPerformance}. Five factors of network topology will be explored, each with three levels: 
\begin{enumerate}
  \item Network size: A larger network provides more capacity for rendering observed outcomes to represent the missing outcomes in the causal estimand. However, it also increases the number of missing outcomes to be imputed. For this simulation study, a reasonable set of sizes are $N=100$, $500$, and $1000$.  
  \item Network density: The density of the influence network directly impacts the amount of exposure to treatments on peers. A denser network will result in stronger peer influence and more structural constraint for the treatment assignment mechanism, by making it harder to isolate exposure to peer treatments. Real-world network densities vary as a function of the network size, and a set of realistic densities for networks with appropriate connectivity are $s=3\log(N)/N$, $6\log(N)/N$, and $9\log(N)/N$.
  \item Number of communities: Similarly, a larger number of smaller communities will make it easier to isolate exposure to peer treatments. However, it will also make it more difficult to observe many outcomes at the specific exposure levels within each community. Realistically, the number of communities increases with the network size $N$ at a particular rate. For this study, a reasonable set of community sizes are $K \approx N^{0.2}$, $N^{0.3}$ and $N^{0.4}$.   
  \item Community structure: How distinct the community structure is on the network (i.e.\ modularity) determines how focused or diffused the peer influence is. Weak community structure may make it harder to isolate exposure to peer treatments, while communities that are very densely connected present challenges to render outcomes at various levels of peer exposure within each community. The strength of community structure can be adjusted through the between-community interactions in the block matrix. Please refer to Section \ref{sec:HMMBModelDescription} for details. For this simulation study, a set of appropriate levels of community structure strength can be achieved by scaling the off-diagonal entries of the baseline block matrix by $0.5$, $1$, and $2$. Smaller scales result in less overlap between communities and, therefore, stronger (i.e.\ more distinct) community structure. The baseline block matrix depends on the number of communities and has larger diagonal entries and sparse off-diagonal entries for realistic networks. A reasonable baseline block matrix for three communities is $\bm{B} = \left[\begin{smallmatrix} 2.3&0.07&0\\ 0.3&2&0\\0&0.3&3 \end{smallmatrix}\right]$, for four communities, $\bm{B} = \left[\begin{smallmatrix} 2.3&0.07&0&0\\ 0.3&2&0&0\\ 0&0&2.5&0.4 \\0&0.3&0&3 \end{smallmatrix}\right]$, and for 6 communities, $\bm{B} = \left[\begin{smallmatrix} 2.3&0.07&0&0&0&0.4\\ 0.3&2&0&0&0&0\\ 0&0&2.5&0.4&0.2&0 \\0&0.3&0&3&0&0 \\0&0&0.25&0&2.5&0 \\0&0.3&0&0&0&2.7 \end{smallmatrix}\right]$.
  \item Node degree distribution: Realistic networks are composed of nodes with diverse degrees, as individuals differ from one another in their level of activity. As real-world networks typically demonstrate Power-Law degree distribution with exponent from $-3$ to $-2$ \citep{Clauset2009}, with a larger exponent corresponding to a heavier tail and more high-degree nodes. More high-degree nodes may make it more difficult to isolate exposure to peer treatments. A reasonable set of exponents for this simulation study are $-\alpha = -2.8$, $-2.4$, and $-2.1$.
\end{enumerate}	

The level of mixed-memberships, represented by the lifestyle pseudocount matrix, is not a factor of interest in this simulation study. It stays the same at a moderate level for each of the community size. The baseline pseudocount matrix has larger diagonal entries and sparse off diagonal entries. A reasonable pseudocount matrix for three communities is $\bm{X} = \left[\begin{smallmatrix} 5&0.1&0\\ 0&5&0.5\\ 0.1&0.2&2\end{smallmatrix}\right]$ , for four communities, $\bm{X} = \left[\begin{smallmatrix} 5&0.1&0&0.3\\ 0&5&0.5&0.1\\ 0.1&0.2&2&0 \\ 0&0.5&0.3&3\end{smallmatrix}\right]$, and for six communities, $\bm{B} = \left[\begin{smallmatrix} 5&0.1&0&0&0&0.5\\ 0.3&2&0&0&0.2&0\\ 0&0&2.5&0.4&0.2&0 \\0&0.3&0&3&0&0.5 \\0.3&0&0.25&0&2.5&0 \\0&0.3&0&0.5&0&2.7 \end{smallmatrix}\right]$. For simplicity, the spread of nodes across different lifestyles is kept even, represented by the uniform vector $\bm{\phi} = \bm{\frac{1}{K}}$ of $K$ entries.

\paragraph{Potential Outcome Model Factors:} 
The true generative model for the potential outcome determines each unit's outcome given its own treatment, its exposure to peer treatments, and the relevant covariates. The potential outcome model will be variations of the additive linear model below with independent Normal variation:
\begin{equation}
\label{eq:truthPOModel}
	Y_i =  \tau Z_i + \gamma \; g_i(\A, \Z) + \sum_{m} {\beta_m h(x_{m,i})} + \mu + \epsilon_i
\end{equation}
The first term, $\tau Z_i$, represents the primary treatment effect. The second term, $\gamma \; g_i(\A, \Z)$, represents the peer effect from exposure to treatments on neighbors, $g_i(\A, \Z)$. In general, the exposure function $g_i(\A, \Z)$ can be of many forms consisted of the influence network $\A$ and the treatment vector $\Z$. The third term $\sum_{m} {\beta_m h(x_{m,i})}$ is the effects of each of the unit covariate $\bm{x}_{m,i}$ transformed by a function $h()$ that describes the relationship between the outcome and the covariate. The fourth term, $\mu$, is the baseline effect on the entire population. The last term, $\epsilon_i \sim \Normal(\mu_\epsilon,\sigma_\epsilon^2)$, gives independent and identically distributed variation for heterogeneity between the units. The specific values of the potential outcome model parameters will be similar to those in the simple simulation study in Section~\ref{sec:confounderExperimentSetup}. For example, $\tau = 5$, $\gamma = 0.1$, $\beta_\lambda = 5$ for the popularity confounder, $\bm{\beta_\pi} = [0, 1, 2, ... , K-1] * \frac{6}{K-1}$ for the community membership confounder with $K$ communities, $\mu = 3$, $\mu_\epsilon = 2$, and $\sigma^2_\epsilon = 1.2$. These parameter values reasonably represent outcomes with a significant primary effect, a much smaller peer effect per unit exposure, significant effects by the confounders, and a moderate variation between units. The baseline effect ($\mu$) and the offset ($\mu_\epsilon$) should not impact the performance of the causal inference procedures, but are included as a sanity check.

The $3 \times 2 \times 2$ factors with a total of $12$ settings below are designed to explore the performance impact from the presence of network confounders and model mis-specification later in the analysis phase:
\begin{enumerate}
  \item Presence of network confounders. The three levels are: 1.) no confounder so the third term in the model above is absent, 2.) popularity confounder by including the network parameter $\lambda_i$ for each unit as covariate in the model, and 3.) community membership confounder by including the network parameter $\bm{\pi}_i$ as covariates in the model. Please refer to Section \ref{sec:HMMBModelDescription} for details on these network parameters.
  \item Identity covariate effect, $h(x)=x$, versus log covariate effect, $h(x)=\log(x)$. Note that when there is no confounder, both levels of this factor produce the same potential outcome model. So in practice, two settings under the $12$ potential outcome settings do not need to be replicated in the simulation study. 
  \item Simple sum peer exposure,  $g_i(\A, \Z) = \sum_{j \in \N_{-i}} A_{ji}Z_j$, for additive peer effect versus log sum exposure, $g_i(\A, \Z) = \log(\sum_{j \in \N_{-i}} A_{ji}Z_j)$, for diminishing peer effect. As mentioned earlier in the description of the causal estimands, this study will consider exposure to treatments on immediate neighbors only, for simplicity. 
\end{enumerate}
\noindent The last two factors introduce variations in the potential outcome model that result in model mis-specification later in the analysis phase for realism. In real-world experiments, one can not assume to know the true potential outcome model. Model mis-specification in the analysis phase could be mitigated by balancing assignment mechanisms in the experimental design phase. 

\subsubsection{Experimental Design Factors}
\label{sec:experimentalDesignFactors}	

Experimental design is a critical part of causal inference. It takes place before the experiment is conducted, where a randomization scheme arrives at a treatment assignment vector (i.e.\ which units are treated) according to a certain criterion or mechanism. A good randomization scheme renders observed outcomes that represent well the missing outcomes in the causal estimand. The factors below include different randomization schemes that will be evaluated in this simulation study, including the novel rerandomization schemes for causal experiments on networks. Overall, the experimental design factors consist of $2 \times 8 = 16$ settings.

\paragraph{Treatment Resource Factor:} This factor has two levels that correspond to low treatment resource and medium treatment resource. In the low resource level, there is only enough resource to treat a small fraction (e.g.\ $3\%$) of the population whereas in the medium resource level, a larger percentage (e.g.\ $10\%$) of the population can receive the treatment. This factor reflects the treatment resource constraint in many of the real-world experiments. Note however, that the randomization scheme may choose not to use all of the treatment resource available if treating fewer units fulfills its criterion or fits its mechanism better. In other words, the treatment resource is available but not necessary to be utilized fully.

\paragraph{Randomization Scheme Factor:} Eight different randomization schemes will be evaluated, corresponding to the eight levels in this factor.
\begin{enumerate}
  \item Complete randomization (CR): This is the classical randomization scheme where the treatment is assigned across units with uniform probability. In the absence of interference (i.e.\ under SUTVA), this randomization produces unconfounded assignments so the classical Neyman estimator for population average treatment effect is unbiased. However, this randomization is likely to be insufficient under network interference especially in the presence of network confounders.
  \item Sequential randomization (SR): This randomization scheme selects eligible units sequentially with uniform probability to put under each of the treatment exposure groups in the causal estimand by treating or controlling it and its neighbors accordingly. As this process continues, many units will become ineligible because of its connection to the assigned units. The objective of this randomization scheme is to observe some outcomes in each of the treatment exposure groups in the causal estimand. However, few units will qualify in the end, due to constraint by the network topology. This scheme is designed specifically for the difference in means estimator under network interference. For a more detailed description of this randomization, please refer to \cite{Toulis2013}.
  \item Insulated neighbors randomization (INR): This is a variant of the sequential randomization, developed in \cite{Toulis2013}. Before starting the sequential assignment process, a adequate fraction (e.g.\ $60\%$) of the shared neighbors of the eligible units are put under control. The intention behind this extra initialization step is to increase the number of observed outcomes in each of the treatment exposure groups at the end, by avoiding treating units that will render many units ineligible.  
  
  \item Rerandomization for covariate balance between treatment exposure groups (RC): Observing balanced covariates between the treatment exposure groups of interest enable imputation of missing outcomes from more representative observed outcomes. Therefore, it may mitigate biased inference from confounding covariates that may not be accounted for entirely during the analysis phase due to model mis-specification \citep{Ho2007}. In the regular case without interference (i.e.\ under SUTVA), there are only two treatment groups consisting of the units under treatment and the units under control. Rerandomization for covariate balance has been developed for causal inference under SUTVA and shown to improve the precision of the causal estimate \citep{Morgan2012}. Basically, random assignment vectors are drawn until one is found that fulfills the covariate balance criterion measured using the Mahalanobis distance between the covariates in the two groups. When there are a large number of covariates, finding treatment assignment vectors that balance all of them can be difficult. Prioritizing the covariates into multiple tiers is a practical solution to this challenge and has been integrated into the rerandomization framework \citep{Morgan2012}.

Under network interference, there are many treatment groups each consisting of units receiving numerous different levels of exposure to peer treatments. However, balancing can still be done between groups with discretized levels of exposure. Under such discretization, the observed outcomes belong to fewer groups, for example: high, medium, and low exposures. An example definition of a discretized treatment exposure group, $\mathscr{T}_g$, is the set of all units with total exposure to peer treatments within a specified range:

\begin{equation}
\label{eq:discretizedTreatmentGroups}
   \mathscr{T}_g = \{i\} \;\; | \;\;  L_g \leq  \sum_{j \in \N_{-i}} A_{ji}Z_j \leq H_g
\end{equation}

Balancing all the possible confounding covariates, such as unit popularity and community memberships, can be difficult across all the treatment exposure groups, so the multi-tier prioritization of the covariates comes in handy. Rerandomization for covariate balance does not cause biased estimate conditional on the covariates, as the treatment mechanism is conditionally independent of the potential outcomes, meeting the unconfounded treatment assignment assumption (Assumption \ref{as:unconfoundedTreatmentAssignmentUnderNetworkInterference}). Lastly, the network covariates are not typically known a-priori, but can be estimated using the a-prior knowledge on the influence network. This form of rerandomization for experiments on networks described here and in the next two schemes is novel.

  \item Rerandomization for balance between treatment exposure groups (RG): Similarly, observing a balanced number of outcomes between the treatment exposure groups of interest enable imputation of missing outcomes from more representative observed outcomes. It may mitigate model mis-specification on the peer exposure during the analysis phase.  This balance criterion can be easily introduced to the rerandomization procedure by enforcing that the largest group can be no bigger than the smallest group by a certain percentage. Again, this criterion does not cause biased estimate, as the treatment mechanism here is driven entirely by the influence network and, therefore, conditionally independent of the potential outcomes, meeting the unconfounded treatment assignment assumption (Assumption \ref{as:unconfoundedTreatmentAssignmentUnderNetworkInterference}).

  \item Rerandomization for both balances above (RCG): Balancing both the covariates as well as the number of observed outcomes between the treatment exposure groups is the most comprehensive rerandomization option considered in this study. With the network constraints and competing balancing objectives, the multi-tier prioritization is practically necessary. Figure~\ref{fig:covariateBalanceSimulatedNetwork} is an example of the resulting treatment exposure groups, as defined in Equation \eqref{eq:discretizedTreatmentGroups}, with much better balance overall through rerandomization. The improvement is especially significant for the popularity covariate. Using complete randomization, units with high activity level will mostly fall under the treatment groups with high exposure to peer treatments, but as seen in Figure~\ref{fig:balanceLayout}, rerandomization is able to find assignment vectors that distribute nodes with varying activity level much more evenly among the different treatment exposure groups. Units with different community memberships, as seen in the four clusters in the network layout, are also distributed more evenly among the different groups. Figure~\ref{fig:balanceChart} shows the covariate balance for each of the six covariates between the treatment groups with high exposure and low exposure. The intervals are computed from independent draws using complete randomization versus rerandomization. Comparing the dotted(top) line against the solid(bottom) line, it is clear that rerandomization improves the balance for each of the covariate. However, rerandomization is still not able to balance every covariate perfectly, due to the constraint imposed by the influence network.

\begin{figure}[t]
    \centering
    \begin{subfigure}[b]{0.38\textwidth}
        \includegraphics[width=\textwidth]{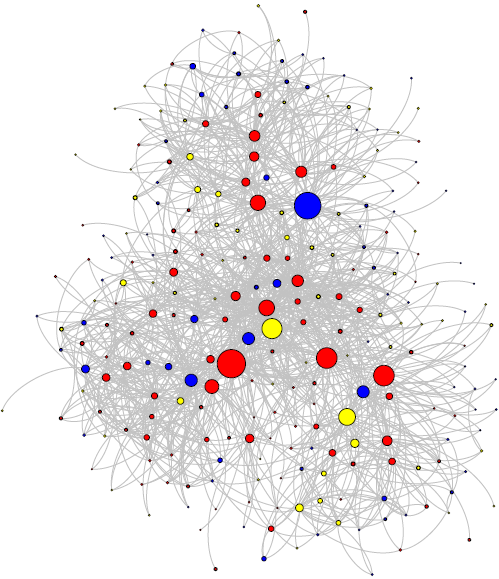}
        \caption{Treatment groups layout}
        \label{fig:balanceLayout}
    \end{subfigure}
    ~ 
    \begin{subfigure}[b]{0.53\textwidth}
        \includegraphics[width=\textwidth]{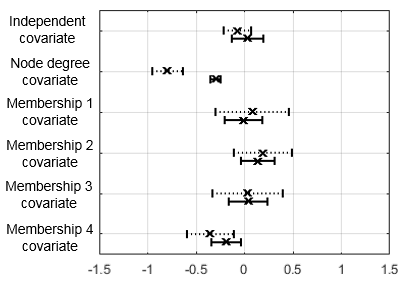}
        \caption{Improved covariate balance}
        \label{fig:balanceChart}
    \end{subfigure}
    \caption[More balanced treatment groups under rerandomization]{An example of improving balance between the treatment groups using rerandomization. For an accepted assignment vector under rerandomization, the figure on the left colors the units by which discretized treatment exposure group they belong to: red for high exposure, yellow for medium, and blue for low. Edges indicate influence and the node size indicates popularity. The figure on the right compares covariate balance between complete randomization and rerandomization, in terms of the difference in standardized covariate mean between the treatment groups with high exposure and low exposure. For each covariate, the dotted(top) line shows the $68\%$ interval of the difference in mean for assignment vectors under complete randomization and the solid(bottom) line under rerandomization. }
    \label{fig:covariateBalanceSimulatedNetwork}
\end{figure}

  \item Preferential treatment on influential units (PT): Many experiments on networks attempt to maximize the overall treatment effect by treating the most influential nodes \citep{Kim2015, Kempe2003}. However, in the context of causal inference, this treatment assignment scheme may lead to biased estimates especially in the presence of the popularity confounder. To investigate this issue, a preferential treatment randomization scheme is included in this study, where units with higher out-degrees in the influence network will be selected with higher probability for treatment. 

  \item Randomization by network cluster (RNC): Randomization by groups of connected individuals, where the whole group is either all treated or all under control, may be employed in experiments for various reasons. Sometimes, it is due to the nature of the treatment dissemination or for convenience \citep{Puffer2005}, which can trace its root back to the classical split plot design \citep{Fisher1925}. For causal inference under network interference, \cite{Ugander2013} proposes cluster randomization in order to compute the treatment exposure probability under specific causal estimands. However, this randomization may lead to large variance in the causal estimate in the presence of community membership confounder.
\end{enumerate}

\subsubsection{Analysis Factor}
\label{sec:analysisFactor}	

The final analysis phase of a causal experiment is the estimation of the causal estimand using outcomes observed from the experiment. The performance of this estimation is better if the observed outcomes represent the missing outcomes well, which is the motivation behind the experimental design phase. However, wherever observed outcomes fall short in representing the missing outcomes, the estimator has an opportunity to account for it. This is typically the case under network interference. The analysis factor in this simulation study consists of $12$ levels, to compare novel estimators that attempt to account for the exposure to peer treatments and the confounding covariates using model-based Bayesian imputation against classical causal estimators using difference in means between observed outcomes in each treatment group:

\begin{enumerate}
 \item Classical Neyman estimator (NM): The Neyman estimator simply takes the difference of means between the treated and the controlled group, as shown in Equation \eqref{eq:neymanEstimator}. It is designed for estimating the population average treatment effect for experiments without interference (i.e.\ SUTVA holds). It also assumes that the observed outcomes represent the missing outcomes well, typically through complete randomization or matching in observational studies. Under interference and in the presence of network confounders, its estimate on the primary effect causal estimand may be biased and may have large variance. Because it assumes away interference, its estimate on the peer effect causal estimand is simply zero.
  \item Difference in means across groups of observed outcomes (DM): This estimator is a straightforward extension of the classical Neyman estimator. It is still based on taking the difference in means between observed outcomes in different treatment groups of the causal estimand. However, each unit's exposure to peer treatments is now considered for a more precise determination on which treatment group it belongs to. The success of this estimator still depends largely on the ability to observe outcomes that represent the missing outcomes well, through tailored randomizations (e.g.\ the sequential randomization and the insulated neighbors randomization). For this simulation study, if there is not enough matching observed outcomes to those in the causal estimand, the closest outcomes will be used in place of the exact matches. This heavy dependence on the randomization may not be practical in the presence of network confounders and when very few observed outcomes will match the peer treatment conditions in the causal estimand, due to the constraint by the network topology. However, its distinct advantage is that it makes no model assumption on the potential outcome. 
  \item [3-12.] Model-based Bayesian imputation: Bayesian imputation of missing outcomes using a potential outcome model is a natural solution in response to the difficulty of observing potential outcomes at numerous peer treatment conditions, especially in the presence of network confounders. With these novel estimators, all of the observed outcomes may be used to estimate the parameters of the potential outcome model, from which missing outcomes are imputed. This alleviates the need to observe outcomes that match precisely in the treatment exposure condition  to those outcomes in the causal estimands. However, it is unrealistic to assume perfect knowledge of the true potential outcome model, so some level of model mis-specification should be expected. The ten levels here correspond to each of the potential outcome model variations specified in Section \ref{sec:truthFactors} and Equation \eqref{eq:truthPOModel}, on the presence of different confounders and the relationship between the predictor and the outcome. The cases where the imputation model matches the true potential outcome model represent the best but overly ideal scenarios. The more realistic and interesting cases are when the imputation model attempts to account for the exposure to peer treatments and network confounders, but with model mis-specification from assuming the wrong relationship between the predictor and the outcome. The model mis-specification during the analysis phase can be mitigated by the balancing randomization schemes in the experimental design phase. Each specific Bayesian imputation estimator will be denoted in short hand as BNS (no confounder, sum peer exposure), BIPS (identity popularity confounder, sum peer exposure), BLPS (log popularity confounder, sum peer exposure), BICS (identity community membership confounder, sum peer exposure), BLCS (log community membership confounder, sum peer exposure), BNL (no confounder, log sum peer exposure), BIPL (identity popularity confounder, log sum peer exposure), BLPL (log popularity confounder, log sum peer exposure), BICL (identity community membership confounder, log sum peer exposure), and BLCL (log community membership confounder, log sum peer exposure).  
\end{enumerate}

\section{Simulation Study Analysis}
\label{sec:simulationStudyAnalysis}
The full factorial simulation study described in Section \ref{sec:simulationStudySetup} provides the statistics for rich analysis on which factors and their interactions have significant effects on the causal inference performance, as well as on which specific levels and combinations of levels (e.g.\ a specific combination of the randomization scheme and estimator) produce the best performance overall. Analysis will be performed for both the primary effect causal estimand and the peer effect causal estimand. Analysis of full factorial experiments has its root in classical experimental design literature and is well documented in \cite{Box2005}. The analysis here is on factors that span the aspects of the truth generating process, the randomization scheme, and the causal estimator. Much of the analysis in this simulation study has similar structure to \cite{Rubin1979}.

\subsection{Performance metric}
\label{sec:performanceMetric}

A common performance metric must be defined across all the estimators in Section \ref{sec:analysisFactor}, so that the results of the full factorial study can be compared fairly. Both the Bayesian and point estimators will be evaluated based on their frequentist property over repeated experiments. With the ten replications for each factorial setting and having the true value for the causal estimands, a natural and concise choice of metric is the mean square error (MSE) over the experimental replicates. Computation of the MSE is straightforward for the Neyman and the difference in means estimators, because they produce point estimates. For the Bayesian imputation estimators, the squared error for each experiment replicate should be integrated over the entire posterior density reported by the estimator. This calculation is also used by existing literature \citep{Box2007} for measuring the integrated mean squared error of an estimated density to the true density. In this study, the true density can be interpreted as a point mass at the true value. The integrated squared error represents the expected squared error for each replicate, which is then averaged over the ten replicates. So for the Bayesian estimators, the MSE is computed with expectation over both the posterior density estimate as well as over the experimental replicates. This metric presents a fair comparison between the point estimators and the Bayesian estimators because in both cases, it captures the expected squared error.

\subsection{Analysis of Variance to Identify Significant Factors and Interactions}
\label{sec:ANOVA}

Analysis of variance (ANOVA) on the causal inference performance over all the factors provides an informative first-step analysis in locating all the significant factors and their interactions on the performance. Specifically, ANOVA attempts to explain the total variation in the MSE across all the cases through individual factors and their interactions. If a significant amount of the variation can be explained through a particular factor or interaction of factors, it indicates that it is an important driver for the performance. Given that there are a total of nine factors in the study, results on the full ANOVA may be tedious and unnecessarily detailed. After performing the full ANOVA, it may be prudent to perform ANOVAs with a reduced number of factors by collapsing some (e.g. the truth factors) into one, and marginalizing or conditioning some factors out, to focus explorations on subspaces with meaningful factors and interactions. While each factor is expected to have significant impact on the performance, the following interactions are expected to be significant and meaningful:

\begin{itemize}
  \item Randomization scheme factor and analysis factor: Certain combinations of randomization schemes and estimators are expected to be much more effective than the others, on average over all of the truth settings.
  \item Randomization scheme factor and potential outcome model factors: Certain randomization schemes are able to render more representative observed outcomes to account for interference and the presence of confounders.
  \item Analysis factor and the potential outcome model factors: Certain estimator will be able to better account for interference and the presence of confounder.
  \item Randomization scheme factor and network topology factors: Certain network topology features such as high density and a lack of community structure may present challenges to some randomization schemes more than the others. Also, larger networks should make it easier for some randomization schemes to render more informative observed outcomes.
\end{itemize}

\renewcommand{\tabcolsep}{10pt}
\renewcommand{\arraystretch}{1.2}
\begin{table*}[!t]
\setlength{\belowcaptionskip}{6pt}
\caption[Example analysis of variance table on the factors]{Example analysis of variance table on the factors. For each factor and interaction, the degrees of freedom and mean square indicate the number of settings and its impact on the performance. }
\label{table:exampleANOVA}
\scalebox{1}{
\hspace*{0.9 cm}
\begin{tabular}{l r r l}
\hhline{===}
\multicolumn{1}{c}{Factor and interaction} & \multicolumn{1}{c}{ Degrees of freedom} & \multicolumn{1}{c}{Mean square} \\ \hhline{---}
Network topology (NP) & 7 \rule{1cm}{0cm} & - \rule{1cm}{0cm}& \\ 
Potential outcome model (POM) & 11 \rule{1cm}{0cm} & - \rule{1cm}{0cm}& \\ 
Treatment resource (TR) & 1 \rule{1cm}{0cm}& - \rule{1cm}{0cm}& \\
Randomization scheme (RS) & 7 \rule{1cm}{0cm}& - \rule{1cm}{0cm} & \\  
Analysis (A) & 11 \rule{1cm}{0cm}& - \rule{1cm}{0cm}& \\  
RS $\times$ A & 77 \rule{1cm}{0cm}& - \rule{1cm}{0cm}& \\  
RS $\times$  POM & 77 \rule{1cm}{0cm}& - \rule{1cm}{0cm}& \\  
A $\times$ POM & 121 \rule{1cm}{0cm}& - \rule{1cm}{0cm}& \\  
RS $\times$ NP & 49 \rule{1cm}{0cm}& - \rule{1cm}{0cm}& \\  
RS $\times$ A $\times$ POM & 847 \rule{1cm}{0cm}& - \rule{1cm}{0cm}& \\  
RS $\times$ A $\times$ NP & 539 \rule{1cm}{0cm}& - \rule{1cm}{0cm}& \\  
\hhline{---}
\end{tabular}
}
\bigskip
\bigskip
\end{table*}

\noindent Higher order (e.g. 3-way, 4-way) interactions may be present as well to reveal the interactions between certain combinations of randomization scheme and estimator with the presence of confounders, model mis-specification, and challenging network topology, etc. Table \ref{table:exampleANOVA} shows an example table summarizing the result of an ANOVA indicating all the factors and interactions that have significant impact on the performance. Only the expected interactions are listed here. Many more significant interactions between factors may have significant impacts in the actual simulation study. Also, some of these factors may be combined or marginalized out in the summary for parsimony.

\subsection{Closer Looks on Specific Factors to Identify Best Practices}
\label{sec:closerLooksOnSpecificFactors}

After locating at the high-level which factors and their interactions have significant impact on the performance of the causal inference, we can zoom in on specific factors and levels to identify best practices (i.e. combination of randomization scheme and estimator) and to characterize their performance under realistic truth settings. This can be done through visualizing performance tables that pivot on a few selected factors and levels. The factors not addressed by the specific table may be marginalized or conditioned on (i.e.\ set to reasonable levels). Although the specific tables of interest may be best informed by the ANOVA described above, below are some performance tables that will likely provide insights to the best practices for both the primary effect causal estimand and the peer effect causal estimand:

\renewcommand{\tabcolsep}{4pt}
\renewcommand{\arraystretch}{1.2}
\begin{table*}[!t]
\setlength{\belowcaptionskip}{6pt}
\caption[Combinations of randomizations and estimators over different peer exposure functions in the true outcome model]{Example performance table of different combinations of randomization schemes (columns) and estimators (rows) over different peer exposure functions in the true outcome model (the two sections). The other factors are marginalized out in this table to focus on identifying the most robust randomization and estimator combination for best practice.}
\label{table:exampleSpecificFactors}
\scalebox{0.94}{
\hspace*{-0.3 cm}
\begin{tabular}{l r r r r r r r r r r r r r r r r r l}
\hhline{==================}
& \multicolumn{8}{c}{ Simple sum peer exposure} & \rule{0.5cm}{0cm} & \multicolumn{8}{c}{Log sum peer exposure} \\ \hhline{~--------~--------}
& CR & SR & INR & RC & RG & RCG & PT & RNC & & CR & SR & INR & RC & RG & RCG & PT & RNC & \\ \hhline{------------------}
NM &  -  & -  &  -  & -  &  -  & -  &  -  & -  &    & -  &  -  & -  &  -  & -  &  -  & -  & - &\\ 
DM &  -  & -  &  -  & -  &  -  & -  &  -  & -  &    & -  &  -  & -  &  -  & -  &  -  & -  & - &\\ 
BNS &  -  & -  &  -  & -  &  -  & -  &  -  & -  &    & -  &  -  & -  &  -  & -  &  -  & -  & - &\\ 
BIPS &  -  & -  &  -  & -  &  -  & -  &  -  & -  &    & -  &  -  & -  &  -  & -  &  -  & -  & - &\\ 
BLPS &  -  & -  &  -  & -  &  -  & -  &  -  & -  &    & -  &  -  & -  &  -  & -  &  -  & -  & - &\\ 
BICS &  -  & -  &  -  & -  &  -  & -  &  -  & -  &    & -  &  -  & -  &  -  & -  &  -  & -  & - &\\ 
BLCS &  -  & -  &  -  & -  &  -  & -  &  -  & -  &    & -  &  -  & -  &  -  & -  &  -  & -  & - &\\ 
BNL &  -  & -  &  -  & -  &  -  & -  &  -  & -  &    & -  &  -  & -  &  -  & -  &  -  & -  & - &\\ 
BIPL &  -  & -  &  -  & -  &  -  & -  &  -  & -  &    & -  &  -  & -  &  -  & -  &  -  & -  & - &\\ 
BLPL &  -  & -  &  -  & -  &  -  & -  &  -  & -  &    & -  &  -  & -  &  -  & -  &  -  & -  & - &\\ 
BICL &  -  & -  &  -  & -  &  -  & -  &  -  & -  &    & -  &  -  & -  &  -  & -  &  -  & -  & - &\\
BLCL &  -  & -  &  -  & -  &  -  & -  &  -  & -  &    & -  &  -  & -  &  -  & -  &  -  & -  & - &\\ 
\hhline{------------------}
\end{tabular}
}
\bigskip
\bigskip
\end{table*}

\begin{itemize}
  \item Randomization schemes and estimators on different confounder or peer exposure functions: This performance table should highlight how model mis-specification by the Bayesian imputation estimators can be mitigated by balancing rerandomization schemes. An interesting comparison is against the naive difference in means estimator which does not make any potential outcome model assumption. An example is shown in Table \ref{table:exampleSpecificFactors}.
  \item Randomization schemes and estimators on the presence of confounders: This performance table highlights how the presence of confounders may be mitigated by both balancing rerandomization and including it in the model-based imputation.
  \item Selected combinations of randomization schemes and estimators on network topology factors: Select a few realistic and robust randomization scheme and estimator combinations (e.g.\ insulated neighbors randomization + difference in means estimator, rerandomization for balancing covariates and treatment exposure groups + Bayesian imputation estimator, etc.). Conditional on a potential outcome model setting that introduces model mis-specification, compare performance across the three network topology factors. This table may highlight how dense networks that lack community structure may present challenges for rendering observed outcomes in each treatment exposure group. It may also show the benefit of having a larger network to provide more informative observed outcomes. The table will show how these challenges and benefit affect each of the recommended causal inference procedures under network interference.
\end{itemize}

\section{Conclusion}

Although causal inference under network interference presents unique opportunities for estimating and harnessing the peer influence, it also brings about statistical challenges due to the additional aspects of exposure to peer treatments and network confounders. These challenges are best met through a combination of randomization schemes in the design phase and estimators in the analysis phase, the best practices for a wide range of realistic settings need to be identified. Following the theoretical guidance on the best practices from Chapter \ref{ch:1}, this chapter proposes a full factorial simulation study to identify, demonstrate, and characterize the performance of the best practices. The expectation is that this study will shed light on a complicated problem space with many factors and give practical guidance for causal experiments on networks.

\chapter[Hybrid Mixed-Membership Blockmodels for Influence Networks]{\huge Hybrid Mixed-Membership Blockmodels \\ for Influence Networks}\label{ch:3}
\section{Overview}
\label{sec:overview}

In support of the causal framework in this thesis for experiments on networks, this chapter develops a novel and realistic network model with rigorous inference procedures. It is important for the network model to have the richness and capacity to capture the key attributes that govern social influence between individuals, in order to fulfill its role laid out in Theorem \ref{thm:unconfoundedInfluenceNetworkByConditioning} for meeting the unconfounded influence network assumption (Assumption \ref{as:unconfoundedInfluenceNetworkUnderNetworkInterference}), and to enable more realistic simulated experiments. This chapter proposes the hybrid mixed-membership blockmodels (HMMB) that integrate three canonical network models to capture the characteristics of real-world networks: community-based interactions with mixed-membership, varying node degree, and sparsity. Although the causal framework in this thesis can be implemented with almost any network model, the HMMB provides an option with desirable features. The hybrid model provides the richness needed for realism and enables inference on individual attributes of interest such as mixed-membership and individual popularity. A rigorous inference procedure is developed for estimating the model parameters through iterative Bayesian updates with clever initialization to improve identifiability. For the mixed-membership estimate, a theoretical performance bound is derived from quantifying the information content of the network data through Fisher information analysis. Performance of the inference procedure is evaluated on simulated networks over the realistic range of values on the key model parameters. Results show that inference with HMMB is able to identify the mixed-community memberships and activity level parameters for each node even in the presence of highly overlapping communities.

\section {Introduction}
\label{sec:introduction3}

With an ever rising number of applications and data sources, network science has quickly emerged as a interdisciplinary field in the last fifteen years. Among the various challenges in this field, network modeling and inference remains a critical piece of the puzzle because it provides insights to the structure and formation of the network as well as information on the key attributes at both the individual and network level. Such information which is often not directly observable are valuable in many real-world applications. Much work has been done in network modeling and the closely-related field of community detection. This Section begins with a brief survey of some representative works in these areas of research and ends with a description of the contributions and the overall structure of this chapter.

\subsection{Background}
\label{sec:background}

Early works of network modeling and inference focus on simple models that lead to elegant mathematical properties. \cite{Erdos1960} models the network with a simple sparsity parameter that indicates the probability of any edge to be present, leading to graph percolation theories on the critical phase changes of network connectivity. However, this simple model does not express many of the real-world network characteristics such as the node degree variation and social structures. \cite{Chung2002} and \cite{Aiello2001} model the expected node degrees using a Power-Law distribution, commonly known as the Chung-Lu model. This model satisfies more network realism as it leads to some high degree nodes as well as a small average distance between any two nodes, known as the small-world property. \cite{Hoff2002} embeds each node onto a latent ``social space'' where a distance metric between any two nodes determines their likelihood of interaction (i.e.\ having an edge between them). This model explains the social structure by placing nodes that interact strongly in proximity to each other in the latent space. \cite{Wang1987} explains the social structure through group membership, using the well-known stochastic blockmodels where the blocks represent the groups (i.e.\ communities) and the nodes interact based on their membership to the block and the strength of interactions between the blocks. \cite{Airoldi2008} extends this model to allow mixed-membership for the nodes, known as the mixed-membership stochastic blockmodels. Under this extension, each node may belong to different communities depending on the context of its interaction with another node. The mixed-membership begins to capture the realism of overlapping communities which has received more attention in recent years. To explicitly model the community overlap, \cite{Soufiani2012} and \cite{Yang2012} propose community-based network models where each node may belong to multiple communities. The model in \cite{Soufiani2012} has a nice interpretation as a multiple-scale decomposition referred to as the graphlet decomposition. Some models have attempted to capture more realisms by combining multiple models (i.e.\ hybrid approach). For example, the mixed-membership stochastic blockmodels mentioned earlier incorporates a sparsity term to the blockmodels. \cite{Karrer2011} incorporates the degree correction terms to the blockmodels to capture both the community structure and the varying node degrees. Real-world networks are often comprised of counts of interactions between the individuals. \cite{Perry2013} models such network interactions as counts in point processes.

A closely related and prolific area of research is community detection, with a comprehensive survey in \cite{Fortunato2010}. The most well-known algorithm is probably the spectral method by \cite{Newman2006} where detection is done through the eigenspectrum of the modularity matrix. Most of the existing community detection algorithms work through the principle of network modularity where the network is partitioned into communities (i.e.\ modules) that have much stronger interactions within them than between them. Typically, partitioning is done by maximizing the network modularity \citep{Newman2006modularity}. \cite{Mucha2010} extends the concept of modularity for time-dependent, multiscale, and multiplex networks. \cite{Smith2014} exploits network modularity through random walks on graph, for cued network detection where a few nodes in the community of interest is known a priori. Some recent works perform community partitioning through membership estimates in a generative model. For example, \cite{Ball2011} estimates community membership using the degree corrected stochastic blockmodel mentioned previously, and \cite{Huang2013} proposes a fast mixed-memberships estimation procedure by applying tensor methods to the mixed-membership stochastic blockmodels.

\subsection{Contributions}
This chapter proposes the novel hybrid mixed-membership blockmodels (HMMB) that integrate three canonical network models to capture the characteristics of real-world networks: community-based interactions with mixed-membership, varying node degree, and sparsity. It is the only network model so far to incorporate all of these characteristics. The hybrid model provides the richness needed for realism and enables inference on individual attributes of interest such as mixed-membership and individual popularity. A rigorous inference procedure is developed for estimating the model parameters through iterative Bayesian updates with clever initialization to improve identifiability. For the mixed-membership estimate, a theoretical performance bound is derived from quantifying the information content of the network data through Fisher information analysis. The proposed Bayesian inference procedure is shown with simulation to perform adequately in estimating and identifying the model parameters, even in the presence of significant community overlap.

The chapter proceeds by first proposing the hybrid mixed-membership blockmodels (HMMB) in Section \ref{sec:HMMB}. A Bayesian inference procedure is then developed in Section \ref{sec:inferenceWithHMMB} for model parameter estimation. A derivation of the theoretical performance bound on the mixed-membership estimate is given, followed by a comprehensive performance evaluation on the estimation procedure as the main network features are varied.

\section {Hybrid Mixed-Membership Blockmodels (HMMB)}
\label{sec:HMMB}

This section introduces the proposed hybrid mixed-membership blockmodels (HMMB) for networks representing interactions and influence between individuals. Real-world networks consisting of observed interactions over time are prevalent and diverse, spanning a wide range of applications areas such as communications \citep{Eagle2006, Diesner2005}, social media \citep{Golder2007, Liu2005}, political science \citep{Fowler2006}, and biology \citep{Barabasi2004}, etc. Although not usually directly observable, influence between individuals over time can be constructed or inferred from data on following behaviors and expressed relationships \citep{Bakshy2011, Gomez2010, Cha2010}. For this growing body of data with relevant applications, a desirable statistical model should be rich enough to capture the realism and the fundamental characteristics of the network data. Also, it should be interpretable and intuitive so that inference and estimation of the model parameters can provide useful information on the attributes of the individuals and reveal insights on the formation and structure of the interactions or influence. These are the key motivations for the design of the hybrid mixed-membership blockmodels (HMMB).

\subsection{Model Description: Integrating Three Canonical Models for Realism}
\label{sec:HMMBModelDescription}

The HMMB aims to model real-world networks representing the number of interaction or the units of influence between each pair of individuals. Such networks are typically sparse because interactions only take place between individuals with relationships, which are usually a small fraction of the total possible relationships. In other words, only a small fraction of the possible edges are ``turned-on.'' A widely used model for network sparsity is the Erd\H{o}s-R\'{e}nyi model \citep{Erdos1960}. The frequency of interaction between individuals with a relationship is determined by each individual's level of activity and the roles (i.e.\ community memberships) they play in their interactions. For example, two individuals with a relationship may interact frequently because they are both relatively active and belong to the same communities or communities that have strong interactions. The Chung-Lu model \citep{Chung2002, Aiello2001} is a widely adopted model to capture varying levels of activity for each individual, and the mixed-membership stochastic blockmodels \citep{Airoldi2008, Airoldi2013} are well-known for modeling interactions based on mixed community memberships. The HMMB, defined and described in this section, captures the characteristics of real-world networks by integrating these three canonical models.

Following common notations, a network is defined as $G=(\VSet,\ESet)$, consisting of two sets: the nodes $\VSet$ and the edges $\ESet$ between them. The nodes represent the individuals and the edges represent the interactions and influence between them. Each edge is directed and weighted, representing the direction (e.g.\ sender versus receiver) and the number of interactions or the units of influence between the nodes. Let $N$ be the number of nodes in the network (i.e.\ $N = |\VSet|$). The $N \times N$ adjacency matrix $\A$ provides a convenient and intuitive representation of the network, with each element $a_{ij}$ denoting the number of interactions or units of influence from node $i$ to node $j$. The network here does not imply causal relationships. In other words, it is not a causal graph.

Under the hybrid mixed-membership blockmodels (HMMB), interactions or units of influence between nodes over time are modeled as Poisson point processes where each interaction or unit of influence takes place as random events over time with a certain frequency, similar to the work by Perry and Wolfe \citep{Perry2013}. So each edge, $a_{ij}$ (i.e.\ the count of interaction or units of influence), is modeled with a Poisson distribution with an expected value of $\lambda_{ij}T$:
\begin{equation}
\label{PoissonNetworkEquation}
a_{ij} \sim \Poisson(\lambda_{ij}T)
\end{equation}
$\lambda_{ij}$ is the frequency of interactions or strength of influence from node $i$ to node $j$ measured in number of occurrences or units over time. The symbol $T$ is the time span during which the network data is collected. The unit for time is arbitrary and data set specific. 

For realism, the HMMB integrates three canonical network models in governing the interaction frequency from each node $i$ to each node $j$:
\begin{equation}
\label{rateEquation}
\lambda_{ij} = (\lambda_i \lambda_j)  \times (\bm{\pi}_i^T \bm{B} \bm{\pi}_j) \times I_{ij}
\end{equation}
The entire $N \times N$ frequency matrix $\bm{\Lambda}$ can be expressed in the matrix product below:
\begin{equation}
\label{rateEquationMatrixForm}
\bm{\Lambda} = (\bm{\lambda \lambda}^T)  \circ (\bm{\Pi}^T \bm{B} \bm{\Pi}) \circ \bm{I}
\end{equation}
where $\circ$ denotes the Hadamard product (i.e.\ element-wise product). Equation~\eqref{rateEquation} is the core of the HMMB model, integrating the Chung-Lu term, $\lambda_i\lambda_j$, the mixed-membership stochastic blockmodels term, $\pib_i^T \B \pib_j$, and the Erd\H{o}s-R\'{e}nyi ``on'' and ``off'' switch, $I_{ij}$, for sparsity. The Chung-Lu term, $\bm{\lambda \lambda}^T$, captures the variation in node degrees (i.e.\ activity level), allowing some nodes to be much more active than the others. The frequency of interaction between $i$ and $j$ is proportional to the activity level of both nodes. This well-known model \citep{Chung2002, Aiello2001} is parameterized by $\bm{\lambda}$, a vector of $N$ components and a prior parameter $\alpha$ that determines how quickly the Power-Law distribution for each $\lambda_i$ drops off: $p_0(\lambda_i) \; \propto \; \lambda_i^{-\alpha}$. Modeling the variation in node degrees by a Power-Law prior distribution with its long tail is supported by empirical evidence in many real-world networks with $\alpha$ typically between two and three \citep{Clauset2009}. The mixed-membership stochastic blockmodels term, $\bm{\Pi}^T \bm{B} \bm{\Pi}$, captures the community structure of the network \citep{Airoldi2008, Airoldi2013}. Let $K$ be the number of communities in the network, the mixed-membership stochastic blockmodels are parameterized by $\bm{\Pi} = (\pib_1, \pib_2, ..., \pib_N)$, a $K \times N$ mixed-membership matrix, and $\bm{B}$, a $K \times K$ block matrix. Each $\pib_i$ denotes the mixed-membership of node $i$ to the $K$ communities. The mixed-membership vector $\pib_i$ sums to one (i.e.\ stays inside a simplex) and can be interpreted as the fraction of the time node $i$ acts as a member in each community. The block matrix $\bm{B}$ denotes the strength of interaction within and between all the communities. For example, $b_{11}$ is the strength of within-community interaction for community one, and $b_{12}$ is the strength of interaction from community one to community two. Consequently, $\pib_i^T \B \pib_j$, captures the total strength of interaction from node $i$ to node $j$ based on their memberships in each of the community. Lastly, the sparsity matrix $\bm{I}$ acts as ``on'' and ``off'' switches for each of the edge with distribution:
\begin{equation} 
\label{BernoulliSwitch}
I_{ij} \stackrel{\mathrm{iid}}\sim \Bernoulli(s) 
\end{equation} 
with only one parameter, the sparsity $s$, in this part of the model \citep{Erdos1960}.

It is desirable to generate the mixed-memberships of each node to have different distributions across the communities, according to its ``lifestyle.'' For example, a particular lifestyle may have a tendency to participate mostly in community one and two while another lifestyle may concentrate its participation only in community three. To achieve this diversity, $L$ lifestyles are defined, each with $K$ pseudocounts indicating the lifestyle's tendency towards each community membership, represented by $\bm{X}$, a $K \times L$ mixed-membership pseudocount matrix. The lifestyle indicator vector for each node $i$, $\bm{l}_i$, is then drawn from $\Multinom(1, \bm{\phi})$ where $\bm{\phi}$ is a pre-specified simplex vector of $L$ components indicating the probability of belonging to each lifestyle. Finally, the mixed-membership for each node $i$, $\pib_i$ is drawn from $\Dirichlet(\bm{X l}_i)$ where $\bm{X l}_i$ is the membership pseudocounts of the lifestyle $i$ belongs to. This generative procedure for the mixed-membership is useful for simulating network data, but in real-world data sets, information on the lifestyles is typically unavailable. Figure~\ref{fig:blockDiagram} presents a plate diagram \citep{Bishop2006} laying out the data generation process under HMMB, with pre-specified parameters, parameters drawn using the prior distributions, and the network data drawn using the model distribution.
\begin{figure}[t]
     \centering
     \includegraphics[width=1\textwidth]{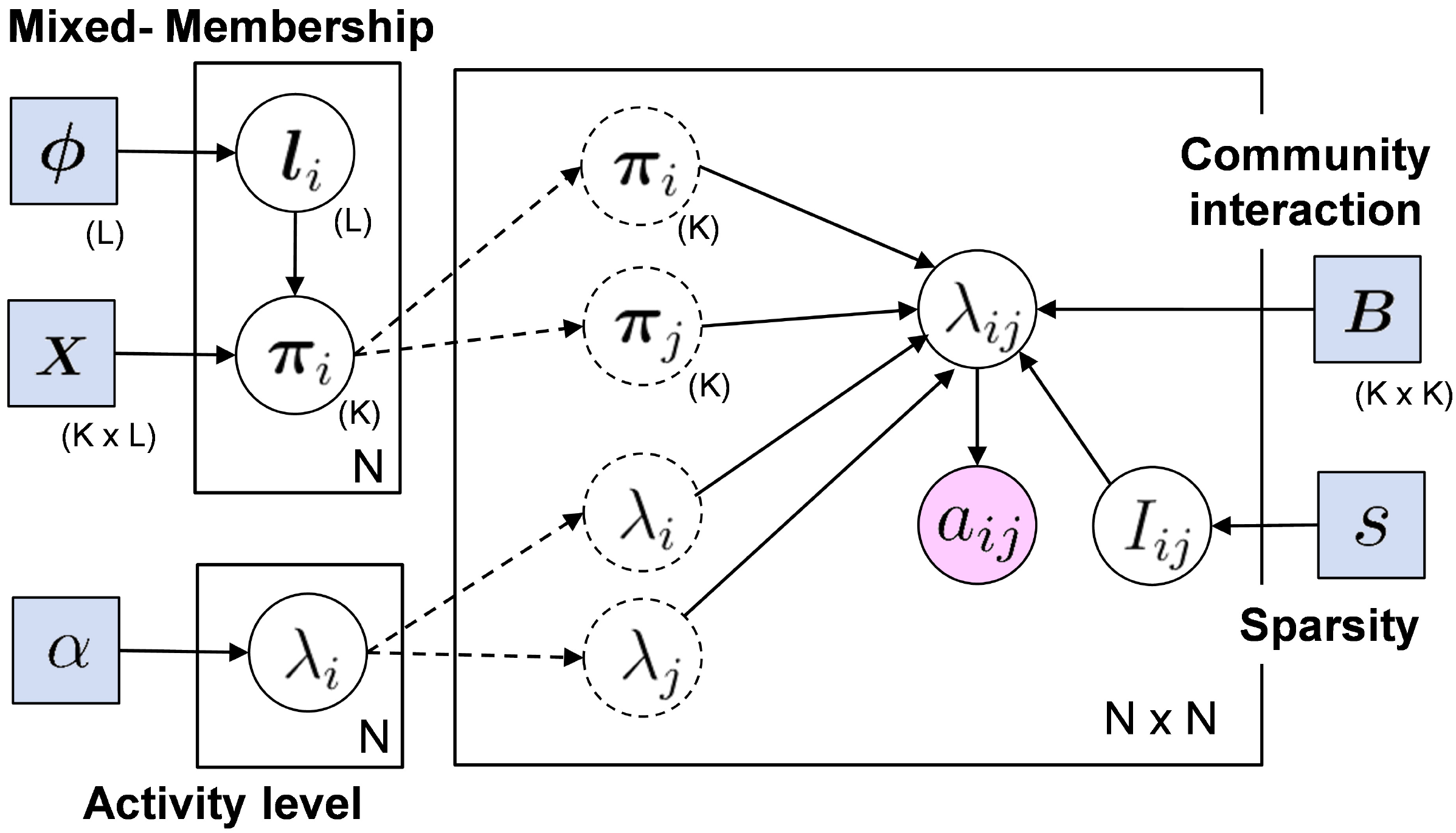}
     \caption[Plate diagram of the full generative process of the hybrid mixed-membership blockmodels]{Plate diagram of the full generative process of the hybrid mixed-membership blockmodels with $K$ communities, $N$ nodes (individuals), $L$ lifestyles, and a time duration of $T$. The blue squares indicate the parameters that are pre-specified. Circles are the parameters that are drawn according to the model distributions. The red circle is the resulting network and the dotted circles indicate the nodal parameters used to draw each edge. The solid boxes specify the number of copies drawn for each parameter.
}
\label{fig:blockDiagram}
\end{figure}

\subsection{Parameter Identifiability}
\label{sec:identifiability}

The HMMB is a composite of three canonical network models and inference will be performed on one model at a time conditional on the parameters from the other two, as shown later in Section~\ref{sec:BayesianParameterEstimation}. Existing works demonstrate identifiability for each of the three canonical network models. The Erd\H{o}s-R\'{e}nyi model only has the sparsity parameter $s$ and each of the edge switch is inferred independently, so it is simply identifiable \citep{Olding2009}. The Chung-Lu model mainly consists of the node degree parameter $\bm{\lambda}$ on each node and has been proven to be identifiable by existing work \citep{Perry2012}. The mixed-membership blockmodels belong to the general class of mixture models known to have symmetric multi-modal likelihood \citep{Buot2007} which may present challenges in identifiability. However, this issue can be mitigated by inference strategies such as clever initialization and multiple starting points, and existing work shows theoretically as well as empirically that parameter estimation in mixed-membership models have good frequentist properties \citep{Mukherjee2009, Airoldi2013, Nguyen2015}. 

Identifiability of the individual model component supports but does not theoretically guarantee identifiability of the joint model. We will instead demonstrate parameter identifiability under the HMMB empirically in Section~\ref{sec:estimationPerformance}, using the proposed inference procedure in Section~\ref{sec:BayesianParameterEstimation}. The proposed inference procedure includes strategies to mitigate possible identifiability issues such as the arbitrary scale shared between the block matrix $\bm{B}$ and the node degree parameter $\bm{\lambda}$. These strategies include the profile likelihoods, clever parameter initializations, and multiple independent inference chains. Identifiability is demonstrated empirically for networks of reasonable sizes (i.e.\ hundreds of nodes). In real-world networks, the number of communities each node participates in, limited by that individual's capacity and interest, does not grow unbounded with the size of the network. So as the size of the network grows, the mixed-membership vectors $\pib_i$ will become increasingly sparse. As the number of communities each individual belongs to is capped at some constant, the effective number of parameters under HMMB is expected to grow at a rate of $O(N)$. The number of observed interactions (i.e.\ data) is expected to grow faster at $O(N\log N)$ even for sparse networks. Therefore, identifiability should scale well with increasing network size.

\section {Inference and Performance Bound with HMMB}
\label{sec:inferenceWithHMMB}

The nodal parameters under HMMB represent valuable information on the individuals in the network. For example, the mixed-memberships offer insights on the community structure and reveal members of each community. Other parameters of interest include individual activity levels (i.e.\ popularity), $\bm{\lambda}$, and the strength of community interaction within and between each other, $\bm{B}$. In applications involving social peer influence, node activity levels and community interaction may provide important information on the amount and the structure of influence. Individual activity levels may also indicate social and economic status. These parameters should be included in the network potential outcome models as laid out in Theorem \ref{thm:unconfoundedInfluenceNetworkByConditioning}. Lastly, knowledge on the model parameters for a particular network of interest enables us to predict future behaviors and generate additional network data with similar characteristics. 

Although the nodal parameters are typically not observable, they can be inferred and estimated using the observed network data, $\bm{A}$. For the causal framework proposed in this thesis, if the influence network is not directly observed, the HMMB parameter estimation can be performed with samples drawn from the network prior or posterior distribution. This section describes the Bayesian joint inference procedure through iterative updates to estimate the model parameters under HMMB. Clever initializations of the parameter estimates prior to the iterative updates are designed to improve identifiability. A fully Bayesian sampling update procedure is proposed, with alternative maximum posterior updates on some of the parameters for faster convergence while sacrificing the property of convergence to the joint posterior distribution. Given the central role of the community mixed-membership in the HMMB, this section characterizes the information content of network data and the theoretical performance bound on mixed-membership estimates. Lastly, estimation performance of the proposed procedures are evaluated and compared across a range of reasonable model parameter settings, demonstrating empirically the parameter identifiability of the HMMB under reasonable settings.

\subsection{Bayesian Parameter Estimation Procedure}
\label{sec:BayesianParameterEstimation}

Parameter estimation in hybrid mixed-membership blockmodels (HMMB) is performed through Bayesian methods \citep{Gelman2014} using iterative Monte Carlo updates. Given the observed network data, $\bm{A}$, the goal is to obtain the posterior distribution on the parameters of interest, namely the mixed-memberships, $\bm{\Pi}$, the activity level (i.e.\ degree), $\bm{\lambda}$, of each node, and the community interaction structure, $\bm{B}$. The posterior distribution captures both the likely values and the uncertainty of the estimates, which is a key advantage of the Bayesian methods. In real-world applications where decisions and analytical claims are being made based on the parameter estimates, having knowledge on the uncertainty of the estimates through the full posterior distribution is especially desirable. The edge switches, $\bm{I}$, may not be of particular interest but still takes part in the posterior distribution. The joint posterior distribution is proportional to the likelihood times the prior:
\begin{equation}
 p(\bm{\Pi}, \bm{\lambda}, \bm{B}, \bm{I} | \bm{A}) \propto \Like(\bm{\Pi}, \bm{\lambda}, \bm{B}, \bm{I} | \bm{A}) \; p_0(\bm{\Pi}, \bm{\lambda}, \bm{B}, \bm{I})
\end{equation}
with the joint likelihood function, according to Equation~\eqref{PoissonNetworkEquation}, \eqref{rateEquation}, and \eqref{BernoulliSwitch}:
\begin{align}
\Like(\bm{\Pi}, \bm{\lambda}, \bm{B}, \bm{I} | \bm{A}) = &\prod_{i,j \in I_{ij}=1} {\Big( \Poisson(a_{ij}; \lambda_{ij}) \; s \Big)}  \prod_{i,j \in I_{ij}=0} {\Big( \delta_{a_{ij}0} \; (1-s) \Big)} \nonumber \\
\propto & \; \exp\bigg(\sum_{i,j \in I_{ij}=1}\Big( a_{ij} \log (\lambda_i\lambda_j\pib_i^T \bm{B} \pib_j) - \lambda_i\lambda_j \ T \ \pib_i^T \bm{B} \pib_j \Big)\bigg) \prod_{i,j \in I_{ij}=0}\delta_{a_{ij}0} \nonumber \\ 
& \; s^{\sum_{ij}{ I_{ij}}} (1-s)^{\sum_{ij}{1-I_{ij}}}
\end{align}
where $\sum_{ij}{ I_{ij}}$ and $\sum_{ij}{1-I_{ij}}$ are the number of ``on'' and ``off'' switches, and the $\delta_{a_{ij}0}$ term makes sure that only edges with no observed interaction can be ``off''. The joint posterior distribution is difficult and inefficient to sample directly. Therefore, the parameters for each component of the model will be iteratively updated one piece at a time, while conditioning on the current sample on the other parameters. This is commonly known as Gibbs sampling which is a type of Markov Chain Monte Carlo (MCMC) method based on conditional sampling for multivariate distributions \citep{Gelman2014}. Figure~\ref{fig:Gibbs} shows each step of the Gibbs update for HMMB. Two candidates of the Gibbs updates are implemented. The first implementation is based entirely on Bayesian sampling using the Markov Chain Monte Carlo (MCMC), which samples from the conditional posterior distribution in each step. The second is a hybrid Bayesian sampling and maximum posterior estimate approach, the Monte Carlo Expectation Maximization (MCEM), which aims to maximize conditional posterior probabilities when updating some of the parameters. The MCMC approach has the theoretical property of eventually converging to the full joint posterior distribution, while the MCEM approach sacrifices this property to gain speed of convergence with the greedy maximization steps. Each parameter update step for both implementations and the governing likelihood function, prior probability, and posterior probability is described in detail in Section~\ref{sec:MCMC} and \ref{sec:MCEM}.  A performance comparison between the MCMC and MCEM procedures is given later in Section~\ref{sec:estimationPerformance} with discussion of results in Table~\ref{coverageTableSimulatedData}.

\begin{figure}
\begin{minipage}{\textwidth}
\centering
\includegraphics[width=0.9\textwidth]{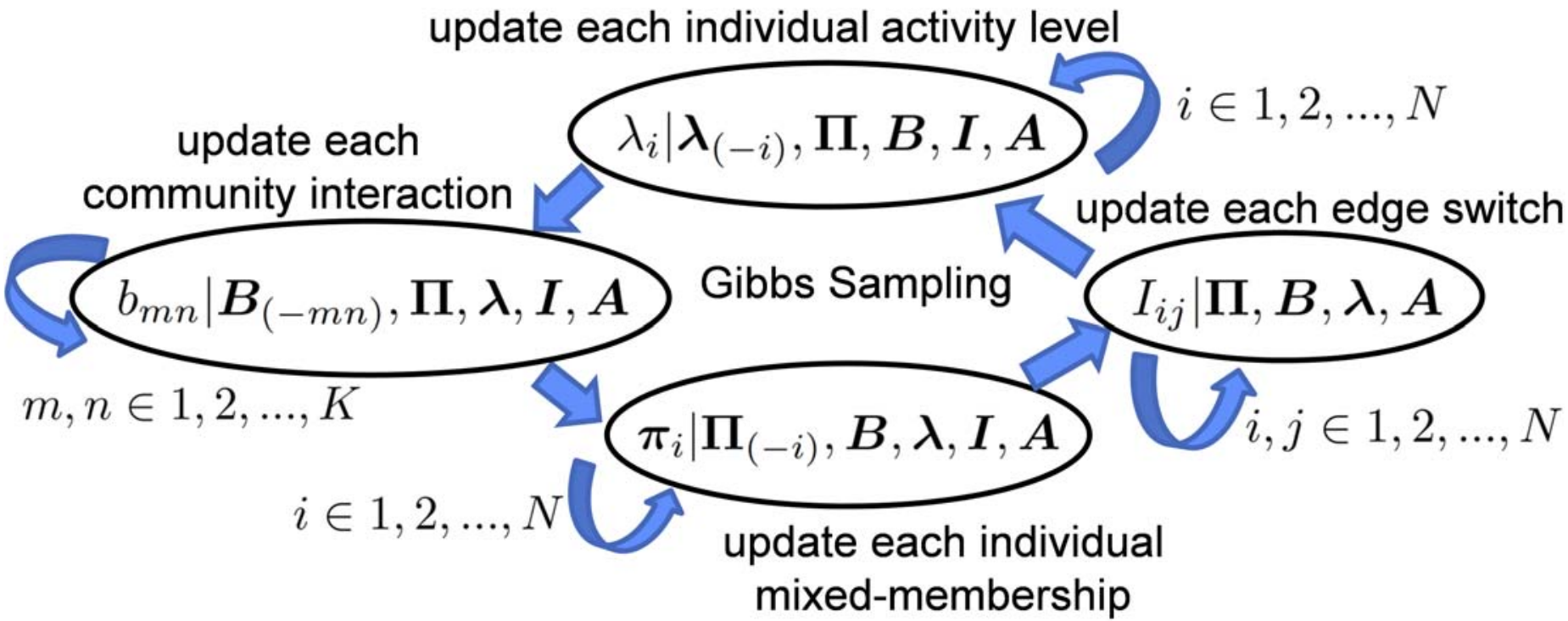}
\captionof{figure}[Bayesian parameter estimation using Gibbs sampling] {The Gibbs sampling procedure for Bayesian parameter estimation iteratively updates the parameters of interest one piece at a time, conditional on the current sample on the other parameters. There are $K$ communities and $N$ nodes in the network. An implementation of this procedure is shown in the algorithm below.}
\label{fig:Gibbs}
\end{minipage}
\end{figure}

Due to the possible challenges in identifiability discussed in Section~\ref{sec:identifiability}, several independent Monte Carlo chains are run as mitigations to the potentially multi-modal likelihood function, to maximize the chance that at least one of the chains will converge to the true mode. At the end of the run, the chain that finishes with the highest joint posterior probability is selected. Additionally, for each run, parameters are initialized cleverly to regions of the parameter space that are likely to be close to the true mode. To remove possible redundancy in the parameter space, a small number of the parameters may be fixed to pre-computed or a-priori values. These mitigation strategies for identifiability is discussed in greater detail in Section~\ref{sec:parameterInit}. Finally, the overall inference procedure is summarized in Algorithm~\ref{InferenceImplementation}. 
\\ \\
\begin{algorithm}[H]\normalsize
\setstretch{1}
\SetKwInOut{Input}{input}\SetKwInOut{Output}{output}
\Input{Network data in adjacency matrix $\bm{A}$}
\Output{Posterior samples on the parameters of interest: $\bm{\Pi}, \bm{\lambda}, \bm{B}$}
~\\
\ForEach {independent inference chain} 
{
    \textbf{initialize} $\bm{\Pi}^0$ to the community assignment by spectral clustering\\
    \textbf{initialize} $\bm{\lambda}^0$ proportional to the observed node degrees\\
    \textbf{initialize} $\bm{\bm{B}}^0$ to a diagonal matrix with small values in off-diagonal entries\\
    \textbf{initialize} $\bm{\bm{I}}^0$ to be ``on'' only for edges with observed interactions\\
    \textbf{set} fixed parameters to precomputed or known values\\
    \Repeat (                              \tcp*[f]{begin the Gibbs sampling updates}) {convergence or maximum iteration reached} 
    {
	\For {$i = 1$ to $N$}  
	{
		\textbf{update} $\lambda_i^{t+1}$ conditional on $\bm{\lambda}_{(-i)}^{t^+}, \bm{\Pi}^t, \bm{B}^t, \bm{I}^t$\\
	}
	\For {$m,n = 1$ to $K$}  
	{
		\textbf{update} $b_{mn}^{t+1}$ conditional on $\B_{(-mn)}^{t^+}, \bm{\lambda}^{t+1}, \bm{\Pi}^t, \bm{I}^t$\\
	}
    	\For {$i = 1$ to $N$}  
	{
		\textbf{update} $\bm{\pi}_i^{t+1}$ conditional on $\bm{\Pi}_{(-i)}^{t^+}, \bm{B}^{t+1}, \bm{\lambda}^{t+1}, \bm{I}^t$\\
	}
	\For {$i, j = 1$ to $N$}  
	{
		\textbf{update} $I_{ij}^{t+1}$ conditional on $\bm{\Pi}^{t+1}, \bm{B}^{t+1}, \bm{\lambda}^{t+1}$\\
	}
	\textbf{store} samples $\bm{\Pi}^{t+1}, \bm{B}^{t+1}, \bm{\lambda}^{t+1}, \bm{I}^{t+1}$ if $t+1 >$ burn-in	
    }
}
\textbf{return} samples from the chain ending with the highest posterior probability
\caption{Joint Inference Procedure Using Gibbs Sampling}
\label{InferenceImplementation}
\end{algorithm}
\ \\

The computational complexity of the proposed inference procedure is $O(N \log N)$, with the following simple run-time analysis. The bottleneck of the proposed procedure lies in the evaluation of the likelihood functions in Equation~\eqref{eq:lambdaLikelihood},~\eqref{eq:BLikelihood}, and ~\eqref{eq:piLikelihood}. To evaluate these likelihood functions in each iteration of the Gibbs sampling, every edge with the switch $I_{ij}$ turned ``on'' requires a matrix product in the mixed-membership blockmodels term, $\pib_i^T\bm{B}\pib_j$. Although the number of communities grow with the size of the network, the number of communities each node takes membership in should not. In other words, in real-world networks, the number of communities each node participates, limited by that individual's capacity and interest, should not grow unbounded with the size of the network. So as the size of the network grows, the mixed-membership vectors $\pib_i$ will become increasingly sparse, making the cost of evaluating $\pib_i^T\bm{B}\pib_j$ at some constant value. So the overall computational complexity scales with the number of ``on'' edges, which is $N \log N$ in real-world networks. Finally, inference with the mixed-membership model has been proven to be scalable and performed on large networks by Gopalan and Blei \citep{Gopalan2013}.

\subsubsection{Parameter Initialization}
\label{sec:parameterInit}

As discussed briefly in Section~\ref{sec:BayesianParameterEstimation} and shown in Algorithm~\ref{InferenceImplementation}, each of the independent inference chain is initialized in a way to increase its chance to converge to the true mode. This is done as a mitigation to the possible challenge in identifiability discussed in Section~\ref{sec:identifiability}. These initialization steps are described in detail here. Clever parameter initialization starts the parameter samples closer to the true mode on the posterior distribution surface. This way, the chance of the Gibbs samples to converge to a false local maximum in a potentially multi-modal posterior distribution is lowered. 

A reasonable initial values of the mixed-membership estimate, $\bm{\Pi}^0$, can be obtained by a fast segmentation algorithm on the network (e.g.\ spectral clustering by \cite{Ng2002}). After the segmentation, nodes belonging to the same cluster are initialized to be members of the same community. This way, nodes with strong interactions will likely be initialized to the same communities, which is a good guess because interactions in real-world networks are typically stronger within communities than between communities. The edge switches are simply initialized to ``on'' for each edge with observed interaction, and ``off'' for all other edges. This initialization, $\bm{I}^0$, is also the maximum posterior estimate. A good initialization of the node degree parameter $\bm{\lambda}^0$ is to simply set it to values proportional to the average weight of the observed in and out edges of each node: 
\begin{equation}
\lambda_i^0 \propto \frac {\sum_{j}{a_{ij}} + \sum_{j}{a_{ji}}}{\sum_{j}I_{ij}^0 + \sum_{j}I_{ji}^0}
\end{equation}
This is a good initial value because it is an unbiased estimator with good asymptotic properties \citep{Arcolano2012} after normalizing out the effect of the edge switches and disregarding the effect of the mixed-membership blockmodels term. The proportionality expression here accommodates an arbitrary relative scaling between $\bm{\lambda}$ and $\bm{B}$ which will be explained shortly. Because within-community interactions are typically stronger than the between-community interactions, the block matrix $\bm{B}^0$ is initialized to roughly diagonal with small values in the off-diagonal entries. 

Although the three component models integrated in HMMB each captures a specific and orthogonal characteristic of the network, there exists one parameter redundancy in the shared relative scale between the block matrix, $\bm{B}$ and the node degree parameters $\bm{\lambda}$. For example, for a subset of the communities $\KSet$ that do not overlap in membership with the rest, one may arbitrarily scale the sub-block-matrix $\B_{\KSet\KSet}$ by $w$ and the degree parameters $\bm{\lambda}_{\VSet_\KSet}$ by $1/w$ on the members of these communities, $\VSet_\KSet$, and arrive at exactly the same likelihood. This issue can be mitigated by the profile likelihood approach where the diagonal entries of the block matrix $\bm{B}$ are fixed to a precomputed value during the Gibbs sampling. It is equivalent to taking a profile slice of the likelihood functions at the fixed values for the diagonal of $\bm{B}$. A reasonable precomputed value for the diagonal of $\bm{B}$ is the maximum posterior estimate computed using the gradient ascent described in Equation~\eqref{BGradientAscent} with uniform $\bm{\lambda}$, after a preliminary run of the Gibbs sampling with an identity block matrix. This forces the model to explain the observed network data using only the within-community interactions (i.e.\ the diagonal of $\bm{B}$). After fixing the diagonal of $\bm{B}$, the iterative Gibbs sampling updates are performed. At the end of the inference procedure, $\bm{B}$ and $\bm{\lambda}$ can be rescaled again to match to the truth before evaluation, because the relative scale between them is arbitrary.

If prior knowledge (e.g.\ the memberships on a few nodes may be known or observed) or a reasonable estimate on some of the parameters is available, one can apply the profile likelihood approach by fixing them to the prior value. An alternative to the profile likelihood approach is to introduce informative prior on these parameters instead of fixing them to a value. 

\subsubsection {Bayesian Sampling Updates (MCMC)}
\label{sec:MCMC}
After the parameters are initialized properly, the main part of the inference computation through iterative updates begins, as shown in Figure~\ref{fig:Gibbs} and in the Gibbs sampling loop in Algorithm~\ref{InferenceImplementation}. This section presents the fully Bayesian approach using the Monte Carlo Markov Chain (MCMC) sampling. Because most of posterior densities in the updates do not have closed-form distributions, the Metropolis-Hasting algorithm is used to propose an update and accept it with a rate based on the posterior probability ratio between the existing sample and the proposed sample. When the proposal distribution is not symmetric, a Hasting correction ratio is added in computing the acceptance rate to maintain the detailed balanced condition of the MCMC. As mentioned previously in Section~\ref{sec:parameterInit}, a few parameters may be fixed in which case they will not participate in the updates described below.

\paragraph{Update $\lambda_i^{t+1}$ conditional on $\bm{\lambda}_{(-i)}^{t^+}, \bm{\Pi}^t, \bm{B}^t, \bm{I}^t$:}
For the node degree (i.e.\ activity level) update on each node $i$ conditioning on the current sample on the other parameters, the governing conditional likelihood function, based on Equation~\eqref{PoissonNetworkEquation}~and~\eqref{rateEquation}, is:
\begin{align}
\label{eq:lambdaLikelihood}
\Like\left(\lambda_i | \bm{\lambda}_{(-i)}, \bm{\Pi}, \bm{B}, \bm{I},\A \right) =  \exp\Bigg( & \sum_{j \in I_{ij}=1}\Big( a_{ij} \log (\lambda_i\lambda_j) - \lambda_i\lambda_j \ T \ \pib_i^T \bm{B} \pib_j \Big) + \nonumber \\ 
& \sum_{j \in I_{ji}=1}\Big( a_{ji} \log (\lambda_i\lambda_j) - \lambda_i\lambda_j \ T \ \pib_j^T \bm{B} \pib_i \Big) \Bigg)
\end{align}
And the posterior distribution is:
\begin{equation}
\label{eq:lambdaPosterior}
p\left(\lambda_i | \bm{\lambda}_{(-i)}, \bm{\Pi}, \bm{B}, \bm{I}, \bm{A}\right) \propto \Like\left(\lambda_i | \bm{\lambda}_{(-i)}, \bm{\Pi}, \bm{B}, \bm{I}, \A\right) \; p_0(\lambda_i )
\end{equation}
Real-world networks often follow Power-Law degree distribution \citep{Chung2002}, which gives a natural prior for $\lambda_i$:
\begin{equation}
p_0(\lambda_i) \propto \lambda_i^{-\alpha}
\end{equation}
Typically, the exponent $\alpha$ is between two and three in real-world networks \citep{Clauset2009}, so one may simply fix $\alpha$ to a reasonable value or update it as another step in the MCMC with a symmetric Normal proposal distribution, $\alpha^{t^+} \sim \text{Normal}\left(\alpha^t, \sigma^2_\alpha\right)$, and posterior distribution, $p(\alpha | \bm{\lambda}) \propto \prod_{i=1}^N {\lambda_i^{-\alpha}}$. The variance, $\sigma^2_\alpha$ of the Normal proposal controls the step size of the proposal from the current sample.

\noindent For updates on $\lambda_i$, a symmetric Normal proposal distribution is used, $\lambda_i^{t^+} \sim \text{Normal}\left(\lambda_i^t, \sigma^2_\lambda\right)$. Again, the variance $\sigma^2_\lambda$ of the Normal proposal controls the step size of the proposal from the current sample.
The acceptance probability, $a_{\lambda}$, of the proposed update is:
\begin{equation}
a_{\lambda} =  \left\{ \begin{array}{cl} \displaystyle \min\left(\frac{p\left(\lambda_i^{t^+} | \bm{\lambda}_{(-i)}, \bm{\Pi}, \bm{B}, \bm{I}, \bm{A}\right)}{p\left(\lambda_i^t | \bm{\lambda}_{(-i)}, \bm{\Pi}, \bm{B}, \bm{I}, \bm{A}\right)} , 1 \right)  & \mbox{if $\lambda^{t^+}_{i} > \epsilon_\lambda$} \\ 0 & \mbox{otherwise} \end{array} \right.
\end{equation}
No Hasting correction ratio is needed because the proposal distribution is symmetric. Note that the new proposal is automatically rejected if it is below the minimal value $\epsilon_\lambda$ typically set at close to zero because nodes can not have negative degrees. So with probability $\alpha_\lambda$, the proposed sample will be accepted (i.e.\ $\lambda_i^{t+1} = \lambda_i^{t^+}$), otherwise, the current sample will be kept (i.e.\ $\lambda_i^{t+1} = \lambda_i^{t}$ ).

\paragraph{Update $b_{mn}^{t+1}$ conditional on $\B_{(-mn)}^{t^+}, \bm{\lambda}^{t+1}, \bm{\Pi}^t, \bm{I}^t$:}
For updating the block matrix one element, $b_{mn}$, at a time conditioning on the current sample on the other parameters, the governing conditional likelihood function, based on Equation~\eqref{PoissonNetworkEquation}~and~\eqref{rateEquation}, is:
\begin{equation}
\label{eq:BLikelihood}
\Like\left(b_{mn} | \bm{B}_{(-mn)}, \bm{\lambda}, \bm{\Pi}, \bm{I},\A \right) =  \exp\Bigg( \sum_{i,j \in I_{ij}=1}\Big( a_{ij} \log (\bm{\pi}_i^T \bm{B}^* \bm{\pi}_j) - \lambda_i\lambda_j \ T \ \bm{\pi}_i^T \bm{B}^* \bm{\pi}_j \Big) \Bigg)
\end{equation}
where $\bm{B}^*$ consists of the fixed elements $\bm{B}_{(-mn)}$ and the element being updated, $b_{mn}$. The posterior distribution is:
\begin{equation}
\label{eq:BPosterior}
p\left(b_{mn} | \bm{B}_{(-mn)}, \bm{\lambda}, \bm{\Pi}, \bm{I}, \bm{A}\right) \propto \Like\left(b_{mn} | \bm{B}_{(-mn)}, \bm{\lambda}, \bm{\Pi}, \bm{I}, \A\right) \; p_0(b_{mn})
\end{equation}
Typically, a flat prior is used for $b_{mn}$. However, given prior knowledge on the structure of the block matrix, one may add an informative prior accordingly (e.g.\ prior with mass at zero on off-diagonal elements for sparse $\bm{B}$). Similar to the node degrees parameter update, a symmetric Normal proposal distribution is used, $b_{mn}^{t^+} \sim \text{Normal}\left(b_{mn}^t, \sigma_B\right)$. With a non-negative constraint, the acceptance probability, $a_{b}$, of the proposed update is:
\begin{equation}
a_b =  \left\{ \begin{array}{cl} \displaystyle\min\left(\frac{p\left(b_{mn}^{t^+} | \bm{B}_{(-mn)}, \bm{\lambda}, \bm{\Pi}, \bm{I}, \bm{A}\right)}{p\left(b_{mn}^t | \bm{B}_{(-mn)}, \bm{\lambda}, \bm{\Pi}, \bm{I}, \bm{A}\right)} , 1 \right)  & \mbox{if $b_{mn}^{t^+} \geq  0$} \\ 0 & \mbox{otherwise} \end{array} \right.
\end{equation}

\paragraph{Update $\bm{\pi}_i^{t+1}$ conditional on $\bm{\Pi}_{(-i)}^{t^+}, \bm{B}^{t+1}, \bm{\lambda}^{t+1}, \bm{I}^t$:}
For the mixed-membership updates on each node $i$, the governing conditional likelihood function, similar to the likelihood for the block matrix is:
\begin{align}
\label{eq:piLikelihood}
\Like(\bm{\pi}_i  | \bm{\Pi}_{(-i)}, \bm{B}, \bm{\lambda}, \bm{I}, \A) = \exp\Bigg( & \sum_{j \in I_{ij}=1}\Big( a_{ij} \log (\bm{\pi}_i^T \bm{B \pi}_j) - {\lambda}_i\lambda_j \ T \ \bm{\pi}_i^T \bm{B \pi}_j \Big) + \nonumber \\
& \sum_{j \in I_{ji}=1}\Big( a_{ji} \log (\bm{\pi}_j^T \bm{B \pi}_i) - \lambda_i\lambda_j \ T \ \bm{\pi}_j^T \bm{B \pi}_i \Big) \Bigg)
\end{align}
And the posterior distribution is:
\begin{equation}
p\left(\bm{\pi}_i | \bm{\Pi}_{(-i)}, \bm{B}, \bm{\lambda}, \bm{I}, \bm{A}\right) \propto \Like\left(\bm{\pi}_i  | \bm{\Pi}_{(-i)}, \bm{B}, \bm{\lambda}, \bm{I}, \A \right) \; p_0(\bm{\pi}_i)
\end{equation}
Without additional information on the different lifestyles of the population, as is typically the case, one can simply use a flat prior $p_0(\bm{\pi}_i) \propto 1$. Because the space for $\bm{\pi}_i$ is a simplex, a logistic-Normal proposal function is used, inspired by the work from \cite{Katz2010}. Let $\bm{\pi}_i^t$ denote the current sample on the mixed-membership of $i$, the proposed update, $\bm{\pi}_i^{t^+}$ is governed by:
\begin{equation}
\bm{\pi}_i^{t^+} \sim \text{logistic}\Bigg(\text{Normal}\Big(\logit\left(\bm{\pi}_i^t \right), \bm{\Sigma_\pib}\Big)\Bigg)
\end{equation}
where the magnitude of a typically diagonal $\bm{\Sigma_\pib}$ controls the step size of the proposal from the current sample. The acceptance probability, $a_{\bm{\pi}}$, of the proposed update is:
\begin{equation}
a_{\bm{\pi}} = \min\left(\frac{p(
\pib_i^{t^+} | \bm{\Pi}_{(-i)}, \bm{B}, \bm{\lambda}, \bm{I}, \bm{A})Q(\pib^t_i; \pib^{t^+}_i)}{p(\pib_i^{t} | \bm{\Pi}_{(-i)}, \bm{B}, \bm{\lambda}, \bm{I}, \bm{A})Q(\pib^{t^+}_i; \pib^t_i)} , 1 \right)
\end{equation}
where $Q(\pib^t_i; \pib^{t^+}_i)$ is the logistic-Normal density function with mean $\pib^{t^+}_i$ and covariance $\bm{\Sigma_\pib}$ evaluated at $\pib^t_i$. The ratio $Q(\pib^t_i; \pib^{t^+}_i) / Q(\pib^{t^+}_i; \pib^{t}_i)$ is the Hasting correction ratio because the logistic-Normal proposal function is not symmetric.

\paragraph{Update $I_{ij}^{t+1}$ conditional on $\bm{\Pi}^{t+1}, \bm{B}^{t+1}, \bm{\lambda}^{t+1}$:}
Lastly, updating the edge switches, $I_{ij}$, is fairly straight forward because each switch is conditionally independent of each other. For edges with observed interactions (i.e.\ $a_{ij} > 0$), $I_{ij}$ must be one. For edges with zero observed interactions (i.e.\ $a_{ij} = 0$), the governing posterior distribution is: 
\begin{align}
p\left(I_{ij} = 1 | \bm{B}, \bm{\lambda}, \bm{\Pi}, \bm{A}\right) &= \frac{p\left(I_{ij}=1 | \bm{B}, \bm{\lambda}, \bm{\Pi}, a_{ij}=0\right) p_0(I_{ij}=1)}{p(a_{ij}=0, I_{ij}=1 |  \bm{B}, \bm{\lambda}, \bm{\Pi}) + p(a_{ij}=0, I_{ij}=0 |  \bm{B}, \bm{\lambda}, \bm{\Pi})} \nonumber \\
&= \frac{\exp(-\lambda_{ij} T) s}{\exp(-\lambda_{ij} T) s + (1-s)}
\end{align}
where $\lambda_{ij}$ is calculated using Equation~\eqref{rateEquation} on the current parameter samples. Updating $I_{ij}^{t+1}$ can simply be done by a Bernoulli draw, $I_{ij}^{t+1} \sim \Bernoulli(\frac{\exp(-\lambda_{ij} \times T) s}{\exp(-\lambda_{ij} \times T) s + (1-s)})$. The sparsity parameter, $s$, can be set to a reasonable value or updated as another step in the MCMC with a closed-form posterior distribution, $s^{t+1} \sim \text{Beta}(\sum_{i,j \in 1:N}{I_{ij}}+1, N^2-\sum_{i,j\in 1:N}{I_{ij}}+1)$. When the network is large, it may become infeasible to update $I_{ij}$ for every edge such that $a_{ij}=0$. A practical implementation is to simply turn all such $I_{ij}$ off which is equivalent to fixing $\bm{I}$ at the maximum posterior estimate. Alternatively, one may selectively update such $I_{ij}$ only on edges likely with low frequency $\lambda_{ij}$.

This completes the description of the MCMC sampling updates for HMMB.

\subsubsection {Alternative: Maximum Posterior Updates (MCEM)}
\label{sec:MCEM}

An alternative to the MCMC procedure is to maximize the posterior probability when updating some of the parameters, instead of sampling from the posterior distribution. This is commonly known as the Monte Carlo Expectation Maximization (MCEM) \citep{Andrieu2003, Wei1990}. MCEM may converge faster due to its greedy steps in maximizing the posterior probability, but does not capture the full posterior distribution and is more prone to being trapped in local modes. Because in this chapter, estimating the posterior distribution of the mixed-memberships is of particular interest, the developed MCEM samples from the mixed-membership posterior distribution, but maximize the posterior probability on the node degree and block matrix parameters. Maximization is done through gradient ascent with appropriate step sizes (i.e.\ greediness) to balance between the speed of convergence and avoiding traps of local maxima. A performance comparison between the MCMC and MCEM procedures is given later in Section~\ref{sec:estimationPerformance} with discussion of results in Table~\ref{coverageTableSimulatedData}.

\paragraph{Alternative update $\lambda_i^{t+1}$ conditional on $\bm{\lambda}_{(-i)}^{t^+}, \bm{\Pi}^t, \bm{B}^t, \bm{I}^t$:}
Maximizing the posterior probability on the node degree parameters, shown in Equation~\eqref{eq:lambdaPosterior}, is done by the following gradient ascent: 
\begin{equation}
\lambda^{t+1}_i = \lambda^{t}_i + \gamma^t_\lambda \Bigg( \sum_{j \in I_{ij}=1}\left(\frac{a_{ij}}{\lambda^t_i} - \lambda^t_j \ T \ \bm{\pi}^T_i \bm{B \pi}_j \right) + \sum_{j \in I_{ji}=1} \left(\frac{a_{ji}}{\lambda^t_i} - \lambda^t_i \ T \ \bm{\pi}^T_j \bm{B \pi}_i \right)  - \frac{\alpha}{\lambda^t_i}\Bigg)
\end{equation}
where $\gamma^t_\lambda$ is the step size, and the gradient is the derivative of the log posterior probability,
 
\noindent $\partial \log \bigl(p\bigl(\lambda_i | \bm{\lambda}_{(-i)}, \bm{\Pi}, \bm{B}, \bm{I}, \bm{A}\bigr)\bigr) / \partial \lambda_i$ evaluated at $\lambda^t_i$. A sensible update strategy is the batch mode where all of the $\lambda_i$ take one gradient ascent step together at a time. This is not only more efficient to compute, but also makes sense for a more balanced climb in the space spanned by $\bm{\lambda}$. 

\paragraph{Alternative update $b_{mn}^{t+1}$ conditional on $\B_{(-mn)}^{t^+}, \bm{\lambda}^{t+1}, \bm{\Pi}^t, \bm{I}^t$:}
Similarly, the posterior probability on the block matrix parameters, shown in Equation~\eqref{eq:BPosterior}, is maximized by:
\begin{equation}
\label{BGradientAscent}
b^{t+1}_{mn} = b^{t}_{mn} + \gamma^t_B \; \sum_{i, j \in I_{ij}=1}\Bigg( \left(\frac{a_{ij}} {\bm{\pi}_i^T \bm{B}^t \bm{\pi}_j} - \lambda_i \lambda_j \ T \right)  \pi_{im}\pi_{jn}\Bigg)
\end{equation}
where $\gamma^t_B$ is the step size, and the gradient is the derivative of the log posterior probability, 

\noindent $\partial \log \bigl(p\bigl(b_{mn} | \bm{B}_{(-mn)}, \bm{\lambda}, \bm{\Pi}, \bm{I}, \bm{A}\bigr)\bigr) / \partial b_{mn}$ evaluated at $b_{mn}^t$. The batch update is again a sensible strategy where all of the $b_{mn}$ are updated together one step at a time.

This completes the description of the proposed MCEM inference procedure for HMMB. Similar to the MCMC, for higher chance of convergence to the global posterior mode, multiple independent MCEM chains can be run then select the one resulting in the highest average joint posterior probability. Similarly, the initializations discussed in Section~\ref{sec:parameterInit} can be used in MCEM. The MCEM implementation proposed here only differs from the proposed MCMC implementation by two update steps.

\subsection{Fisher Information and Performance Bound on Membership}
\label{sec:FisherInfoAndBound}

Given the central role of the community mixed-membership in the HMMB, this section presents a theoretical membership estimation performance bound through Fisher information analysis. This theoretical result reveals what makes certain network data more informative than the others. It also provides the best case estimation performance given a specific network data. For any data set or experiment, it is useful to characterize the information content and the best case performance one can expect. This theoretical result may help guide researchers in collecting enough data and the most informative data, to achieve a certain level of estimation performance. 

The theoretical results presented here consist of the Fisher information matrix, Cram\'{e}r-Rao bound, and the asymptotic posterior distribution of the mixed-membership estimate $\pib_i$ for each node $i$, conditional on the other parameters. This is a best case scenario because in practice the estimates on the other parameters have uncertainty and possibly bias as well, but nevertheless this theoretical result sheds light on the information content of the network data as well as the minimal variance one can expect to achieve in estimating the mixed-memberships. Fisher information analysis of the data on the mixed-membership begins with the log likelihood function:
\begin{align}
\ell(\bm{\pi}_i  | \bm{\Pi}_{(-i)}, \bm{B}, \bm{\lambda}, \bm{I}, \A) = & \sum_{j \in I_{ij}=1}\Big( a_{ij} \log (\bm{\pi}_i^T \bm{B \pi}_j) - \lambda_i \lambda_j \ T \ \bm{\pi}_i^T \bm{B \pi}_j \Big) + \nonumber \\
& \sum_{j \in I_{ji}=1}\Big( a_{ji} \log (\bm{\pi}_j^T \bm{B \pi}_i) - \lambda_i \lambda_j \ T \ \bm{\pi}_j^T \bm{B \pi}_i \Big)
\end{align}

Because $\pib_i$ stays inside a $(K-1)$ simplex, the redundant dimension is removed through the following reparameterization:
\begin{equation}
\pib_i = \left[  \pi_{i,1} \  \ \pi_{i,2} \  \ ... \  \ \pi_{i,K-1} \ \ \ \ 1-\sum_{k=1}^K \pi_{ik}  \right]
\end{equation}
where the $K{\text{th}}$ component is entirely determined by the others. This $K{\text{th}}$ component represents the membership to a reference community $K$. The reference community can be any one of the $K$ communities. The Fisher information matrix on the reparameterized $\pib_i$ is $(K-1) \times (K-1)$ in size, and each of the $mn{\text{th}}$ element is:
\begin{align}
\label{eq:FisherInformationMatrix3}
\FisherInfo(\bm{\pi}_i)_{mn} &= - E_{\A} \left( \frac{\partial^2 \ell(\bm{\pi}_i  | \bm{\Pi}_{(-i)}, \bm{B}, \bm{\lambda}, \bm{I}, \A)}{\partial \pi_{i,m} \partial \pi_{i,n}} \right) \nonumber \\
&= T \Bigg( \sum_{j\in I_{ij}=1} \frac{\lambda_i\lambda_j \left( \bm{B}_{m\cdot}\bm{\pi}_j - \bm{B}_{K\cdot}\bm{\pi}_j \right) \left( \bm{B}_{n\cdot}\bm{\pi}_j - \bm{B}_{K\cdot}\bm{\pi}_j \right)} {\bm{\pi}_i^T \bm{B \pi}_j} \; + \nonumber \\
& \;\;\;\;\;\;\;\;\;\; \sum_{j\in I_{ji}=1} \frac{\lambda_i\lambda_j \left( \bm{\pi}_j^T\bm{B}_{\cdot m} - \bm{\pi}_j^T\bm{B}_{\cdot  K} \right) \left( \bm{\pi}_j^T\bm{B}_{\cdot n} - \bm{\pi}_j ^T\bm{B}_{\cdot K}\right)} {\bm{\pi}_j^T \bm{B \pi}_i} \Bigg)
\end{align}
where $m,n \in 1:(K-1)$. While the precise information content on each membership estimate is captured by the entire Fisher information matrix, each of the $m^\text{th}$ diagonal element of the Fisher information matrix, $\FisherInfo(\bm{\pi}_i)_{mm}$, gives an interpretable first order indication on the information content of the data on node $i$'s membership to community $m$:
\begin{align}
\FisherInfo(\bm{\pi}_i)_{mm} = T \Bigg( & \sum_{j\in I_{ij}=1} \frac{\lambda_i\lambda_j \left( \bm{B}_{m.}\bm{\pi}_j - \bm{B}_{K.}\bm{\pi}_j \right)^2} {\bm{\pi}_i^T \bm{B \pi}_j} \; + \nonumber \\
& \sum_{j\in I_{ji}=1} \frac{\lambda_i\lambda_j \left( \bm{\pi}_j^T \bm{B}_{.m} - \bm{\pi}_j^T \bm{B}_{.K}\right)^2} {\bm{\pi}_j^T \bm{B \pi}_i}\Bigg)
\end{align}
Agreeing with intuition, this quantity shows that information content is proportional to the amount of time, $T$, during which data is collected. It also accumulates over all the incoming and outgoing edges to $i$. The information content of each such edge is proportional to the node degrees product $\lambda_i \lambda_j$ (i.e.\ the Chung-Lu term), and is maximized when either most of the interactions between $i$ and $j$ can be explained through community $m$ or none of such interactions can be explained through community $m$ (i.e.\ making the squared difference term large). Lastly, the information content of each edge is normalized by the blockmodel term. This result is interesting because it reveals the kind of edges (i.e.\ network data) that contributes the most to the information content towards estimating each community membership of a given node.

With the Fisher information matrix, one can specify the asymptotic posterior distribution of the mixed-membership, $\pib_i$, of each node $i$, through the Bernstein-von Mises Theorem \citep{VanDerVaart2000}, assuming standard regularity conditions:

\begin{theorem}[Asymptotic Posterior Distribution of Mixed-Membership]
\label{thm:asymptoticPosteriorOfMixedMembership}
Conditional on the other parameters, the posterior distribution of the mixed-membership $\pib_i$ of node $i$ converges asymptotically, with increasing network data, in distribution to a Normal with mean at the true value and covariance equal to the inverse of the Fisher information matrix derived in Equation~\eqref{eq:FisherInformationMatrix3}: 
\begin{equation}
\bm{\pi_i} \: \vert \: \bm{\Pi}_{(-i)},\bm{\lambda,I,B,\bm{A}} \: \stackrel{D}{\longrightarrow} \: \text{Normal}(\bm{\pi}_i, \FisherInfo(\pib_i)^{-1})
\end{equation}
\end{theorem}

\noindent The inverse of the Fisher information matrix, $\FisherInfo(\pib_i)^{-1}$, is also commonly known as the Cram\'{e}r-Rao lower bound, which gives the minimal covariance of the maximum likelihood estimate. The Bernstein-von Mises Theorem provides a Bayesian interpretation of the Cram\'{e}r-Rao lower bound in Theorem \ref{thm:asymptoticPosteriorOfMixedMembership}. While this is still a best case analysis, conditioning on the other parameters is more justified asymptotically as the parameter estimates converge toward their true values with an increasing amount of data. This result is useful for characterizing how well one may do in inferring the mixed-membership parameters given the amount of network data available. Practically, this informs researchers on how much data to collect in order to meet a certain desired estimation performance.

\subsection{Parameter Estimation Performance Characterization}
\label{sec:estimationPerformance}

This section characterizes the performance of the estimation procedure described above, on networks simulated over a range of reasonable and realistic values on the key model parameters. Overall, the proposed procedure with clever parameter initialization performs adequately across a reasonable range of parameter settings. This demonstrates empirically the parameter identifiability of the HMMB. In challenging settings where the community structure is no longer distinct due to strong between-community interactions or highly mixed memberships, the procedure's performance drops off gracefully and remains adequate, highlighting the strength of inference with HMMB for networks with highly overlapping communities. 

The estimation procedure is evaluated across the four main network features as they are varied by sweeping through the corresponding model parameters. As described in Section~\ref{sec:HMMB} and summarized in Figure~\ref{fig:blockDiagram}, the HMMB represents a diverse range of networks by varying these four features: 1.) strength of between-community interactions governed by the block matrix $\bm{B}$, 2.) level of mixed-memberships governed by the mixed-membership pseudocount matrix $\bm{X}$, 3.) number of high degree nodes governed by the Power-Law exponent $\alpha$, and 4.) network density governed by the sparsity parameter $s$. Figure~\ref{fig:networkLayout} provides visualization of network samples across these features. Nodes are sized by their degrees and colored by their mixed-memberships to four communities each color-coded by red, green, blue, and black (e.g.\ a dark purple node has membership in the red, blue, and black communities).  Figure~\ref{fig:baselineNetwork} shows the baseline setting with moderate levels of between-community interaction and mixed-memberships, a small fraction of high degree nodes, and medium network density. Like many real-world networks, although there is significant overlap between the communities, each of the four communities still has a distinct structure observable in its own cluster of interactions.  Figure~\ref{fig:strongInterCommunityInteraction} shows a network sample with strong between-community interactions. Due to the large number of interactions between the communities, each community no longer has a distinct structure and the network looks closer to one big cluster of interactions. Similarly,  Figure~\ref{fig:highlyMixedMembership} shows a network sample that has very little distinct community structure due to a high level of mixed-memberships, which can be seen in the mixed-colors of most nodes. Figure~\ref{fig:sparseNetwork} shows a network sample with low network density (i.e.\ sparse) where each node interacts with much fewer nodes. For brevity, a visualization with more high degree nodes is skipped here, because it looks just like the baseline case with more high degree nodes.

\begin{figure}
    \centering
    \begin{subfigure}[b]{0.5\textwidth}
        \includegraphics[width=\textwidth]{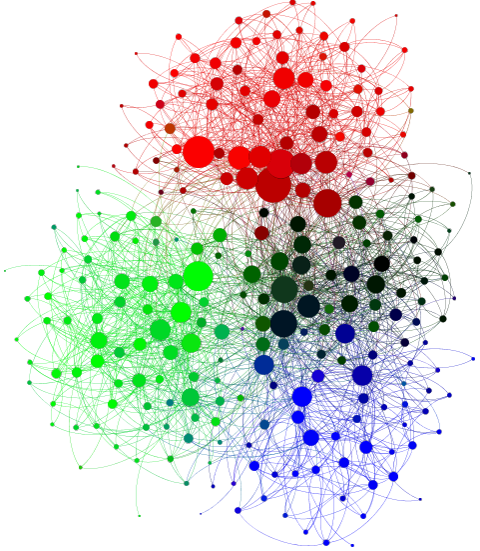}
        \caption{Baseline}
        \label{fig:baselineNetwork}
    \end{subfigure}
    ~ 
    \begin{subfigure}[b]{0.47\textwidth}
        \includegraphics[width=\textwidth]{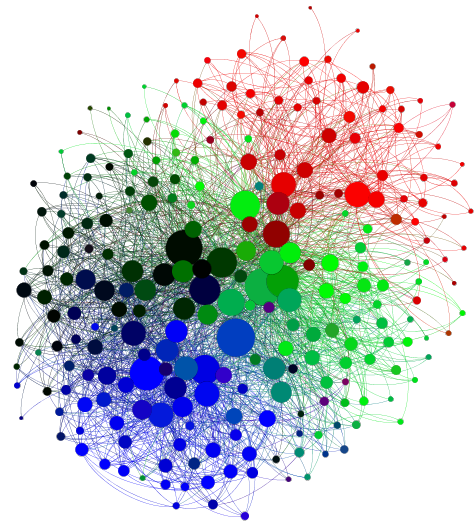}
        \caption{Strong between-community interactions}
        \label{fig:strongInterCommunityInteraction}
    \end{subfigure}
    \begin{subfigure}[b]{0.47\textwidth}
        \includegraphics[width=\textwidth]{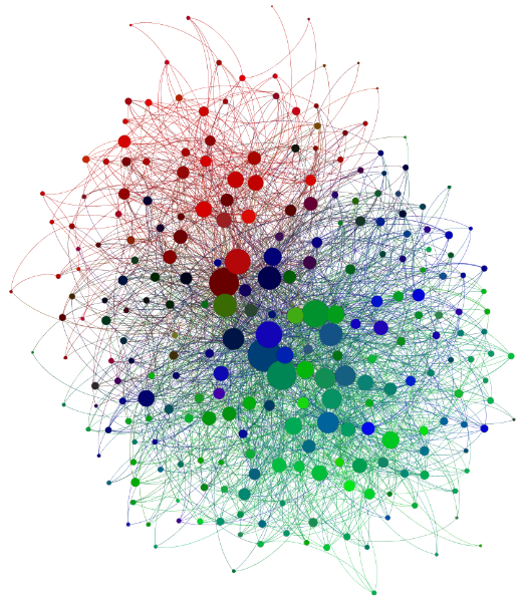}
        \caption{Highly mixed memberships}
        \label{fig:highlyMixedMembership}
    \end{subfigure}
    \begin{subfigure}[b]{0.47\textwidth}
        \includegraphics[width=\textwidth]{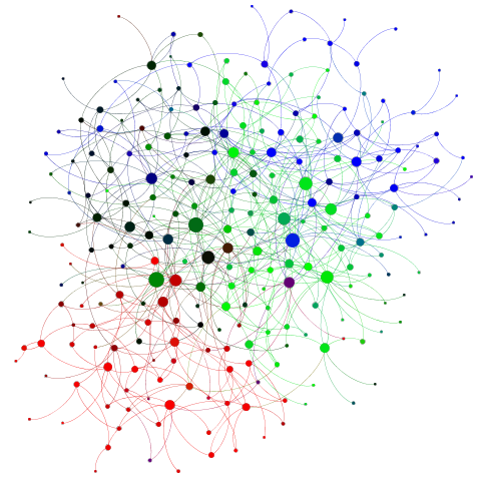}
        \caption{Sparse}
        \label{fig:sparseNetwork}
    \end{subfigure}
    \caption[Networks layouts with varying levels of between-community interactions, mixed-memberships, and sparsity] {Networks layouts with varying levels of between-community interactions, mixed-memberships, and sparsity. The node color represents the mixed-memberships to four communities color-coded in red, green, blue, and black. Larger nodes are those with higher degrees (i.e.\ more interactions)}
    \label{fig:networkLayout}
\end{figure}

A typical result of the Bayesian estimation procedure is the posterior interval that covers $p\%$ of the posterior distribution. A desirable posterior interval captures the true value of the parameter $p\%$ of the time in the narrowest interval possible. Therefore, the common performance evaluation criteria are their frequentist coverage properties over repeated experiments, specifically truth coverage probability by the posterior interval and the interval width \citep{Gelman2014}. For each network setting, the results in this section are reported in these two measures of performance averaged over $25$ independent simulations where a new network is drawn each time. The parameter estimates of interest are the mixed-memberships, $\bm{\Pi}$, the node degrees (i.e.\ individual activity levels), $\bm{\lambda}$, and the block matrix (i.e.\ strength of community interaction), $\bm{B}$. For brevity, results are reported as the average over all the nodes and communities. As discussed in Section~\ref{sec:parameterInit}, there exists a arbitrary scaling between the diagonal of the block matrix and the node degree parameters, which is dealt with by rescaling them to match the truth before evaluation. Because the block matrix diagonal is scaled according to the truth, it is not included in the evaluation. 

\renewcommand{\tabcolsep}{3pt}
\renewcommand{\arraystretch}{1.6}
\begin{table*}[t]
\begin{centering}
\setlength{\belowcaptionskip}{6pt}
\caption[Estimation performance analysis with truth coverage]{Truth coverage of the $90\%$ posterior interval and the interval width in parentheses}
\scalebox{0.9}{
\hspace*{0.0cm}
\begin{tabular}{|cc|c|c|c|c|l}
\cline{1-6}
 \multicolumn{2}{|c|}{} & \multicolumn{4}{c|}{Estimation strategy} \\  
\multicolumn{1}{| p{2.3cm}} {\centering \phantom \newline Network setting} &  \multicolumn{1}{p{2cm}|}{\phantom \newline \centering Parameter of interest} & \multicolumn{1}{p{3cm}|}{\centering MCMC with full init strategy} & \multicolumn{1}{p{3cm}|}{\centering MCMC with parameter init without fixing block diag} & \multicolumn{1}{p{3cm}|}{\centering MCMC with no init strategy} & \multicolumn{1}{p{3cm}|} {\centering MCEM with full init strategy} \\ \cline{1-6}
\multicolumn{1}{|c|}{\multirow{3}{60pt}{\centering Baseline network}} &
\multicolumn{1}{c|}{$\bm{\Pi}$} & 83.0\% (0.089) & 69.4\% (0.098) &  57.7\% (0.171) & 82.3\% (0.082) &    \\ \cline{2-6}
\multicolumn{1}{|c|}{}                        &
\multicolumn{1}{c|}{$\bm{\lambda}$} & 85.9\% (0.046) & 92.4\% (0.095) & 51.5\% (0.148) & - &  \\ \cline{2-6}
\multicolumn{1}{|c|}{}                        &
\multicolumn{1}{c|}{$\bm{B}$} & 70.7\% (0.049) & 67.0\% (0.026) & 62.3\% (0.007) & - &   \\ \cline{1-6}

\multicolumn{1}{|c|}{\multirow{3}{60pt}{\centering Strong between-community interactions}} &
\multicolumn{1}{c|}{$\bm{\Pi}$} &  78.4\% (0.102) & 71.0\% (0.112) &  61.1\% (0.148) & 74.8\% (0.095) &    \\ \cline{2-6}
\multicolumn{1}{|c|}{}                        &
\multicolumn{1}{c|}{$\bm{\lambda}$} & 84.0\% (0.043) & 91.5\% (0.095) & 42.7\% (0.118) & - &    \\ \cline{2-6}
\multicolumn{1}{|c|}{}                        &
\multicolumn{1}{c|}{$\bm{B}$} & 74.0\% (0.066) & 66.7\% (0.056) & 64.7\% (0.013) & - &   \\ \cline{1-6}
\end{tabular}
\label{coverageTableSimulatedData}
}
\end{centering}
\end{table*}

The first result is shown in Table~\ref{coverageTableSimulatedData}, a comparison between variants of the proposed estimation procedure, on the baseline network setting and a more challenging setting with strong between-community interactions. The variants include the MCMC procedure described in Section~\ref{sec:MCMC} with no initialization strategy, partial initialization strategy (without fixing the block matrix diagonal), and full initialization strategy as described in Section~\ref{sec:parameterInit}. The MCEM procedure described in Section~\ref{sec:MCEM} is also included as a comparison against the MCMC procedure. The baseline network setting, with a sample visualization in Figure \ref{fig:baselineNetwork}, has $256$ nodes, four communities, $20\%$ density (i.e.\ $s=0.2$), a Power-Law exponent ($\alpha$) of $2.9$, a time span of $100$ units, a block matrix with moderate between-community interaction, $\bm{B} = \left[\begin{smallmatrix} 2.3&0.07&0&0\\ 0.3&2&0&0\\ 0&0&2.5&0.4 \\ 0&0.3&0&3\end{smallmatrix}\right]$, a pseudocount matrix with moderate mixed-memberships, $\bm{X} = \left[\begin{smallmatrix} 5&0.1&0.1&1\\ 0.1&5&1&0.1\\ 0.1&0.1&2&0.5 \\ 0.1&1&0.3&3\end{smallmatrix}\right]$, with each node belonging to one of the four lifestyles (i.e.\ the four rows in $\bm{X}$) with equal probability, $\bm{\phi} = \left[0.25, \; 0.25, \; 0.25, \; 0.25\right]$. The model parameter values under this baseline setting are picked to be reasonable for real-world networks. The more challenging setting with strong between-community interactions has the same parameter values as the baseline setting, except the block matrix now has off-diagonal entries three times as large. A sample network under this setting is visualized in Figure~\ref{fig:strongInterCommunityInteraction}.

The first three columns of results in Table~\ref{coverageTableSimulatedData} show that the full initialization strategy described in Section~\ref{sec:parameterInit} improves the estimation performance, evident in the better truth coverage and generally narrower interval. The clever parameter initialization started the MCMC in a more feasible region of the large parameter space, increasing the likelihood of converging upon the true posterior mode. Fixing the block matrix diagonal to a precomputed value effectively addressed the identifiability issue from the arbitrary scale between the block matrix diagonal and the node degrees. The only exception appears to be the coverage for $\bm{\lambda}$ under partial initialization strategy, which is a few percentage points better than the coverage under full initialization. However, this is at the expense of doubling the posterior interval width. Without fixing the block matrix diagonal, the MCMC visits a wider range of values for $\bm{\lambda}$ due to the arbitrary scale, resulting in a wide posterior interval that more likely covers the truth. This does not make the posterior interval for $\bm{\lambda}$ better under partial initialization. 

Comparing the first and the last columns in the result shows that the MCEM procedure performs roughly on par with the MCMC procedure, with slightly lower coverage and narrower intervals. This is not surprising due to the more greedy nature of the MCEM procedure in maximizing the posterior probability in updating $\bm{\lambda}$ and $\bm{B}$. As a result, MCEM does not render posterior distributions on $\bm{\lambda}$ and $\bm{B}$, which is why the results on those parameters is left blank under MCEM. Under the baseline setting, the MCMC and the MCEM procedures found the posterior mode in roughly the same number of iterations ($100$ and $97$ iterations on average), so there does not seem to be a gain in the speed of convergence for MCEM. Note that in both procedures, more iterations are needed after reaching the posterior mode for the samples to mix well enough to capture the posterior distribution, while the number of iterations to reach posterior mode simply gives an indication on the convergence speed. Under the more challenging setting with strong between-community interactions, MCMC took on average $22\%$ more iterations to reach the posterior mode than MCEM ($168$ and $138$ iterations), so in more challenging settings, the MCEM procedure seems to converge faster than the MCMC procedure, although the difference may not justify sacrificing the nice property of capturing the full posterior distribution.

As the MCMC procedure with full initialization strategy provides the best performance and posterior intervals on all the parameters, it will be the default estimation procedure for the remaining results going forward. Focusing on the first column of the result shows that the $90\%$ posterior interval slightly under-covers the true $\bm{\Pi}$ and $\bm{\lambda}$ by a few percentage points in the baseline setting. This is caused by fixing the block matrix diagonal to pre-computed values which have some errors from the true values. This is a small price to pay for getting around the arbitrary scale between the block matrix diagonal and $\bm{\lambda}$. Coverage on $\bm{B}$ is worse but reasonable, again, due to the error on the fixed diagonal. In the more challenging setting under strong between-community interactions, the coverage on $\bm{\Pi}$ and $\bm{\lambda}$ drop slightly. This shows that the estimation procedure functions reasonably well even when the community structure is no longer coherent (see Figure~\ref{fig:strongInterCommunityInteraction}), highlighting the strength of modeling and inference under HMMB for highly overlapping communities. Moreover, the mixed-membership estimates nearly achieve the precision of the theoretical lower bound derived in Section~\ref{sec:FisherInfoAndBound}. The Cram\'{e}r-Rao bound for the baseline setting here gives a lower bound on the posterior interval to have on average a minimum width of $0.075$ whereas the estimated posterior interval has an average width of $0.089$. For the setting with strong between-community interactions, the Cram\'{e}r-Rao bound gives on average a minimum posterior width of $0.095$, and the estimated posterior interval a width of $0.102$. The closeness in reaching the bound indicates the estimation procedure is fairly efficient statistically. In reality, this bound may not be reached by any unbiased estimator, because it is derived conditional on all other parameters (see Theorem \ref{thm:asymptoticPosteriorOfMixedMembership}) so the uncertainties of the other parameters are not accounted for, making this bound a best-case analysis.

\begin{figure}
    \centering
    \begin{subfigure}[b]{0.6\textwidth}
        \includegraphics[width=\textwidth]{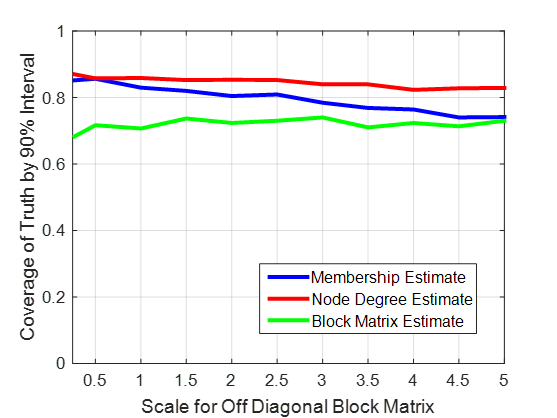}
        \caption{Truth coverage}
        \label{fig:interCommunityInteractionCoveragePlot}
    \end{subfigure}
    ~ 
    \begin{subfigure}[b]{0.6\textwidth}
        \includegraphics[width=\textwidth]{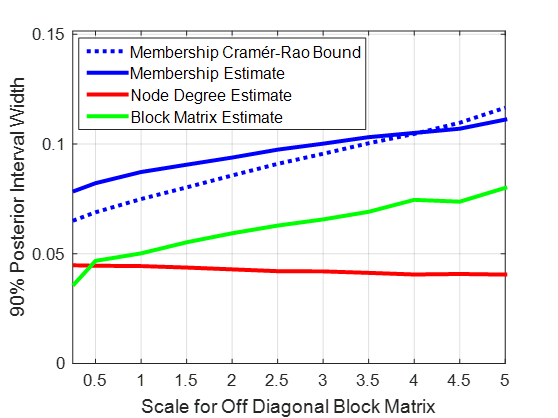}
        \caption{Posterior interval width}
        \label{fig:interCommunityInteractionWidthPlot}
    \end{subfigure}
    \caption{Estimation performance with varying strength of between-community interactions.}
    \label{fig:interCommunityInteractionPerformance}
\end{figure}

Having established that the MCMC procedure with full initialization performs reasonably well under the baseline setting, we now evaluate its performance as the four main features of the HMMB are varied. Figure~\ref{fig:interCommunityInteractionPerformance} shows the estimation performance as the scale of the block matrix off-diagonal entries varies from $0.25$ to five. This scale corresponds to the strength of between-community interactions, with a scale of one being the baseline setting in Table~\ref{coverageTableSimulatedData} and Figure~\ref{fig:baselineNetwork}, and a scale of three being the challenging setting in Table~\ref{coverageTableSimulatedData} and Figure~\ref{fig:strongInterCommunityInteraction}. As the scale increases, estimation becomes more challenging as the network loses more and more of its community structure and eventually becomes one big cluster of interactions. Estimation is particularly difficult in this setting, because without a coherent community structure in the interactions, it is difficult for an estimator to recover the communities and each node's memberships to them. There can be many ways to explain the observed network, making the posterior distribution multi-modal. However, the estimation performance drops off gracefully as seen in Figure~\ref{fig:interCommunityInteractionCoveragePlot}, with reasonable truth coverage even with very strong between-community interactions. This highlights the strength of modeling and inference under HMMB for highly overlapping communities. The posterior interval widths for $\bm{\Pi}$ and $\bm{\lambda}$ also increase gracefully as shown in Figure~\ref{fig:interCommunityInteractionWidthPlot}. The node degree posterior interval becomes more narrow due to more total interactions overall. It is worth noting that the mixed-membership interval width stays slightly above the Cram\'{e}r-Rao lower bound and eventually falls below the bound at $\text{scale}=4$. This is likely due to biased mixed-membership estimates as the community structure loses coherence. In statistical inference, more bias in the estimates often leads to a reduction in variance, known as the bias-variance trade-off \citep{Blitzstein2014}. The Cram\'{e}r-Rao lower bound rises with increasingly strong between-community interactions. This is due to a decrease in the Fisher information content of the interactions because they can be explained through multiple communities. This matches the theoretical result in Equation~\eqref{eq:FisherInformationMatrix3} and the intuition discussed below it.  

\begin{figure}
    \centering
    \begin{subfigure}[b]{0.6\textwidth}
        \includegraphics[width=\textwidth]{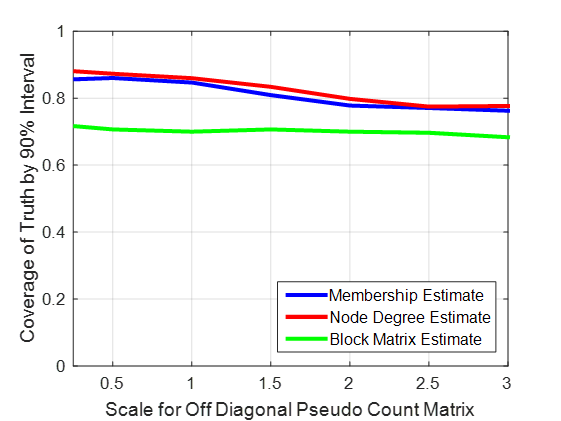}
        \caption{Truth coverage}
        \label{fig:mixedMembershipCoveragePlot}
    \end{subfigure}
    ~ 
    \begin{subfigure}[b]{0.6\textwidth}
        \includegraphics[width=\textwidth]{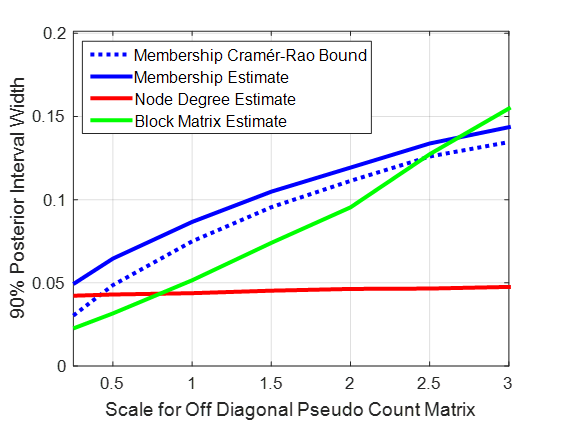}
        \caption{Posterior interval width}
        \label{fig:mixedMembershipWidthPlot}
    \end{subfigure}
    \caption{Estimation performance with varying level of mixed-memberships.}
    \label{fig:mixedMembershipPerformance}
\end{figure}

Going onto the next network feature, Figure~\ref{fig:mixedMembershipPerformance} shows the estimation performance as the scale of the pseudocount matrix ($\bm{X}$) off-diagonal entries varies from $0.25$ to $3$. This scale corresponds to the level of mixed-memberships, with a scale of one being the baseline setting shown in Figure~\ref{fig:baselineNetwork}, and a scale of three being the highly mixed memberships setting shown in Figure~\ref{fig:highlyMixedMembership}. The results here are very similar to the previous case on varying the strength of between-community interactions. As the scale increases, estimation becomes more challenging and the Fisher information content decreases, for all the same reasons as the previous case. Again, the estimation performance drops off gracefully and the mixed-membership interval width tracks the Cram\'{e}r-Rao lower bound nicely, indicating the estimation procedure to be statistically efficient.

\begin{figure}
    \centering
    \begin{subfigure}[b]{0.6\textwidth}
        \includegraphics[width=\textwidth]{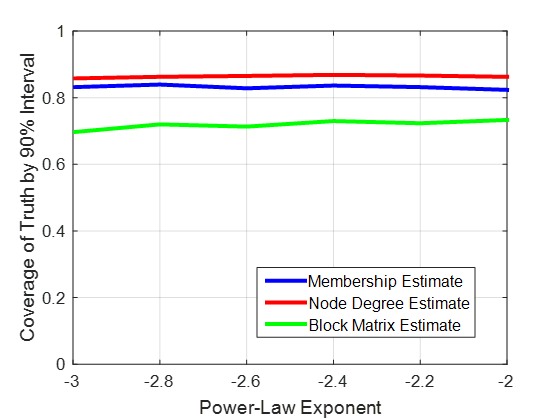}
        \caption{Truth coverage}
        \label{fig:powerLawExponentCoveragePlot}
    \end{subfigure}
    ~ 
    \begin{subfigure}[b]{0.6\textwidth}
        \includegraphics[width=\textwidth]{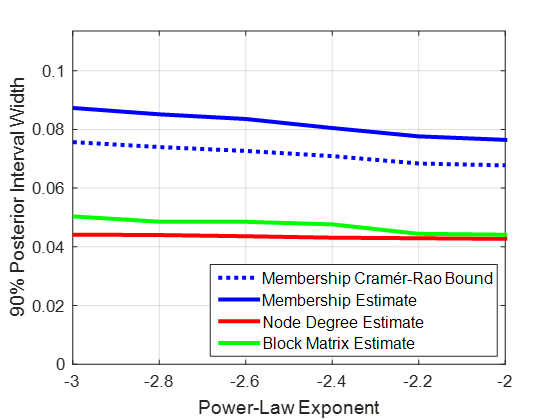}
        \caption{Posterior interval width}
        \label{fig:powerLawExponentWidthPlot}
    \end{subfigure}
    \caption{Estimation performance with varying number of high-degree nodes.}
    \label{fig:powerLawExponentPerformance}
\end{figure}

The next network feature to be varied is the number of high degree nodes. Figure~\ref{fig:powerLawExponentPerformance} shows the estimation performance as the negative Power-Law exponent ($-\alpha$) varies from $-3$ to $-2$, which is the range of values observed in real-world networks \citep{Clauset2009}. Less negative exponent leads to a higher number of high degree nodes, as the Power-Law distribution decays more slowly. Truth coverage remains steady and the posterior interval widths decrease gradually. Again, the mixed-membership interval width tracks the Cram\'{e}r-Rao lower bound nicely, indicating statistical efficiency. The decreasing Cram\'{e}r-Rao lower bound matches the theoretical result in Equation~\eqref{eq:FisherInformationMatrix3} which indicates that an increase of high degree nodes leads to higher Fisher information content.  

\begin{figure}
    \centering
    \begin{subfigure}[b]{0.6\textwidth}
        \includegraphics[width=\textwidth]{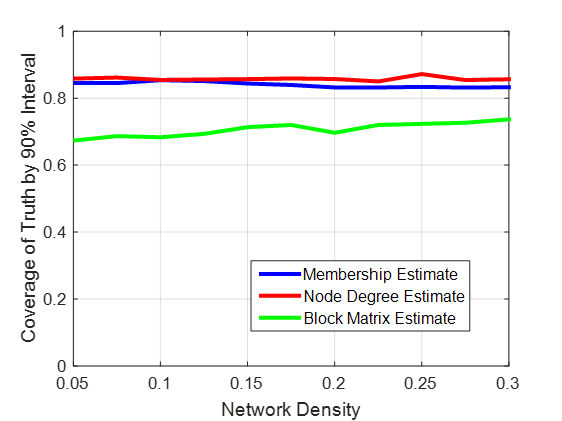}
        \caption{Truth coverage}
        \label{fig:sparsityCoveragePlot}
    \end{subfigure}
    ~ 
    \begin{subfigure}[b]{0.6\textwidth}
        \includegraphics[width=\textwidth]{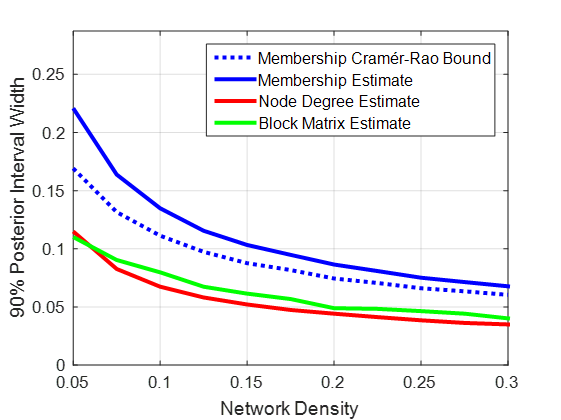}
        \caption{Posterior interval width}
        \label{fig:sparsityWidthPlot}
    \end{subfigure}
    \caption{Estimation performance with varying network density (i.e.\ sparsity).}
    \label{fig:sparsityPerformance}
\end{figure}

The last network feature to be varied is the network density. Figure~\ref{fig:sparsityPerformance} shows the estimation performance as the network density (i.e. the sparsity parameter $s$) varies from $0.05$ to $0.3$. A visualization of a sparse network with $s=0.05$ is shown in Figure~\ref{fig:sparseNetwork}. Truth coverage remains steady and the posterior interval widths decrease with increasing network density. Higher density increases the total number of interacting edges, making estimation easier with higher Fisher information content as shown in Equation~\eqref{eq:FisherInformationMatrix3}. Accordingly, the mixed-membership interval width gets closer to the decreasing Cram\'{e}r-Rao lower bound with higher network density.

Overall, the proposed MCMC procedure with full initialization strategy is able to perform adequately across the ranges of reasonable network settings.

\section{Conclusion}
This chapter presents a rich hybrid network model, the hybrid mixed-membership blockmodels (HMMB), that incorporates several key characteristics of realistic networks: community-based interactions with mixed-membership, varying node degree, and sparsity. The proposed Bayesian inference procedure is shown to perform adequately in estimating and identifying the model parameters as the main network features are varied under simulation, even in the presence of significant community overlaps. These parameter estimates reveal the key individual and network attributes that should be conditioned on for the Bayesian imputation of missing network potential outcomes, in accordance with Theorem \ref{thm:unconfoundedInfluenceNetworkByConditioning}. In practice, these parameter estimates are included as covariates of the network potential outcome models as a critical part of the causal framework proposed in this thesis.



\begin{singlespacing}
  \renewcommand{\bibname}{References}

  \bibliographystyle{ecca}
  \bibliography{references_thesis}
\end{singlespacing}


\end{document}